\documentclass[longauth]{aa}

\usepackage{graphicx}
\usepackage{natbib}
\usepackage{scalerel}
\usepackage{csquotes}
\usepackage{lscape}
\usepackage{multirow}

\usepackage[table]{xcolor}

\usepackage{txfonts}
\usepackage[pdfencoding=auto,psdextra]{hyperref}
\hypersetup{
    colorlinks=true,
    linkcolor=blue,
    filecolor=magenta,      
    urlcolor=blue,
    citecolor=blue,
    breaklinks=true
}
\makeatletter
\renewcommand\aa@pageof{, page \thepage{} of \pageref*{LastPage}}
\makeatother

\usepackage[utf8]{inputenc}

\usepackage[switch, modulo]{lineno}
\linenumbers

\usepackage{orcidlink}

\usepackage{euclid}

\def\Planck{\textit{Planck}}
\def\Herschel{\textit{Herschel}}
\def\Spitzer{\textit{Spitzer}}
\newcommand{\zphz}{\ensuremath{z_{\rm\scriptscriptstyle\tt PHZ }}}
\newcommand{\znnpz}{\ensuremath{z_{\rm\scriptscriptstyle\tt NNPZ }}}

\newcommand{\Mh}[0]{M_{\rm h}}
\newcommand{\pd}{\phantom{0}}
\newcommand{\pa}{\phantom{a}}

\defcitealias{behroozi2013}{B13}
\defcitealias{legrand2019}{L19}
\defcitealias{shuntov2022}{S22}

\begin{document}

\title{Euclid Quick Data Release (Q1)}
\subtitle{The first \Euclid view of \Planck\ galaxy protocluster candidates at cosmic noon}    

\newcommand{\orcid}[1]{}

\author{Euclid Collaboration: T.~Dusserre\thanks{\email{tanguy.dusserre@universite-paris-saclay.fr}}\inst{\ref{aff1},\ref{aff184}}
\and H.~Dole\orcid{0000-0002-9767-3839}\thanks{\email{Herve.Dole@universite-paris-saclay.fr}}\inst{\ref{aff1}}
\and F.~Sarron\orcid{0000-0001-8376-0360}\inst{\ref{aff2},\ref{aff3}}
\and G.~Castignani\orcid{0000-0001-6831-0687}\inst{\ref{aff4}}
\and N.~Ramos-Chernenko\inst{\ref{aff5},\ref{aff6}}
\and N.~Mai\inst{\ref{aff86},\ref{aff87}}
\and M.~Langer\orcid{0000-0002-9088-2718}\inst{\ref{aff78},\ref{aff1}}
\and N.~Aghanim\orcid{0000-0002-6688-8992}\inst{\ref{aff1}}
\and A.~Garic\inst{\ref{aff185},\ref{aff1}}
\and I.-E.~Mellouki\inst{\ref{aff1}}
\and N.~Dagoneau\orcid{0000-0002-1361-2562}\inst{\ref{aff7}}
\and O.~Chapuis\orcid{0000-0003-0632-3700}\inst{\ref{aff8}}
\and B.~L.~Frye\orcid{0000-0003-1625-8009}\inst{\ref{aff9}}
\and M.~Polletta\orcid{0000-0001-7411-5386}\inst{\ref{aff10},\ref{aff11}}
\and H.~Dannerbauer\orcid{0000-0001-7147-3575}\inst{\ref{aff6}}
\and L.~Maurin\orcid{0000-0002-8406-0857}\inst{\ref{aff1}}
\and E.~Soubrie\orcid{0000-0001-9295-1863}\inst{\ref{aff1}}
\and A.~Biviano\orcid{0000-0002-0857-0732}\inst{\ref{aff12},\ref{aff13}}
\and A.~Enia\orcid{0000-0002-0200-2857}\inst{\ref{aff19},\ref{aff4}}
\and F.~Gentile\orcid{}\inst{\ref{aff4},\ref{aff147}}
\and E.~Daddi\orcid{}\inst{\ref{aff99}}
\and S.~Mei\orcid{0000-0002-2849-559X}\inst{\ref{aff86},\ref{aff87}}
\and J.~Pérez-Martinez\inst{\ref{aff5},\ref{aff6}}
\and B.~Altieri\orcid{0000-0003-3936-0284}\inst{\ref{aff14}}
\and A.~Amara\inst{\ref{aff15}}
\and S.~Andreon\orcid{0000-0002-2041-8784}\inst{\ref{aff16}}
\and N.~Auricchio\orcid{0000-0003-4444-8651}\inst{\ref{aff4}}
\and C.~Baccigalupi\orcid{0000-0002-8211-1630}\inst{\ref{aff13},\ref{aff12},\ref{aff17},\ref{aff18}}
\and M.~Baldi\orcid{0000-0003-4145-1943}\inst{\ref{aff19},\ref{aff4},\ref{aff20}}
\and A.~Balestra\orcid{0000-0002-6967-261X}\inst{\ref{aff21}}
\and S.~Bardelli\orcid{0000-0002-8900-0298}\inst{\ref{aff4}}
\and P.~Battaglia\orcid{0000-0002-7337-5909}\inst{\ref{aff4}}
\and A.~Bonchi\orcid{0000-0002-2667-5482}\inst{\ref{aff22}}
\and D.~Bonino\orcid{0000-0002-3336-9977}\inst{\ref{aff23}}
\and E.~Branchini\orcid{0000-0002-0808-6908}\inst{\ref{aff24},\ref{aff25},\ref{aff16}}
\and M.~Brescia\orcid{0000-0001-9506-5680}\inst{\ref{aff26},\ref{aff27}}
\and J.~Brinchmann\orcid{0000-0003-4359-8797}\inst{\ref{aff28},\ref{aff29}}
\and S.~Camera\orcid{0000-0003-3399-3574}\inst{\ref{aff30},\ref{aff31},\ref{aff23}}
\and V.~Capobianco\orcid{0000-0002-3309-7692}\inst{\ref{aff23}}
\and C.~Carbone\orcid{0000-0003-0125-3563}\inst{\ref{aff10}}
\and J.~Carretero\orcid{0000-0002-3130-0204}\inst{\ref{aff32},\ref{aff33}}
\and S.~Casas\orcid{0000-0002-4751-5138}\inst{\ref{aff34}}
\and M.~Castellano\orcid{0000-0001-9875-8263}\inst{\ref{aff35}}
\and S.~Cavuoti\orcid{0000-0002-3787-4196}\inst{\ref{aff27},\ref{aff36}}
\and K.~C.~Chambers\orcid{0000-0001-6965-7789}\inst{\ref{aff37}}
\and A.~Cimatti\inst{\ref{aff38}}
\and C.~Colodro-Conde\inst{\ref{aff5}}
\and G.~Congedo\orcid{0000-0003-2508-0046}\inst{\ref{aff39}}
\and C.~J.~Conselice\orcid{0000-0003-1949-7638}\inst{\ref{aff40}}
\and L.~Conversi\orcid{0000-0002-6710-8476}\inst{\ref{aff41},\ref{aff14}}
\and Y.~Copin\orcid{0000-0002-5317-7518}\inst{\ref{aff42}}
\and A.~Costille\inst{\ref{aff43}}
\and F.~Courbin\orcid{0000-0003-0758-6510}\inst{\ref{aff44},\ref{aff45}}
\and H.~M.~Courtois\orcid{0000-0003-0509-1776}\inst{\ref{aff46}}
\and M.~Cropper\orcid{0000-0003-4571-9468}\inst{\ref{aff47}}
\and A.~Da~Silva\orcid{0000-0002-6385-1609}\inst{\ref{aff48},\ref{aff49}}
\and H.~Degaudenzi\orcid{0000-0002-5887-6799}\inst{\ref{aff50}}
\and G.~De~Lucia\orcid{0000-0002-6220-9104}\inst{\ref{aff12}}
\and A.~M.~Di~Giorgio\orcid{0000-0002-4767-2360}\inst{\ref{aff51}}
\and C.~Dolding\orcid{0009-0003-7199-6108}\inst{\ref{aff47}}
\and F.~Dubath\orcid{0000-0002-6533-2810}\inst{\ref{aff50}}
\and C.~A.~J.~Duncan\orcid{0009-0003-3573-0791}\inst{\ref{aff40}}
\and X.~Dupac\inst{\ref{aff14}}
\and S.~Dusini\orcid{0000-0002-1128-0664}\inst{\ref{aff52}}
\and A.~Ealet\orcid{0000-0003-3070-014X}\inst{\ref{aff42}}
\and S.~Escoffier\orcid{0000-0002-2847-7498}\inst{\ref{aff53}}
\and M.~Farina\orcid{0000-0002-3089-7846}\inst{\ref{aff51}}
\and R.~Farinelli\inst{\ref{aff4}}
\and F.~Faustini\orcid{0000-0001-6274-5145}\inst{\ref{aff22},\ref{aff35}}
\and S.~Ferriol\inst{\ref{aff42}}
\and F.~Finelli\orcid{0000-0002-6694-3269}\inst{\ref{aff4},\ref{aff54}}
\and P.~Fosalba\orcid{0000-0002-1510-5214}\inst{\ref{aff55},\ref{aff56}}
\and S.~Fotopoulou\orcid{0000-0002-9686-254X}\inst{\ref{aff57}}
\and M.~Frailis\orcid{0000-0002-7400-2135}\inst{\ref{aff12}}
\and E.~Franceschi\orcid{0000-0002-0585-6591}\inst{\ref{aff4}}
\and M.~Fumana\orcid{0000-0001-6787-5950}\inst{\ref{aff10}}
\and S.~Galeotta\orcid{0000-0002-3748-5115}\inst{\ref{aff12}}
\and K.~George\orcid{0000-0002-1734-8455}\inst{\ref{aff58}}
\and B.~Gillis\orcid{0000-0002-4478-1270}\inst{\ref{aff39}}
\and C.~Giocoli\orcid{0000-0002-9590-7961}\inst{\ref{aff4},\ref{aff20}}
\and P.~G\'omez-Alvarez\orcid{0000-0002-8594-5358}\inst{\ref{aff59},\ref{aff14}}
\and J.~Gracia-Carpio\inst{\ref{aff60}}
\and A.~Grazian\orcid{0000-0002-5688-0663}\inst{\ref{aff21}}
\and F.~Grupp\inst{\ref{aff60},\ref{aff58}}
\and L.~Guzzo\orcid{0000-0001-8264-5192}\inst{\ref{aff61},\ref{aff16},\ref{aff62}}
\and S.~Gwyn\orcid{0000-0001-8221-8406}\inst{\ref{aff63}}
\and S.~V.~H.~Haugan\orcid{0000-0001-9648-7260}\inst{\ref{aff64}}
\and J.~Hoar\inst{\ref{aff14}}
\and W.~Holmes\inst{\ref{aff65}}
\and F.~Hormuth\inst{\ref{aff66}}
\and A.~Hornstrup\orcid{0000-0002-3363-0936}\inst{\ref{aff67},\ref{aff68}}
\and P.~Hudelot\inst{\ref{aff69}}
\and K.~Jahnke\orcid{0000-0003-3804-2137}\inst{\ref{aff70}}
\and M.~Jhabvala\inst{\ref{aff71}}
\and E.~Keih\"anen\orcid{0000-0003-1804-7715}\inst{\ref{aff72}}
\and S.~Kermiche\orcid{0000-0002-0302-5735}\inst{\ref{aff53}}
\and A.~Kiessling\orcid{0000-0002-2590-1273}\inst{\ref{aff65}}
\and B.~Kubik\orcid{0009-0006-5823-4880}\inst{\ref{aff42}}
\and K.~Kuijken\orcid{0000-0002-3827-0175}\inst{\ref{aff73}}
\and M.~K\"ummel\orcid{0000-0003-2791-2117}\inst{\ref{aff58}}
\and M.~Kunz\orcid{0000-0002-3052-7394}\inst{\ref{aff74}}
\and H.~Kurki-Suonio\orcid{0000-0002-4618-3063}\inst{\ref{aff75},\ref{aff76}}
\and Q.~Le~Boulc'h\inst{\ref{aff77}}
\and A.~M.~C.~Le~Brun\orcid{0000-0002-0936-4594}\inst{\ref{aff78}}
\and D.~Le~Mignant\orcid{0000-0002-5339-5515}\inst{\ref{aff43}}
\and S.~Ligori\orcid{0000-0003-4172-4606}\inst{\ref{aff23}}
\and P.~B.~Lilje\orcid{0000-0003-4324-7794}\inst{\ref{aff64}}
\and V.~Lindholm\orcid{0000-0003-2317-5471}\inst{\ref{aff75},\ref{aff76}}
\and I.~Lloro\orcid{0000-0001-5966-1434}\inst{\ref{aff79}}
\and G.~Mainetti\orcid{0000-0003-2384-2377}\inst{\ref{aff77}}
\and D.~Maino\inst{\ref{aff61},\ref{aff10},\ref{aff62}}
\and E.~Maiorano\orcid{0000-0003-2593-4355}\inst{\ref{aff4}}
\and O.~Mansutti\orcid{0000-0001-5758-4658}\inst{\ref{aff12}}
\and S.~Marcin\inst{\ref{aff80}}
\and O.~Marggraf\orcid{0000-0001-7242-3852}\inst{\ref{aff81}}
\and M.~Martinelli\orcid{0000-0002-6943-7732}\inst{\ref{aff35},\ref{aff82}}
\and N.~Martinet\orcid{0000-0003-2786-7790}\inst{\ref{aff43}}
\and F.~Marulli\orcid{0000-0002-8850-0303}\inst{\ref{aff83},\ref{aff4},\ref{aff20}}
\and R.~Massey\orcid{0000-0002-6085-3780}\inst{\ref{aff84}}
\and S.~Maurogordato\inst{\ref{aff85}}
\and E.~Medinaceli\orcid{0000-0002-4040-7783}\inst{\ref{aff4}}
\and M.~Melchior\inst{\ref{aff88}}
\and Y.~Mellier\inst{\ref{aff89},\ref{aff69}}
\and M.~Meneghetti\orcid{0000-0003-1225-7084}\inst{\ref{aff4},\ref{aff20}}
\and E.~Merlin\orcid{0000-0001-6870-8900}\inst{\ref{aff35}}
\and G.~Meylan\inst{\ref{aff90}}
\and A.~Mora\orcid{0000-0002-1922-8529}\inst{\ref{aff91}}
\and M.~Moresco\orcid{0000-0002-7616-7136}\inst{\ref{aff83},\ref{aff4}}
\and L.~Moscardini\orcid{0000-0002-3473-6716}\inst{\ref{aff83},\ref{aff4},\ref{aff20}}
\and C.~Neissner\orcid{0000-0001-8524-4968}\inst{\ref{aff92},\ref{aff33}}
\and R.~C.~Nichol\orcid{0000-0003-0939-6518}\inst{\ref{aff15}}
\and S.-M.~Niemi\inst{\ref{aff93}}
\and J.~W.~Nightingale\orcid{0000-0002-8987-7401}\inst{\ref{aff94}}
\and C.~Padilla\orcid{0000-0001-7951-0166}\inst{\ref{aff92}}
\and S.~Paltani\orcid{0000-0002-8108-9179}\inst{\ref{aff50}}
\and F.~Pasian\orcid{0000-0002-4869-3227}\inst{\ref{aff12}}
\and K.~Pedersen\inst{\ref{aff95}}
\and W.~J.~Percival\orcid{0000-0002-0644-5727}\inst{\ref{aff96},\ref{aff97},\ref{aff98}}
\and V.~Pettorino\inst{\ref{aff93}}
\and S.~Pires\orcid{0000-0002-0249-2104}\inst{\ref{aff99}}
\and G.~Polenta\orcid{0000-0003-4067-9196}\inst{\ref{aff22}}
\and M.~Poncet\inst{\ref{aff100}}
\and L.~A.~Popa\inst{\ref{aff101}}
\and L.~Pozzetti\orcid{0000-0001-7085-0412}\inst{\ref{aff4}}
\and F.~Raison\orcid{0000-0002-7819-6918}\inst{\ref{aff60}}
\and R.~Rebolo\orcid{0000-0003-3767-7085}\inst{\ref{aff5},\ref{aff102},\ref{aff103}}
\and A.~Renzi\orcid{0000-0001-9856-1970}\inst{\ref{aff104},\ref{aff52}}
\and J.~Rhodes\orcid{0000-0002-4485-8549}\inst{\ref{aff65}}
\and G.~Riccio\inst{\ref{aff27}}
\and E.~Romelli\orcid{0000-0003-3069-9222}\inst{\ref{aff12}}
\and M.~Roncarelli\orcid{0000-0001-9587-7822}\inst{\ref{aff4}}
\and R.~Saglia\orcid{0000-0003-0378-7032}\inst{\ref{aff58},\ref{aff60}}
\and Z.~Sakr\orcid{0000-0002-4823-3757}\inst{\ref{aff105},\ref{aff106},\ref{aff107}}
\and A.~G.~S\'anchez\orcid{0000-0003-1198-831X}\inst{\ref{aff60}}
\and D.~Sapone\orcid{0000-0001-7089-4503}\inst{\ref{aff108}}
\and B.~Sartoris\orcid{0000-0003-1337-5269}\inst{\ref{aff58},\ref{aff12}}
\and M.~Schirmer\orcid{0000-0003-2568-9994}\inst{\ref{aff70}}
\and P.~Schneider\orcid{0000-0001-8561-2679}\inst{\ref{aff81}}
\and T.~Schrabback\orcid{0000-0002-6987-7834}\inst{\ref{aff109}}
\and A.~Secroun\orcid{0000-0003-0505-3710}\inst{\ref{aff53}}
\and G.~Seidel\orcid{0000-0003-2907-353X}\inst{\ref{aff70}}
\and S.~Serrano\orcid{0000-0002-0211-2861}\inst{\ref{aff55},\ref{aff110},\ref{aff56}}
\and P.~Simon\inst{\ref{aff81}}
\and C.~Sirignano\orcid{0000-0002-0995-7146}\inst{\ref{aff104},\ref{aff52}}
\and G.~Sirri\orcid{0000-0003-2626-2853}\inst{\ref{aff20}}
\and L.~Stanco\orcid{0000-0002-9706-5104}\inst{\ref{aff52}}
\and J.~Steinwagner\orcid{0000-0001-7443-1047}\inst{\ref{aff60}}
\and P.~Tallada-Cresp\'{i}\orcid{0000-0002-1336-8328}\inst{\ref{aff32},\ref{aff33}}
\and A.~N.~Taylor\inst{\ref{aff39}}
\and H.~I.~Teplitz\orcid{0000-0002-7064-5424}\inst{\ref{aff111}}
\and I.~Tereno\inst{\ref{aff48},\ref{aff112}}
\and S.~Toft\orcid{0000-0003-3631-7176}\inst{\ref{aff113},\ref{aff114}}
\and R.~Toledo-Moreo\orcid{0000-0002-2997-4859}\inst{\ref{aff115}}
\and F.~Torradeflot\orcid{0000-0003-1160-1517}\inst{\ref{aff33},\ref{aff32}}
\and I.~Tutusaus\orcid{0000-0002-3199-0399}\inst{\ref{aff106}}
\and L.~Valenziano\orcid{0000-0002-1170-0104}\inst{\ref{aff4},\ref{aff54}}
\and J.~Valiviita\orcid{0000-0001-6225-3693}\inst{\ref{aff75},\ref{aff76}}
\and T.~Vassallo\orcid{0000-0001-6512-6358}\inst{\ref{aff58},\ref{aff12}}
\and G.~Verdoes~Kleijn\orcid{0000-0001-5803-2580}\inst{\ref{aff116}}
\and A.~Veropalumbo\orcid{0000-0003-2387-1194}\inst{\ref{aff16},\ref{aff25},\ref{aff24}}
\and Y.~Wang\orcid{0000-0002-4749-2984}\inst{\ref{aff111}}
\and J.~Weller\orcid{0000-0002-8282-2010}\inst{\ref{aff58},\ref{aff60}}
\and A.~Zacchei\orcid{0000-0003-0396-1192}\inst{\ref{aff12},\ref{aff13}}
\and G.~Zamorani\orcid{0000-0002-2318-301X}\inst{\ref{aff4}}
\and F.~M.~Zerbi\inst{\ref{aff16}}
\and I.~A.~Zinchenko\orcid{0000-0002-2944-2449}\inst{\ref{aff58}}
\and E.~Zucca\orcid{0000-0002-5845-8132}\inst{\ref{aff4}}
\and V.~Allevato\orcid{0000-0001-7232-5152}\inst{\ref{aff27}}
\and M.~Ballardini\orcid{0000-0003-4481-3559}\inst{\ref{aff117},\ref{aff118},\ref{aff4}}
\and M.~Bolzonella\orcid{0000-0003-3278-4607}\inst{\ref{aff4}}
\and E.~Bozzo\orcid{0000-0002-8201-1525}\inst{\ref{aff50}}
\and C.~Burigana\orcid{0000-0002-3005-5796}\inst{\ref{aff119},\ref{aff54}}
\and R.~Cabanac\orcid{0000-0001-6679-2600}\inst{\ref{aff106}}
\and A.~Cappi\inst{\ref{aff4},\ref{aff85}}
\and D.~Di~Ferdinando\inst{\ref{aff20}}
\and J.~A.~Escartin~Vigo\inst{\ref{aff60}}
\and G.~Fabbian\orcid{0000-0002-3255-4695}\inst{\ref{aff120}}
\and L.~Gabarra\orcid{0000-0002-8486-8856}\inst{\ref{aff121}}
\and M.~Huertas-Company\orcid{0000-0002-1416-8483}\inst{\ref{aff5},\ref{aff6},\ref{aff122},\ref{aff123}}
\and J.~Mart\'{i}n-Fleitas\orcid{0000-0002-8594-569X}\inst{\ref{aff91}}
\and S.~Matthew\orcid{0000-0001-8448-1697}\inst{\ref{aff39}}
\and M.~Maturi\orcid{0000-0002-3517-2422}\inst{\ref{aff105},\ref{aff124}}
\and N.~Mauri\orcid{0000-0001-8196-1548}\inst{\ref{aff38},\ref{aff20}}
\and R.~B.~Metcalf\orcid{0000-0003-3167-2574}\inst{\ref{aff83},\ref{aff4}}
\and A.~Pezzotta\orcid{0000-0003-0726-2268}\inst{\ref{aff125},\ref{aff60}}
\and M.~P\"ontinen\orcid{0000-0001-5442-2530}\inst{\ref{aff75}}
\and C.~Porciani\orcid{0000-0002-7797-2508}\inst{\ref{aff81}}
\and I.~Risso\orcid{0000-0003-2525-7761}\inst{\ref{aff126}}
\and V.~Scottez\inst{\ref{aff89},\ref{aff127}}
\and M.~Sereno\orcid{0000-0003-0302-0325}\inst{\ref{aff4},\ref{aff20}}
\and M.~Tenti\orcid{0000-0002-4254-5901}\inst{\ref{aff20}}
\and M.~Viel\orcid{0000-0002-2642-5707}\inst{\ref{aff13},\ref{aff12},\ref{aff18},\ref{aff17},\ref{aff128}}
\and M.~Wiesmann\orcid{0009-0000-8199-5860}\inst{\ref{aff64}}
\and Y.~Akrami\orcid{0000-0002-2407-7956}\inst{\ref{aff129},\ref{aff130}}
\and S.~Alvi\orcid{0000-0001-5779-8568}\inst{\ref{aff117}}
\and I.~T.~Andika\orcid{0000-0001-6102-9526}\inst{\ref{aff131},\ref{aff132}}
\and S.~Anselmi\orcid{0000-0002-3579-9583}\inst{\ref{aff52},\ref{aff104},\ref{aff133}}
\and M.~Archidiacono\orcid{0000-0003-4952-9012}\inst{\ref{aff61},\ref{aff62}}
\and F.~Atrio-Barandela\orcid{0000-0002-2130-2513}\inst{\ref{aff134}}
\and K.~Benson\inst{\ref{aff47}}
\and P.~Bergamini\orcid{0000-0003-1383-9414}\inst{\ref{aff61},\ref{aff4}}
\and D.~Bertacca\orcid{0000-0002-2490-7139}\inst{\ref{aff104},\ref{aff21},\ref{aff52}}
\and M.~Bethermin\orcid{0000-0002-3915-2015}\inst{\ref{aff135}}
\and A.~Blanchard\orcid{0000-0001-8555-9003}\inst{\ref{aff106}}
\and L.~Blot\orcid{0000-0002-9622-7167}\inst{\ref{aff136},\ref{aff133}}
\and H.~B\"ohringer\orcid{0000-0001-8241-4204}\inst{\ref{aff60},\ref{aff137},\ref{aff138}}
\and S.~Borgani\orcid{0000-0001-6151-6439}\inst{\ref{aff139},\ref{aff13},\ref{aff12},\ref{aff17},\ref{aff128}}
\and M.~L.~Brown\orcid{0000-0002-0370-8077}\inst{\ref{aff40}}
\and S.~Bruton\orcid{0000-0002-6503-5218}\inst{\ref{aff140}}
\and A.~Calabro\orcid{0000-0003-2536-1614}\inst{\ref{aff35}}
\and B.~Camacho~Quevedo\orcid{0000-0002-8789-4232}\inst{\ref{aff55},\ref{aff56}}
\and F.~Caro\inst{\ref{aff35}}
\and C.~S.~Carvalho\inst{\ref{aff112}}
\and T.~Castro\orcid{0000-0002-6292-3228}\inst{\ref{aff12},\ref{aff17},\ref{aff13},\ref{aff128}}
\and F.~Cogato\orcid{0000-0003-4632-6113}\inst{\ref{aff83},\ref{aff4}}
\and T.~Contini\orcid{0000-0003-0275-938X}\inst{\ref{aff106}}
\and A.~R.~Cooray\orcid{0000-0002-3892-0190}\inst{\ref{aff141}}
\and M.~Costanzi\orcid{0000-0001-8158-1449}\inst{\ref{aff139},\ref{aff12},\ref{aff13}}
\and O.~Cucciati\orcid{0000-0002-9336-7551}\inst{\ref{aff4}}
\and S.~Davini\orcid{0000-0003-3269-1718}\inst{\ref{aff25}}
\and F.~De~Paolis\orcid{0000-0001-6460-7563}\inst{\ref{aff142},\ref{aff143},\ref{aff144}}
\and G.~Desprez\orcid{0000-0001-8325-1742}\inst{\ref{aff116}}
\and A.~D\'iaz-S\'anchez\orcid{0000-0003-0748-4768}\inst{\ref{aff145}}
\and J.~J.~Diaz\inst{\ref{aff6},\ref{aff5}}
\and S.~Di~Domizio\orcid{0000-0003-2863-5895}\inst{\ref{aff24},\ref{aff25}}
\and J.~M.~Diego\orcid{0000-0001-9065-3926}\inst{\ref{aff146}}
\and P.-A.~Duc\orcid{0000-0003-3343-6284}\inst{\ref{aff135}}
\and Y.~Fang\inst{\ref{aff58}}
\and A.~G.~Ferrari\orcid{0009-0005-5266-4110}\inst{\ref{aff20}}
\and P.~G.~Ferreira\orcid{0000-0002-3021-2851}\inst{\ref{aff121}}
\and A.~Finoguenov\orcid{0000-0002-4606-5403}\inst{\ref{aff75}}
\and A.~Franco\orcid{0000-0002-4761-366X}\inst{\ref{aff143},\ref{aff142},\ref{aff144}}
\and K.~Ganga\orcid{0000-0001-8159-8208}\inst{\ref{aff86}}
\and J.~Garc\'ia-Bellido\orcid{0000-0002-9370-8360}\inst{\ref{aff129}}
\and T.~Gasparetto\orcid{0000-0002-7913-4866}\inst{\ref{aff12}}
\and V.~Gautard\inst{\ref{aff147}}
\and E.~Gaztanaga\orcid{0000-0001-9632-0815}\inst{\ref{aff56},\ref{aff55},\ref{aff148}}
\and F.~Giacomini\orcid{0000-0002-3129-2814}\inst{\ref{aff20}}
\and F.~Gianotti\orcid{0000-0003-4666-119X}\inst{\ref{aff4}}
\and A.~H.~Gonzalez\orcid{0000-0002-0933-8601}\inst{\ref{aff149}}
\and G.~Gozaliasl\orcid{0000-0002-0236-919X}\inst{\ref{aff150},\ref{aff75}}
\and A.~Gruppuso\orcid{0000-0001-9272-5292}\inst{\ref{aff4},\ref{aff20}}
\and M.~Guidi\orcid{0000-0001-9408-1101}\inst{\ref{aff19},\ref{aff4}}
\and C.~M.~Gutierrez\orcid{0000-0001-7854-783X}\inst{\ref{aff151}}
\and A.~Hall\orcid{0000-0002-3139-8651}\inst{\ref{aff39}}
\and W.~G.~Hartley\inst{\ref{aff50}}
\and S.~Hemmati\orcid{0000-0003-2226-5395}\inst{\ref{aff152}}
\and C.~Hern\'andez-Monteagudo\orcid{0000-0001-5471-9166}\inst{\ref{aff103},\ref{aff5}}
\and H.~Hildebrandt\orcid{0000-0002-9814-3338}\inst{\ref{aff153}}
\and J.~Hjorth\orcid{0000-0002-4571-2306}\inst{\ref{aff95}}
\and J.~J.~E.~Kajava\orcid{0000-0002-3010-8333}\inst{\ref{aff154},\ref{aff155}}
\and Y.~Kang\orcid{0009-0000-8588-7250}\inst{\ref{aff50}}
\and V.~Kansal\orcid{0000-0002-4008-6078}\inst{\ref{aff156},\ref{aff157}}
\and D.~Karagiannis\orcid{0000-0002-4927-0816}\inst{\ref{aff117},\ref{aff158}}
\and K.~Kiiveri\inst{\ref{aff72}}
\and C.~C.~Kirkpatrick\inst{\ref{aff72}}
\and S.~Kruk\orcid{0000-0001-8010-8879}\inst{\ref{aff14}}
\and J.~Le~Graet\orcid{0000-0001-6523-7971}\inst{\ref{aff53}}
\and L.~Legrand\orcid{0000-0003-0610-5252}\inst{\ref{aff159},\ref{aff160}}
\and M.~Lembo\orcid{0000-0002-5271-5070}\inst{\ref{aff117},\ref{aff118}}
\and F.~Lepori\orcid{0009-0000-5061-7138}\inst{\ref{aff161}}
\and G.~Leroy\orcid{0009-0004-2523-4425}\inst{\ref{aff162},\ref{aff84}}
\and G.~F.~Lesci\orcid{0000-0002-4607-2830}\inst{\ref{aff83},\ref{aff4}}
\and J.~Lesgourgues\orcid{0000-0001-7627-353X}\inst{\ref{aff34}}
\and L.~Leuzzi\orcid{0009-0006-4479-7017}\inst{\ref{aff83},\ref{aff4}}
\and T.~I.~Liaudat\orcid{0000-0002-9104-314X}\inst{\ref{aff163}}
\and S.~J.~Liu\orcid{0000-0001-7680-2139}\inst{\ref{aff51}}
\and A.~Loureiro\orcid{0000-0002-4371-0876}\inst{\ref{aff164},\ref{aff165}}
\and J.~Macias-Perez\orcid{0000-0002-5385-2763}\inst{\ref{aff166}}
\and G.~Maggio\orcid{0000-0003-4020-4836}\inst{\ref{aff12}}
\and M.~Magliocchetti\orcid{0000-0001-9158-4838}\inst{\ref{aff51}}
\and E.~A.~Magnier\orcid{0000-0002-7965-2815}\inst{\ref{aff37}}
\and F.~Mannucci\orcid{0000-0002-4803-2381}\inst{\ref{aff167}}
\and R.~Maoli\orcid{0000-0002-6065-3025}\inst{\ref{aff168},\ref{aff35}}
\and C.~J.~A.~P.~Martins\orcid{0000-0002-4886-9261}\inst{\ref{aff169},\ref{aff28}}
\and M.~Migliaccio\inst{\ref{aff170},\ref{aff171}}
\and M.~Miluzio\inst{\ref{aff14},\ref{aff172}}
\and P.~Monaco\orcid{0000-0003-2083-7564}\inst{\ref{aff139},\ref{aff12},\ref{aff17},\ref{aff13}}
\and C.~Moretti\orcid{0000-0003-3314-8936}\inst{\ref{aff18},\ref{aff128},\ref{aff12},\ref{aff13},\ref{aff17}}
\and G.~Morgante\inst{\ref{aff4}}
\and S.~Nadathur\orcid{0000-0001-9070-3102}\inst{\ref{aff148}}
\and K.~Naidoo\orcid{0000-0002-9182-1802}\inst{\ref{aff148}}
\and A.~Navarro-Alsina\orcid{0000-0002-3173-2592}\inst{\ref{aff81}}
\and S.~Nesseris\orcid{0000-0002-0567-0324}\inst{\ref{aff129}}
\and F.~Passalacqua\orcid{0000-0002-8606-4093}\inst{\ref{aff104},\ref{aff52}}
\and K.~Paterson\orcid{0000-0001-8340-3486}\inst{\ref{aff70}}
\and L.~Patrizii\inst{\ref{aff20}}
\and A.~Philippon\inst{\ref{aff1}}
\and A.~Pisani\orcid{0000-0002-6146-4437}\inst{\ref{aff53},\ref{aff173}}
\and D.~Potter\orcid{0000-0002-0757-5195}\inst{\ref{aff161}}
\and S.~Quai\orcid{0000-0002-0449-8163}\inst{\ref{aff83},\ref{aff4}}
\and M.~Radovich\orcid{0000-0002-3585-866X}\inst{\ref{aff21}}
\and G.~Rodighiero\orcid{0000-0002-9415-2296}\inst{\ref{aff104},\ref{aff21}}
\and S.~Sacquegna\orcid{0000-0002-8433-6630}\inst{\ref{aff142},\ref{aff143},\ref{aff144}}
\and M.~Sahl\'en\orcid{0000-0003-0973-4804}\inst{\ref{aff174}}
\and D.~B.~Sanders\orcid{0000-0002-1233-9998}\inst{\ref{aff37}}
\and E.~Sarpa\orcid{0000-0002-1256-655X}\inst{\ref{aff18},\ref{aff128},\ref{aff17}}
\and A.~Schneider\orcid{0000-0001-7055-8104}\inst{\ref{aff161}}
\and D.~Sciotti\orcid{0009-0008-4519-2620}\inst{\ref{aff35},\ref{aff82}}
\and E.~Sellentin\inst{\ref{aff175},\ref{aff73}}
\and F.~Shankar\orcid{0000-0001-8973-5051}\inst{\ref{aff176}}
\and L.~C.~Smith\orcid{0000-0002-3259-2771}\inst{\ref{aff177}}
\and S.~A.~Stanford\orcid{0000-0003-0122-0841}\inst{\ref{aff178}}
\and K.~Tanidis\orcid{0000-0001-9843-5130}\inst{\ref{aff121}}
\and G.~Testera\inst{\ref{aff25}}
\and R.~Teyssier\orcid{0000-0001-7689-0933}\inst{\ref{aff173}}
\and S.~Tosi\orcid{0000-0002-7275-9193}\inst{\ref{aff24},\ref{aff126}}
\and A.~Troja\orcid{0000-0003-0239-4595}\inst{\ref{aff104},\ref{aff52}}
\and M.~Tucci\inst{\ref{aff50}}
\and C.~Valieri\inst{\ref{aff20}}
\and A.~Venhola\orcid{0000-0001-6071-4564}\inst{\ref{aff179}}
\and D.~Vergani\orcid{0000-0003-0898-2216}\inst{\ref{aff4}}
\and G.~Verza\orcid{0000-0002-1886-8348}\inst{\ref{aff180}}
\and P.~Vielzeuf\orcid{0000-0003-2035-9339}\inst{\ref{aff53}}
\and N.~A.~Walton\orcid{0000-0003-3983-8778}\inst{\ref{aff177}}
\and J.~R.~Weaver\orcid{0000-0003-1614-196X}\inst{\ref{aff181}}
\and J.~G.~Sorce\orcid{0000-0002-2307-2432}\inst{\ref{aff182},\ref{aff1}}
\and D.~Scott\orcid{0000-0002-6878-9840}\inst{\ref{aff183}}}
										   
\institute{Universit\'e Paris-Saclay, CNRS, Institut d'astrophysique spatiale, 91405, Orsay, France\label{aff1}
\and
Institut de Recherche en Informatique de Toulouse (IRIT), Universit\'e de Toulouse, CNRS, Toulouse INP, UT3, 31062 Toulouse, France\label{aff2}
\and
Laboratoire MCD, Centre de Biologie Int\'egrative (CBI), Universit\'e de Toulouse, CNRS, UT3, 31062 Toulouse, France\label{aff3}
\and
INAF-Osservatorio di Astrofisica e Scienza dello Spazio di Bologna, Via Piero Gobetti 93/3, 40129 Bologna, Italy\label{aff4}
\and
Instituto de Astrof\'{\i}sica de Canarias, V\'{\i}a L\'actea, 38205 La Laguna, Tenerife, Spain\label{aff5}
\and
Departamento de Astrof\'isica, Universidad de La Laguna (ULL), 38206, La Laguna, Tenerife, Spain\label{aff6}
\and
Universit\'e Paris-Saclay, CEA, D\'epartement d'\'Electronique des D\'etecteurs et d'Informatique pour la Physique, 91191, Gif-sur-Yvette, France\label{aff7}
\and
Universit\'e Paris-Saclay, CNRS, Inria, LISN, 91191, Gif-sur-Yvette, France\label{aff8}
\and
Department of Astronomy/Steward Observatory, University of Arizona, 933 N. Cherry Avenue, Tucson, AZ 85721, USA\label{aff9}
\and
INAF-IASF Milano, Via Alfonso Corti 12, 20133 Milano, Italy\label{aff10}
\and
Department of Astronomy \& Astrophysics, University of California at San Diego, 9500 Gilman Drive, La Jolla, CA 92093, USA\label{aff11}
\and
INAF-Osservatorio Astronomico di Trieste, Via G. B. Tiepolo 11, 34143 Trieste, Italy\label{aff12}
\and
IFPU, Institute for Fundamental Physics of the Universe, via Beirut 2, 34151 Trieste, Italy\label{aff13}
\and
ESAC/ESA, Camino Bajo del Castillo, s/n., Urb. Villafranca del Castillo, 28692 Villanueva de la Ca\~nada, Madrid, Spain\label{aff14}
\and
School of Mathematics and Physics, University of Surrey, Guildford, Surrey, GU2 7XH, UK\label{aff15}
\and
INAF-Osservatorio Astronomico di Brera, Via Brera 28, 20122 Milano, Italy\label{aff16}
\and
INFN, Sezione di Trieste, Via Valerio 2, 34127 Trieste TS, Italy\label{aff17}
\and
SISSA, International School for Advanced Studies, Via Bonomea 265, 34136 Trieste TS, Italy\label{aff18}
\and
Dipartimento di Fisica e Astronomia, Universit\`a di Bologna, Via Gobetti 93/2, 40129 Bologna, Italy\label{aff19}
\and
INFN-Sezione di Bologna, Viale Berti Pichat 6/2, 40127 Bologna, Italy\label{aff20}
\and
INAF-Osservatorio Astronomico di Padova, Via dell'Osservatorio 5, 35122 Padova, Italy\label{aff21}
\and
Space Science Data Center, Italian Space Agency, via del Politecnico snc, 00133 Roma, Italy\label{aff22}
\and
INAF-Osservatorio Astrofisico di Torino, Via Osservatorio 20, 10025 Pino Torinese (TO), Italy\label{aff23}
\and
Dipartimento di Fisica, Universit\`a di Genova, Via Dodecaneso 33, 16146, Genova, Italy\label{aff24}
\and
INFN-Sezione di Genova, Via Dodecaneso 33, 16146, Genova, Italy\label{aff25}
\and
Department of Physics "E. Pancini", University Federico II, Via Cinthia 6, 80126, Napoli, Italy\label{aff26}
\and
INAF-Osservatorio Astronomico di Capodimonte, Via Moiariello 16, 80131 Napoli, Italy\label{aff27}
\and
Instituto de Astrof\'isica e Ci\^encias do Espa\c{c}o, Universidade do Porto, CAUP, Rua das Estrelas, PT4150-762 Porto, Portugal\label{aff28}
\and
Faculdade de Ci\^encias da Universidade do Porto, Rua do Campo de Alegre, 4150-007 Porto, Portugal\label{aff29}
\and
Dipartimento di Fisica, Universit\`a degli Studi di Torino, Via P. Giuria 1, 10125 Torino, Italy\label{aff30}
\and
INFN-Sezione di Torino, Via P. Giuria 1, 10125 Torino, Italy\label{aff31}
\and
Centro de Investigaciones Energ\'eticas, Medioambientales y Tecnol\'ogicas (CIEMAT), Avenida Complutense 40, 28040 Madrid, Spain\label{aff32}
\and
Port d'Informaci\'{o} Cient\'{i}fica, Campus UAB, C. Albareda s/n, 08193 Bellaterra (Barcelona), Spain\label{aff33}
\and
Institute for Theoretical Particle Physics and Cosmology (TTK), RWTH Aachen University, 52056 Aachen, Germany\label{aff34}
\and
INAF-Osservatorio Astronomico di Roma, Via Frascati 33, 00078 Monteporzio Catone, Italy\label{aff35}
\and
INFN section of Naples, Via Cinthia 6, 80126, Napoli, Italy\label{aff36}
\and
Institute for Astronomy, University of Hawaii, 2680 Woodlawn Drive, Honolulu, HI 96822, USA\label{aff37}
\and
Dipartimento di Fisica e Astronomia "Augusto Righi" - Alma Mater Studiorum Universit\`a di Bologna, Viale Berti Pichat 6/2, 40127 Bologna, Italy\label{aff38}
\and
Institute for Astronomy, University of Edinburgh, Royal Observatory, Blackford Hill, Edinburgh EH9 3HJ, UK\label{aff39}
\and
Jodrell Bank Centre for Astrophysics, Department of Physics and Astronomy, University of Manchester, Oxford Road, Manchester M13 9PL, UK\label{aff40}
\and
European Space Agency/ESRIN, Largo Galileo Galilei 1, 00044 Frascati, Roma, Italy\label{aff41}
\and
Universit\'e Claude Bernard Lyon 1, CNRS/IN2P3, IP2I Lyon, UMR 5822, Villeurbanne, F-69100, France\label{aff42}
\and
Aix-Marseille Universit\'e, CNRS, CNES, LAM, Marseille, France\label{aff43}
\and
Institut de Ci\`{e}ncies del Cosmos (ICCUB), Universitat de Barcelona (IEEC-UB), Mart\'{i} i Franqu\`{e}s 1, 08028 Barcelona, Spain\label{aff44}
\and
Instituci\'o Catalana de Recerca i Estudis Avan\c{c}ats (ICREA), Passeig de Llu\'{\i}s Companys 23, 08010 Barcelona, Spain\label{aff45}
\and
UCB Lyon 1, CNRS/IN2P3, IUF, IP2I Lyon, 4 rue Enrico Fermi, 69622 Villeurbanne, France\label{aff46}
\and
Mullard Space Science Laboratory, University College London, Holmbury St Mary, Dorking, Surrey RH5 6NT, UK\label{aff47}
\and
Departamento de F\'isica, Faculdade de Ci\^encias, Universidade de Lisboa, Edif\'icio C8, Campo Grande, PT1749-016 Lisboa, Portugal\label{aff48}
\and
Instituto de Astrof\'isica e Ci\^encias do Espa\c{c}o, Faculdade de Ci\^encias, Universidade de Lisboa, Campo Grande, 1749-016 Lisboa, Portugal\label{aff49}
\and
Department of Astronomy, University of Geneva, ch. d'Ecogia 16, 1290 Versoix, Switzerland\label{aff50}
\and
INAF-Istituto di Astrofisica e Planetologia Spaziali, via del Fosso del Cavaliere, 100, 00100 Roma, Italy\label{aff51}
\and
INFN-Padova, Via Marzolo 8, 35131 Padova, Italy\label{aff52}
\and
Aix-Marseille Universit\'e, CNRS/IN2P3, CPPM, Marseille, France\label{aff53}
\and
INFN-Bologna, Via Irnerio 46, 40126 Bologna, Italy\label{aff54}
\and
Institut d'Estudis Espacials de Catalunya (IEEC),  Edifici RDIT, Campus UPC, 08860 Castelldefels, Barcelona, Spain\label{aff55}
\and
Institute of Space Sciences (ICE, CSIC), Campus UAB, Carrer de Can Magrans, s/n, 08193 Barcelona, Spain\label{aff56}
\and
School of Physics, HH Wills Physics Laboratory, University of Bristol, Tyndall Avenue, Bristol, BS8 1TL, UK\label{aff57}
\and
Universit\"ats-Sternwarte M\"unchen, Fakult\"at f\"ur Physik, Ludwig-Maximilians-Universit\"at M\"unchen, Scheinerstrasse 1, 81679 M\"unchen, Germany\label{aff58}
\and
FRACTAL S.L.N.E., calle Tulip\'an 2, Portal 13 1A, 28231, Las Rozas de Madrid, Spain\label{aff59}
\and
Max Planck Institute for Extraterrestrial Physics, Giessenbachstr. 1, 85748 Garching, Germany\label{aff60}
\and
Dipartimento di Fisica "Aldo Pontremoli", Universit\`a degli Studi di Milano, Via Celoria 16, 20133 Milano, Italy\label{aff61}
\and
INFN-Sezione di Milano, Via Celoria 16, 20133 Milano, Italy\label{aff62}
\and
NRC Herzberg, 5071 West Saanich Rd, Victoria, BC V9E 2E7, Canada\label{aff63}
\and
Institute of Theoretical Astrophysics, University of Oslo, P.O. Box 1029 Blindern, 0315 Oslo, Norway\label{aff64}
\and
Jet Propulsion Laboratory, California Institute of Technology, 4800 Oak Grove Drive, Pasadena, CA, 91109, USA\label{aff65}
\and
Felix Hormuth Engineering, Goethestr. 17, 69181 Leimen, Germany\label{aff66}
\and
Technical University of Denmark, Elektrovej 327, 2800 Kgs. Lyngby, Denmark\label{aff67}
\and
Cosmic Dawn Center (DAWN), Denmark\label{aff68}
\and
Institut d'Astrophysique de Paris, UMR 7095, CNRS, and Sorbonne Universit\'e, 98 bis boulevard Arago, 75014 Paris, France\label{aff69}
\and
Max-Planck-Institut f\"ur Astronomie, K\"onigstuhl 17, 69117 Heidelberg, Germany\label{aff70}
\and
NASA Goddard Space Flight Center, Greenbelt, MD 20771, USA\label{aff71}
\and
Department of Physics and Helsinki Institute of Physics, Gustaf H\"allstr\"omin katu 2, 00014 University of Helsinki, Finland\label{aff72}
\and
Leiden Observatory, Leiden University, Einsteinweg 55, 2333 CC Leiden, The Netherlands\label{aff73}
\and
Universit\'e de Gen\`eve, D\'epartement de Physique Th\'eorique and Centre for Astroparticle Physics, 24 quai Ernest-Ansermet, CH-1211 Gen\`eve 4, Switzerland\label{aff74}
\and
Department of Physics, P.O. Box 64, 00014 University of Helsinki, Finland\label{aff75}
\and
Helsinki Institute of Physics, Gustaf H{\"a}llstr{\"o}min katu 2, University of Helsinki, Helsinki, Finland\label{aff76}
\and
Centre de Calcul de l'IN2P3/CNRS, 21 avenue Pierre de Coubertin 69627 Villeurbanne Cedex, France\label{aff77}
\and
Laboratoire d'etude de l'Univers et des phenomenes eXtremes, Observatoire de Paris, Universit\'e PSL, Sorbonne Universit\'e, CNRS, 92190 Meudon, France\label{aff78}
\and
SKA Observatory, Jodrell Bank, Lower Withington, Macclesfield, Cheshire SK11 9FT, UK\label{aff79}
\and
University of Applied Sciences and Arts of Northwestern Switzerland, School of Computer Science, 5210 Windisch, Switzerland\label{aff80}
\and
Universit\"at Bonn, Argelander-Institut f\"ur Astronomie, Auf dem H\"ugel 71, 53121 Bonn, Germany\label{aff81}
\and
INFN-Sezione di Roma, Piazzale Aldo Moro, 2 - c/o Dipartimento di Fisica, Edificio G. Marconi, 00185 Roma, Italy\label{aff82}
\and
Dipartimento di Fisica e Astronomia "Augusto Righi" - Alma Mater Studiorum Universit\`a di Bologna, via Piero Gobetti 93/2, 40129 Bologna, Italy\label{aff83}
\and
Department of Physics, Institute for Computational Cosmology, Durham University, South Road, Durham, DH1 3LE, UK\label{aff84}
\and
Universit\'e C\^{o}te d'Azur, Observatoire de la C\^{o}te d'Azur, CNRS, Laboratoire Lagrange, Bd de l'Observatoire, CS 34229, 06304 Nice cedex 4, France\label{aff85}
\and
Universit\'e Paris Cit\'e, CNRS, Astroparticule et Cosmologie, 75013 Paris, France\label{aff86}
\and
CNRS-UCB International Research Laboratory, Centre Pierre Binetruy, IRL2007, CPB-IN2P3, Berkeley, USA\label{aff87}
\and
University of Applied Sciences and Arts of Northwestern Switzerland, School of Engineering, 5210 Windisch, Switzerland\label{aff88}
\and
Institut d'Astrophysique de Paris, 98bis Boulevard Arago, 75014, Paris, France\label{aff89}
\and
Institute of Physics, Laboratory of Astrophysics, Ecole Polytechnique F\'ed\'erale de Lausanne (EPFL), Observatoire de Sauverny, 1290 Versoix, Switzerland\label{aff90}
\and
Aurora Technology for European Space Agency (ESA), Camino bajo del Castillo, s/n, Urbanizacion Villafranca del Castillo, Villanueva de la Ca\~nada, 28692 Madrid, Spain\label{aff91}
\and
Institut de F\'{i}sica d'Altes Energies (IFAE), The Barcelona Institute of Science and Technology, Campus UAB, 08193 Bellaterra (Barcelona), Spain\label{aff92}
\and
European Space Agency/ESTEC, Keplerlaan 1, 2201 AZ Noordwijk, The Netherlands\label{aff93}
\and
School of Mathematics, Statistics and Physics, Newcastle University, Herschel Building, Newcastle-upon-Tyne, NE1 7RU, UK\label{aff94}
\and
DARK, Niels Bohr Institute, University of Copenhagen, Jagtvej 155, 2200 Copenhagen, Denmark\label{aff95}
\and
Waterloo Centre for Astrophysics, University of Waterloo, Waterloo, Ontario N2L 3G1, Canada\label{aff96}
\and
Department of Physics and Astronomy, University of Waterloo, Waterloo, Ontario N2L 3G1, Canada\label{aff97}
\and
Perimeter Institute for Theoretical Physics, Waterloo, Ontario N2L 2Y5, Canada\label{aff98}
\and
Universit\'e Paris-Saclay, Universit\'e Paris Cit\'e, CEA, CNRS, AIM, 91191, Gif-sur-Yvette, France\label{aff99}
\and
Centre National d'Etudes Spatiales -- Centre spatial de Toulouse, 18 avenue Edouard Belin, 31401 Toulouse Cedex 9, France\label{aff100}
\and
Institute of Space Science, Str. Atomistilor, nr. 409 M\u{a}gurele, Ilfov, 077125, Romania\label{aff101}
\and
Consejo Superior de Investigaciones Cientificas, Calle Serrano 117, 28006 Madrid, Spain\label{aff102}
\and
Universidad de La Laguna, Departamento de Astrof\'{\i}sica, 38206 La Laguna, Tenerife, Spain\label{aff103}
\and
Dipartimento di Fisica e Astronomia "G. Galilei", Universit\`a di Padova, Via Marzolo 8, 35131 Padova, Italy\label{aff104}
\and
Institut f\"ur Theoretische Physik, University of Heidelberg, Philosophenweg 16, 69120 Heidelberg, Germany\label{aff105}
\and
Institut de Recherche en Astrophysique et Plan\'etologie (IRAP), Universit\'e de Toulouse, CNRS, UPS, CNES, 14 Av. Edouard Belin, 31400 Toulouse, France\label{aff106}
\and
Universit\'e St Joseph; Faculty of Sciences, Beirut, Lebanon\label{aff107}
\and
Departamento de F\'isica, FCFM, Universidad de Chile, Blanco Encalada 2008, Santiago, Chile\label{aff108}
\and
Universit\"at Innsbruck, Institut f\"ur Astro- und Teilchenphysik, Technikerstr. 25/8, 6020 Innsbruck, Austria\label{aff109}
\and
Satlantis, University Science Park, Sede Bld 48940, Leioa-Bilbao, Spain\label{aff110}
\and
Infrared Processing and Analysis Center, California Institute of Technology, Pasadena, CA 91125, USA\label{aff111}
\and
Instituto de Astrof\'isica e Ci\^encias do Espa\c{c}o, Faculdade de Ci\^encias, Universidade de Lisboa, Tapada da Ajuda, 1349-018 Lisboa, Portugal\label{aff112}
\and
Cosmic Dawn Center (DAWN)\label{aff113}
\and
Niels Bohr Institute, University of Copenhagen, Jagtvej 128, 2200 Copenhagen, Denmark\label{aff114}
\and
Universidad Polit\'ecnica de Cartagena, Departamento de Electr\'onica y Tecnolog\'ia de Computadoras,  Plaza del Hospital 1, 30202 Cartagena, Spain\label{aff115}
\and
Kapteyn Astronomical Institute, University of Groningen, PO Box 800, 9700 AV Groningen, The Netherlands\label{aff116}
\and
Dipartimento di Fisica e Scienze della Terra, Universit\`a degli Studi di Ferrara, Via Giuseppe Saragat 1, 44122 Ferrara, Italy\label{aff117}
\and
Istituto Nazionale di Fisica Nucleare, Sezione di Ferrara, Via Giuseppe Saragat 1, 44122 Ferrara, Italy\label{aff118}
\and
INAF, Istituto di Radioastronomia, Via Piero Gobetti 101, 40129 Bologna, Italy\label{aff119}
\and
School of Physics and Astronomy, Cardiff University, The Parade, Cardiff, CF24 3AA, UK\label{aff120}
\and
Department of Physics, Oxford University, Keble Road, Oxford OX1 3RH, UK\label{aff121}
\and
Universit\'e PSL, Observatoire de Paris, Sorbonne Universit\'e, CNRS, LERMA, 75014, Paris, France\label{aff122}
\and
Universit\'e Paris-Cit\'e, 5 Rue Thomas Mann, 75013, Paris, France\label{aff123}
\and
Zentrum f\"ur Astronomie, Universit\"at Heidelberg, Philosophenweg 12, 69120 Heidelberg, Germany\label{aff124}
\and
INAF - Osservatorio Astronomico di Brera, via Emilio Bianchi 46, 23807 Merate, Italy\label{aff125}
\and
INAF-Osservatorio Astronomico di Brera, Via Brera 28, 20122 Milano, Italy, and INFN-Sezione di Genova, Via Dodecaneso 33, 16146, Genova, Italy\label{aff126}
\and
ICL, Junia, Universit\'e Catholique de Lille, LITL, 59000 Lille, France\label{aff127}
\and
ICSC - Centro Nazionale di Ricerca in High Performance Computing, Big Data e Quantum Computing, Via Magnanelli 2, Bologna, Italy\label{aff128}
\and
Instituto de F\'isica Te\'orica UAM-CSIC, Campus de Cantoblanco, 28049 Madrid, Spain\label{aff129}
\and
CERCA/ISO, Department of Physics, Case Western Reserve University, 10900 Euclid Avenue, Cleveland, OH 44106, USA\label{aff130}
\and
Technical University of Munich, TUM School of Natural Sciences, Physics Department, James-Franck-Str.~1, 85748 Garching, Germany\label{aff131}
\and
Max-Planck-Institut f\"ur Astrophysik, Karl-Schwarzschild-Str.~1, 85748 Garching, Germany\label{aff132}
\and
Laboratoire Univers et Th\'eorie, Observatoire de Paris, Universit\'e PSL, Universit\'e Paris Cit\'e, CNRS, 92190 Meudon, France\label{aff133}
\and
Departamento de F{\'\i}sica Fundamental. Universidad de Salamanca. Plaza de la Merced s/n. 37008 Salamanca, Spain\label{aff134}
\and
Universit\'e de Strasbourg, CNRS, Observatoire astronomique de Strasbourg, UMR 7550, 67000 Strasbourg, France\label{aff135}
\and
Center for Data-Driven Discovery, Kavli IPMU (WPI), UTIAS, The University of Tokyo, Kashiwa, Chiba 277-8583, Japan\label{aff136}
\and
Ludwig-Maximilians-University, Schellingstrasse 4, 80799 Munich, Germany\label{aff137}
\and
Max-Planck-Institut f\"ur Physik, Boltzmannstr. 8, 85748 Garching, Germany\label{aff138}
\and
Dipartimento di Fisica - Sezione di Astronomia, Universit\`a di Trieste, Via Tiepolo 11, 34131 Trieste, Italy\label{aff139}
\and
California Institute of Technology, 1200 E California Blvd, Pasadena, CA 91125, USA\label{aff140}
\and
Department of Physics \& Astronomy, University of California Irvine, Irvine CA 92697, USA\label{aff141}
\and
Department of Mathematics and Physics E. De Giorgi, University of Salento, Via per Arnesano, CP-I93, 73100, Lecce, Italy\label{aff142}
\and
INFN, Sezione di Lecce, Via per Arnesano, CP-193, 73100, Lecce, Italy\label{aff143}
\and
INAF-Sezione di Lecce, c/o Dipartimento Matematica e Fisica, Via per Arnesano, 73100, Lecce, Italy\label{aff144}
\and
Departamento F\'isica Aplicada, Universidad Polit\'ecnica de Cartagena, Campus Muralla del Mar, 30202 Cartagena, Murcia, Spain\label{aff145}
\and
Instituto de F\'isica de Cantabria, Edificio Juan Jord\'a, Avenida de los Castros, 39005 Santander, Spain\label{aff146}
\and
CEA Saclay, DFR/IRFU, Service d'Astrophysique, Bat. 709, 91191 Gif-sur-Yvette, France\label{aff147}
\and
Institute of Cosmology and Gravitation, University of Portsmouth, Portsmouth PO1 3FX, UK\label{aff148}
\and
Department of Astronomy, University of Florida, Bryant Space Science Center, Gainesville, FL 32611, USA\label{aff149}
\and
Department of Computer Science, Aalto University, PO Box 15400, Espoo, FI-00 076, Finland\label{aff150}
\and
Instituto de Astrof\'\i sica de Canarias, c/ Via Lactea s/n, La Laguna 38200, Spain. Departamento de Astrof\'\i sica de la Universidad de La Laguna, Avda. Francisco Sanchez, La Laguna, 38200, Spain\label{aff151}
\and
Caltech/IPAC, 1200 E. California Blvd., Pasadena, CA 91125, USA\label{aff152}
\and
Ruhr University Bochum, Faculty of Physics and Astronomy, Astronomical Institute (AIRUB), German Centre for Cosmological Lensing (GCCL), 44780 Bochum, Germany\label{aff153}
\and
Department of Physics and Astronomy, Vesilinnantie 5, 20014 University of Turku, Finland\label{aff154}
\and
Serco for European Space Agency (ESA), Camino bajo del Castillo, s/n, Urbanizacion Villafranca del Castillo, Villanueva de la Ca\~nada, 28692 Madrid, Spain\label{aff155}
\and
ARC Centre of Excellence for Dark Matter Particle Physics, Melbourne, Australia\label{aff156}
\and
Centre for Astrophysics \& Supercomputing, Swinburne University of Technology,  Hawthorn, Victoria 3122, Australia\label{aff157}
\and
Department of Physics and Astronomy, University of the Western Cape, Bellville, Cape Town, 7535, South Africa\label{aff158}
\and
DAMTP, Centre for Mathematical Sciences, Wilberforce Road, Cambridge CB3 0WA, UK\label{aff159}
\and
Kavli Institute for Cosmology Cambridge, Madingley Road, Cambridge, CB3 0HA, UK\label{aff160}
\and
Department of Astrophysics, University of Zurich, Winterthurerstrasse 190, 8057 Zurich, Switzerland\label{aff161}
\and
Department of Physics, Centre for Extragalactic Astronomy, Durham University, South Road, Durham, DH1 3LE, UK\label{aff162}
\and
IRFU, CEA, Universit\'e Paris-Saclay 91191 Gif-sur-Yvette Cedex, France\label{aff163}
\and
Oskar Klein Centre for Cosmoparticle Physics, Department of Physics, Stockholm University, Stockholm, SE-106 91, Sweden\label{aff164}
\and
Astrophysics Group, Blackett Laboratory, Imperial College London, London SW7 2AZ, UK\label{aff165}
\and
Univ. Grenoble Alpes, CNRS, Grenoble INP, LPSC-IN2P3, 53, Avenue des Martyrs, 38000, Grenoble, France\label{aff166}
\and
INAF-Osservatorio Astrofisico di Arcetri, Largo E. Fermi 5, 50125, Firenze, Italy\label{aff167}
\and
Dipartimento di Fisica, Sapienza Universit\`a di Roma, Piazzale Aldo Moro 2, 00185 Roma, Italy\label{aff168}
\and
Centro de Astrof\'{\i}sica da Universidade do Porto, Rua das Estrelas, 4150-762 Porto, Portugal\label{aff169}
\and
Dipartimento di Fisica, Universit\`a di Roma Tor Vergata, Via della Ricerca Scientifica 1, Roma, Italy\label{aff170}
\and
INFN, Sezione di Roma 2, Via della Ricerca Scientifica 1, Roma, Italy\label{aff171}
\and
HE Space for European Space Agency (ESA), Camino bajo del Castillo, s/n, Urbanizacion Villafranca del Castillo, Villanueva de la Ca\~nada, 28692 Madrid, Spain\label{aff172}
\and
Department of Astrophysical Sciences, Peyton Hall, Princeton University, Princeton, NJ 08544, USA\label{aff173}
\and
Theoretical astrophysics, Department of Physics and Astronomy, Uppsala University, Box 515, 751 20 Uppsala, Sweden\label{aff174}
\and
Mathematical Institute, University of Leiden, Einsteinweg 55, 2333 CA Leiden, The Netherlands\label{aff175}
\and
School of Physics \& Astronomy, University of Southampton, Highfield Campus, Southampton SO17 1BJ, UK\label{aff176}
\and
Institute of Astronomy, University of Cambridge, Madingley Road, Cambridge CB3 0HA, UK\label{aff177}
\and
Department of Physics and Astronomy, University of California, Davis, CA 95616, USA\label{aff178}
\and
Space physics and astronomy research unit, University of Oulu, Pentti Kaiteran katu 1, FI-90014 Oulu, Finland\label{aff179}
\and
Center for Computational Astrophysics, Flatiron Institute, 162 5th Avenue, 10010, New York, NY, USA\label{aff180}
\and
Department of Astronomy, University of Massachusetts, Amherst, MA 01003, USA\label{aff181}
\and
Univ. Lille, CNRS, Centrale Lille, UMR 9189 CRIStAL, 59000 Lille, France\label{aff182}
\and
Department of Physics and Astronomy, University of British Columbia, Vancouver, BC V6T 1Z1, Canada\label{aff183}  
\and
Centre National d'Etudes Spatiales (CNES), 2, Place Maurice Quentin, 75039, Paris, France\label{aff184}
\and 
Université Paris-Saclay, CentraleSupélec, 8-10 rue Joliot-Curie, Gif-sur-Yvette, France\label{aff185}
}




\abstract{
A large catalogue of candidate galaxy protoclusters with high star-formation rates was produced by the \Planck\ collaboration. 
We search, in the first data release (Q1) of the \Euclid survey, for the visible and infrared counterparts of the eight \Planck\ galaxy protocluster candidates in \Euclid Q1, expected to be above $z > 1.5$. Our goal is to investigate the optical nature of these overdensities previously detected in the submillimetre wavelength range.
We make use of the catalogues delivered by the \Euclid Science Ground Segment (SGS), especially the positions, photometry, photometric redshifts, and derived physical parameters. We complete these catalogues with refined estimations of redshifts and physical parameters using photometric data from \Spitzer. After a galaxy selection on the \HE magnitude and on the photometric redshift quality in the \Euclid data, we search for overdensities using the \texttt{DETECTIFz} algorithm, an overdensity finder based on Delaunay tessellation that uses photometric redshift probability distributions through Monte Carlo simulations. 
Focusing our search on the eight high-star forming \Planck\ protocluster candidates, we find that two of them have one \Euclid counterpart, and six have between two and four \Euclid counterparts, which amounts to a total of 20 \Euclid\ counterparts. This illustrates the taxonomy of \Planck\ candidates, made of multiple structures along the line of sight as well as single structures.
These \Euclid counterparts lie at photometric redshifts $1.4<z_{\rm ph} < 2.7$ and 12 of them also have partial \Herschel\ coverage. All detections have also been confirmed by at least one other independent protocluster detection algorithm, namely the Poisson Probability Method and Detection Algorithm for NAscent Structures. We study the colours, derived stellar masses and star-formation rates (SFRs) of the detected member galaxies of those protocluster candidate counterparts. We also estimate the total stellar masses, SFRs, and the halo mass lower limits for all \Euclid\ protocluster candidates using stellar mass-halo mass relations from the literature. 
We find that in the dark matter halo mass ($\Mh$) / redshift plane, these \Planck\ and \Euclid overdense regions lie in the region $12.6 <\logten (\Mh/M_\odot)< 13.4$, $1.4<z< 2.7$. This means that the halos of our objects are expected to have experienced a transition between cold flows in hot media to accretion of hot material.

}


   \titlerunning{The first \Euclid view on \Planck\ galaxy protocluster candidates at cosmic noon}
   
   \authorrunning{Euclid Collaboration: T. Dusserre et al.}
   
   \maketitle
   \nolinenumbers
   
\section{\label{sect:Intro}Introduction}

Large-scale structure in the Universe forms hierarchically via the gravitational collapse of initially low-amplitude density perturbations. Mass is distributed in the form of a complex cosmic web network of filaments, walls, voids, and nodes. The most massive nodes, at the intersection of filaments, are the sites of clusters of galaxies and of their high-redshift progenitors, usually called `protoclusters' \citep{Overzier2016,alberts22,remus2023}. Understanding the transition of the present largest gravitationally bound structures, namely clusters of galaxies, from an early protocluster stage is key if we are to understand the full process of matter assembly in the Universe.

The most common definition for a protocluster is an overdensity in the matter field that should reach a virial mass of at least $10^{14} M_\odot$ at $z=0$ under the spherical collapse model \citep{chiang2013,muldrew2015,Overzier2016,chiang2017,alberts22,remus2023}. However, this definition is difficult to use with observational data. Indeed, protoclusters have mainly been detected as galaxy overdensities, whereas their most massive components are dark matter and baryonic gas, that account for about 95\% of the total mass \citep{fukugita2004,ettori2003,gonzalez2013,eckert2016,chiu2016}. Fortunately, it has been shown that star-forming galaxies act as good tracers for protoclusters \citep{casey2016}. Using this definition, we primarily detect highly star-forming protoclusters. This leads to another definition more relevant for observations: \cite{gouin2022} showed from simulations that overdensities at redshift $z>1.3$ containing at least seven gravitationally bound star-forming galaxies have a 92\% chance of being protoclusters. 

\begin{figure}[htbp!]
    \centering
    \includegraphics[width=\linewidth]{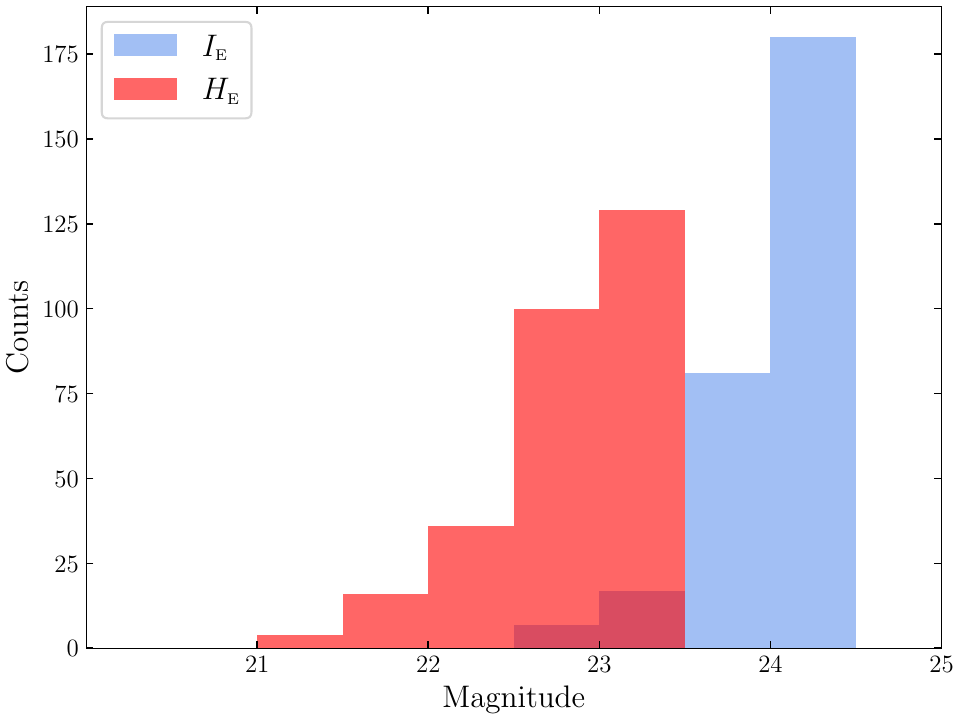}
    \caption{Histograms of \IE and \HE magnitudes for all galaxy members. Our cut at $\HE \leq 23.5$ is clearly visible.}
    \label{fig:histo_magnitudes_I_H}
\end{figure}

Protoclusters are not only pivotal for understanding structure formation; they also contain crucial information about the evolution of galaxies and gas-heating processes \citep[for a recent review, see][]{alberts22}. In this context, a particularly crucial, but still not fully understood mechanism, is the transition from star-forming to quiescent galaxies \citep[quenching,][]{martig2009,schawinski2014}. This marks the end of the peak epoch for the star formation history of the Universe, called the Cosmic Noon \citep[$z\,{\simeq}\,1.5$--$3$,][]{madau14}. Simulations \citep{dekel06} and observational data \citep[e.g.,][]{daddi2021} also suggest the existence of a transition in gas feeding of the structures at $z \simeq 2$. Since important episodes of star-formation occurred in protoclusters \citep[e.g.,][]{kodama2007}, several approaches have been proposed to capture this epoch, such as the search for overdensities of H$\alpha$ and Ly$\alpha$ emitters \citep{steidel2000,shi2019,koyama13} or of dusty star-forming galaxies and sub-millimetre (submm) galaxies \citep{chapman09, casey15,wang2016, miller18, koyama2021, hill2024}.

The largest catalogue of protocluster candidates was delivered by \Planck\footnote{\Planck\ mission: \cite{planck2020_I}}. It contains 2151 high-redshift submm sources over 28\% of the sky \citep{planck16}. The \Planck\ protocluster candidates \citep{planck16} were selected by colour in the High-Frequency Instrument (HFI) submm-cleaned images, targeting $z \simeq2$ groups of rest-frame far-IR galaxies having significant star-formation activity. This sample is thus different from the Sunyaev--Zeldovich (SZ) cluster catalogue \citep{plancksz_cat}. We refer the reader to \cite{planck16} for details on the construction of the \Planck\ catalogue of protocluster candidates\footnote{Called `\texttt{PHZ}' for the \Planck\ catalogue of high-$z$ candidates. We shall avoid use of this label here because of confusion with the ‘\texttt{PHZ}' (photometric redshifts) of \Euclid.}. \cite{gouin2022} used the IllustrisTNG simulations \citep{pillepich2018,nelson2019} to reproduce the \Planck\ selection, confirming the contamination of star-forming sources along the line of sight \citep[like][]{negrello2017} but also showing that more than 70\% of the \Planck\ protocluster candidates are expected to evolve into galaxy clusters by $z = 0$.

Follow-up observations and physical characterisation of hundreds of \Planck\ candidates were also performed by \Herschel\ \citep{planck15} and \Spitzer\ \citep{martinache2018} in addition to studies of individual candidates \citep{flores-cacho2016,kneissl2019,polletta2021,koyama2021,polletta2022,hill2024}. Observations from \Planck\ and \Herschel\ were realised in the submm and far-infrared ranges respectively, which trace efficiently the SFR \citep{kennicutt1998}. The main limitation is the angular resolution of these telescopes (from \ang{;;30} to \ang{;5;}), the full width at half maximum of which can correspond to several times the typical angular size of a protocluster. Using the James Webb Space Telescope, \citet{polletta2024} also conducted a follow-up of one particular candidate. However, the number of confirmed protoclusters is still small and it is only via large surveys conducted at optical and near-infrared wavelengths that the situation is expected to drastically change. In particular, stage IV surveys, such as Rubin/LSST\footnote{\url{https://rubinobservatory.org}}, \textit{Roman}\footnote{\url{https://roman.gsfc.nasa.gov}} or \Euclid \citep{EuclidSkyOverview} in the context of this work offer unique possibilities to observe high-redshift galaxies and identify tens of thousands of overdense regions that could be associated with distant clusters and/or protoclusters \citep[e.g.,][]{EP-Boehringer}.

\begin{figure}
    \centering
    \includegraphics[width=\hsize,trim={0 0 0 1.5cm},clip]{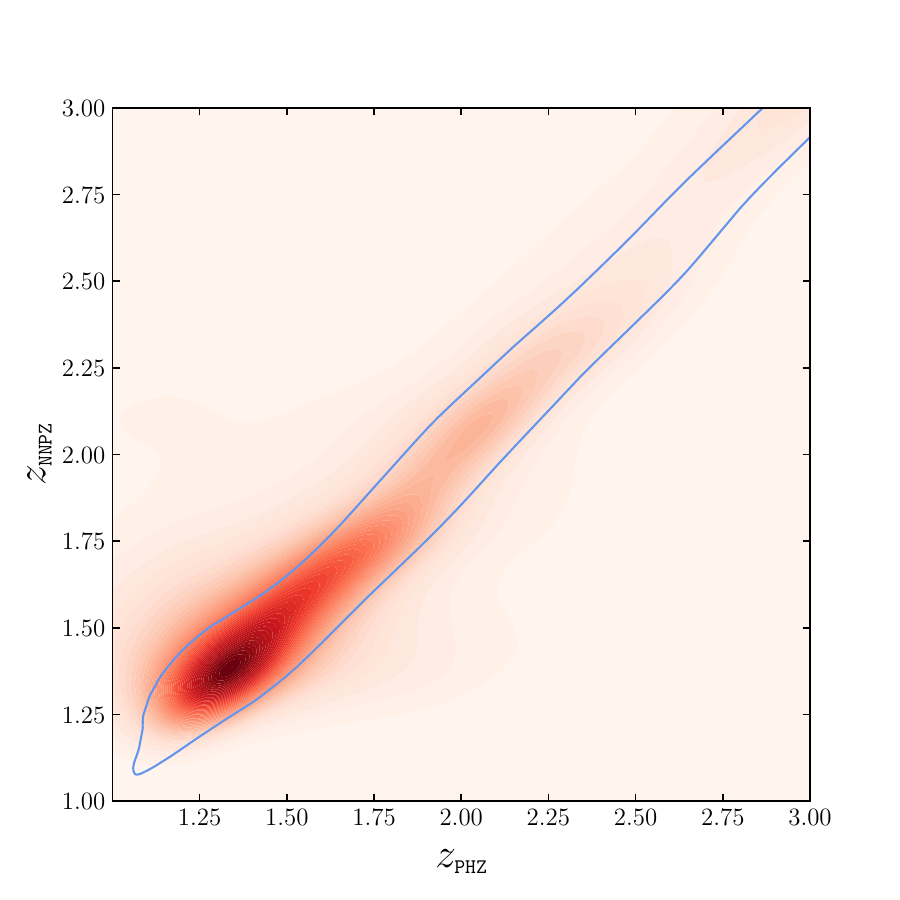}
    \caption{
     \texttt{NNPZ}-estimated redshift (\znnpz) versus \texttt{PHZ}-estimated photometric redshift (\zphz) plane for $z_{\rm \texttt{NNPZ}}>1.3$. The background shows the distribution of 100 000 sources around the eight \Planck\ protocluster candidates in the EDF-S and EDF-F fields. Darker tones represent higher densities of sources. The blue line is the iso-density contour inside which 68\% of sources lie. 
    }
    \label{fig:nnpz_vs_phz}
\end{figure}

\begin{figure*}[htbp!]
    \hspace{-2.2cm}
    \includegraphics[width=1.2\linewidth]{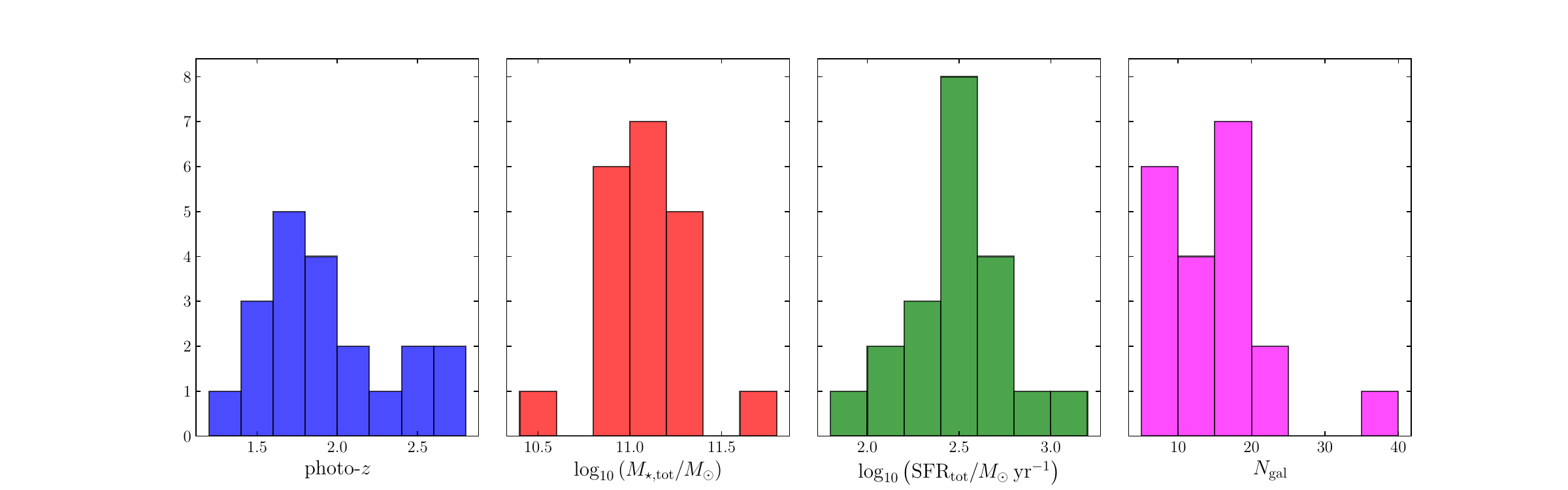}
    \caption{Histograms of the four main physical quantities (derived with \texttt{NNPZ}) of the 20 \Euclid\ counterparts of the eight \Planck\ protocluster candidates: photometric redshift; total stellar mass; total star-formation rate; and number of member galaxies.}
    \label{fig:histos}
\end{figure*}

In this study, we focus on the \Planck\ high-redshift protocluster candidates only. We detect and resolve them with \Euclid, then quantify their physical properties including stellar mass ($M_*$) and star-formation rate (SFR). This allows us to assess the evolutionary state they are in. Section~\ref{sect:ObservationsSamples} provides a description of the \Euclid observations and fields used in the study, specifically the tiles matching the positions of our eight \Planck\ sources. In Sect.~\ref{sect:Algorithms}, we describe the methods used to detect galaxy overdensities in the \Euclid\ data. In Sect.~\ref{sect:DetectionsPlanck} we highlight the 20 \Euclid overdensity detections as counterparts of eight \Planck\ protocluster candidates, and explore their physical properties. We discuss our findings in Sect.~\ref{sect:Discussion}, in particular the level to which our detections match the expectations of protoclusters and their evolutionary states. Finally, we conclude in Sect.~\ref{sect:conclusion} and we formulate the open questions that this study raises, for instance about the morphology of protoclusters. Throughout this paper, we use AB magnitudes, an IMF from \cite{chabrier03}, and the \Planck\ 2018 flat $\Lambda$CDM cosmology \citep{planck2020_I,planck2020_VI} with $H_0 = 67.66\,{\rm km\,s^{-1}\,Mpc^{-1}}$, $\Omega_{\rm b} = 0.0489$, $\Omega_{\rm m} = 0.3111$, and $\Omega_{\Lambda} = 0.6889$.

\section{\label{sect:ObservationsSamples}\Euclid observations, data processing, and selected sky areas in the Q1 footprint}

\subsection{\Euclid observations and data processing}

\Euclid has observed the three \Euclid\ Deep Fields (EDFs), and the Q1 data release \citep{Q1-TP001} corresponds to the approximate depth of the \Euclid\ Wide Survey (EWS). Processing was performed by the science ground segment (SGS) from images delivered by the VIS \citep{EuclidSkyVIS,Q1-TP002} and NISP instruments \citep{EuclidSkyNISP,Q1-TP003}. The multiwavelength photometric catalogue was delivered by the `merged' processing function MER \citep{Q1-TP004} and the `photometric redshift' processing function (PF), or \texttt{PHZ} for short \citep{Q1-TP005}. 

\subsection{\label{sect:nnpz}\Euclid data}

We use the catalogues obtained from the \Euclid VIS (see Appendix~\ref{appendix:VIS_images_planck}) and NIR stacked tiles in bands \IE, \YE, \JE, and \HE delivered by MER. We use photometry and the photometric redshift probability distribution functions (PDFs) from \cite{Q1-SP031}. This extends the MER and \texttt{PHZ} catalogues by including \Spitzer\ infrared bands and other ground-based data. \cite{Q1-SP031} use the \texttt{NNPZ} algorithm \citep{tanaka2018,Desprez-EP10} to derive the photometric redshifts. Adding \Spitzer\ data allows estimates of the photo-$z$ to be based on more photometric points than \texttt{PHZ}, and the catalogue and associated physical parameters are thus more relevant for our study of galaxies potentially located at $z>1.5$. For reference, the spectral bands from \Euclid's VIS to \Spitzer's IRAC 2 correspond to a rest-frame wavelength range of $200 - 1500 \,  {\rm nm}$ at $z=2$. Stellar masses and SFRs are also derived from the photometric data, using the IMF from \cite{chabrier03}.

\subsection{Selected sky areas: the \Euclid\ field of interest for eight \Planck\ protocluster candidates}

Eight \Planck\ protocluster candidates fall in \Euclid Q1 fields, all of them being in EDF-S and EDF-F. The footprint of the EDF-F and EDF-S Q1 fields are shown in Fig.~\ref{fig:edf}, together with the positions of the \Planck\ protocluster candidates of \cite{planck16}. In Table~\ref{tab:planck_fields} we report the \Euclid\ MER tile number, names, and coordinates for the eight \Planck\ targets.

\texttt{DETECTIFz} assumes the flat sky approximation to compute the source density. For each \Planck\ protocluster candidate, we select a circle of \ang{1;} diameter around it to estimate the background density. The maximum error in angular separation due to the flat sky approximation is then \ang{;;0.04}, about one order of magnitude lower than the angular resolution of \Euclid. In three cases, such selection contains two different \Planck\ candidates, some of them being separated by only a few arcminutes, as can be seen in Fig.~\ref{fig:edf}.

\begin{table*}
\setlength{\tabcolsep}{3pt}
\centering

\caption{\Euclid protoclusters detected with \texttt{DETECTIFz} at the location of \Planck\ protocluster candidates.}
\label{tab:selected_protoclusters_top}
\begin{tabular}{l@{\hskip 1em}ccccccccccc}
\hline\hline
\noalign{\vskip 2pt}
\Planck\ + \Euclid  & RA & Dec  & $z_{\rm ph}$ &  $N_{\rm gal}$ & S/N   & Radius  & $R_{200}$   & $M_{\ast,{\rm tot}}$      &  \multicolumn{3}{c}{SFR$_{\rm tot}$ [$\logten({\rm SFR}/M_{\odot}\,\rm{yr}^{-1})$]}  \\ 
 overdensity name  & [deg]   &  [deg]  &           &      &      & [arcmin] & [arcmin] & [$\logten(M_*/M_{\odot})$] &  \Euclid  &  \Herschel  & \Planck \\
   \noalign{\vskip 2pt}
\hline

\noalign{\vskip 2pt}
G221.09\_EUC\_1 	 & 	52.7334 	 & 	$-$26.4440 	 & 	$1.73^{+0.06}_{-0.06}$ 	 & 	15 	 & 	3.6 	 & 	2.14 	 & 	0.48 	 & 	$11.15^{+0.09}_{-0.10}$ 	 & 	$2.57^{+0.18}_{-0.36}$ 	 & 3.38$^{+0.08}_{-0.06}$ &  4.57$^{+0.08}_{-0.14}$ \\ 
 \noalign{\vskip 2pt}
G221.09\_EUC\_2 	 & 	52.6780 	 & 	$-$26.3657 	 & 	$1.98^{+0.06}_{-0.04}$ 	 & 	8 	 & 	3.1 	 & 	2.64 	 & 	0.37 	 & 	$10.95^{+0.12}_{-0.12}$ 	 & 	$2.33^{+0.15}_{-0.15}$ & 3.63$^{+0.07}_{-0.05}$ &  4.57$^{+0.08}_{-0.14}$ 	 \\ 
 \noalign{\vskip 2pt}
G221.09\_EUC\_3 	 & 	52.6976 	 & 	$-$26.4105 	 & 	$1.98^{+0.07}_{-0.09}$ 	 & 	12 	 & 	3.1 	 & 	2.19 	 & 	0.40 	 & 	$10.97^{+0.12}_{-0.13}$ 	 & 	$2.49^{+0.19}_{-0.19}$ & 3.71$^{+0.06}_{-0.04}$ &  4.57$^{+0.08}_{-0.14}$	 \\ 
\noalign{\vskip 2pt}
\hline
 \noalign{\vskip 2pt}
G222.05\_EUC\_1 	 & 	53.1945 	 & 	$-$26.8829 	 & 	$1.44^{+0.03}_{-0.05}$ 	 & 	9 	 & 	3.1 	 & 	1.64 	 & 	0.45 	 & 	$10.92^{+0.06}_{-0.09}$ 	 & 	$2.06^{+0.20}_{-0.14}$ 	& 2.86$^{+0.08}_{-0.06}$ &  4.41$^{+0.11}_{-0.21}$ \\ 
 \noalign{\vskip 2pt}
G222.05\_EUC\_2 	 & 	53.2056 	 & 	$-$26.9356 	 & 	$1.56^{+0.05}_{-0.04}$ 	 & 	16 	 & 	3.8 	 & 	2.74 	 & 	0.50 	 & 	$11.02^{+0.10}_{-0.09}$ 	 & 	$2.39^{+0.14}_{-0.20}$ 	& 3.29$^{+0.07}_{-0.05}$ &  4.41$^{+0.11}_{-0.21}$ \\ 
 \noalign{\vskip 2pt}
G222.05\_EUC\_3 	 & 	53.1627 	 & 	$-$26.9806 	 & 	$1.79^{+0.08}_{-0.06}$ 	 & 	11 	 & 	5.5 	 & 	2.42 	 & 	0.42 	 & 	$10.95^{+0.10}_{-0.11}$ 	 & 	$2.18^{+0.20}_{-0.17}$ 	& 3.09$^{+0.10}_{-0.08}$ &  4.41$^{+0.11}_{-0.21}$ \\ 
 \noalign{\vskip 2pt}
G222.05\_EUC\_4 	 & 	53.1602 	 & 	$-$26.8222 	 & 	$2.60^{+0.09}_{-0.08}$ 	 & 	16 	 & 	5.1 	 & 	6.33 	 & 	0.41 	 & 	$11.29^{+0.11}_{-0.12}$ 	 & 	$2.84^{+0.12}_{-0.16}$ 	& 4.49$^{+0.03}_{-0.02}$ &  4.87$^{+0.08}_{-0.11}$ \\ 
\noalign{\vskip 2pt}
\hline
 \noalign{\vskip 2pt}
G222.75\_EUC\_1 	 & 	51.3510 	 & 	$-$27.5813 	 & 	$1.66^{+0.14}_{-0.06}$ 	 & 	17 	 & 	3.4 	 & 	1.86 	 & 	0.51 	 & 	$11.15^{+0.10}_{-0.12}$ 	 & 	$2.45^{+0.22}_{-0.16}$ 	& 3.13$^{+0.07}_{-0.06}$ &  4.17$^{+0.10}_{-0.18}$ \\ 
\noalign{\vskip 2pt}
\hline
 \noalign{\vskip 2pt}
G223.18\_EUC\_1 	 & 	52.5416 	 & 	$-$27.7002 	 & 	$1.97^{+0.13}_{-0.01}$ 	 & 	9 	 & 	3.2 	 & 	2.90 	 & 	0.38 	 & 	$11.00^{+0.10}_{-0.12}$ 	 & 	$2.45^{+0.15}_{-0.15}$ 	& 3.77$^{+0.05}_{-0.03}$ &  4.22$^{+0.12}_{-0.36}$ \\ 
 \noalign{\vskip 2pt}
G223.18\_EUC\_2 	 & 	52.6231 	 & 	$-$27.6509 	 & 	$2.39^{+0.17}_{-0.14}$ 	 & 	36 	 & 	5.5 	 & 	4.69 	 & 	0.56 	 & 	$11.64^{+0.12}_{-0.12}$ 	 & 	$3.00^{+0.14}_{-0.17}$ 	& 4.48$^{+0.02}_{-0.02}$ &  4.50$^{+0.11}_{-0.28}$ \\ 
\noalign{\vskip 2pt}
\hline
 \noalign{\vskip 2pt}
G254.49\_EUC\_1 	 & 	60.6815 	 & 	$-$47.2287 	 & 	$1.59^{+0.02}_{-0.08}$ 	 & 	13 	 & 	3.6 	 & 	1.78 	 & 	0.49 	 & 	$11.12^{+0.11}_{-0.11}$ 	 & 	$2.40^{+0.27}_{-0.37}$ 	& ... &  4.09$^{+0.12}_{-0.20}$ \\ 
 \noalign{\vskip 2pt}
G254.49\_EUC\_2 	 & 	60.7235 	 & 	$-$47.2912 	 & 	$1.64^{+0.08}_{-0.04}$ 	 & 	20 	 & 	4.8 	 & 	2.93 	 & 	0.56 	 & 	$11.32^{+0.09}_{-0.11}$ 	 & 	$2.65^{+0.19}_{-0.24}$ 	& ...                    &  4.09$^{+0.12}_{-0.20}$ \\ 
 \noalign{\vskip 2pt}
G254.49\_EUC\_3 	 & 	60.7705 	 & 	$-$47.3329 	 & 	$2.13^{+0.08}_{-0.10}$ 	 & 	16 	 & 	4.7 	 & 	3.13 	 & 	0.43 	 & 	$11.16^{+0.12}_{-0.11}$ 	 & 	$2.52^{+0.19}_{-0.26}$ 	& ...                    &  4.09$^{+0.12}_{-0.20}$ \\ 
 \noalign{\vskip 2pt}
G254.49\_EUC\_4 	 & 	60.7472 	 & 	$-$47.2737 	 & 	$2.61^{+0.11}_{-0.12}$ 	 & 	15 	 & 	4.2 	 & 	4.51 	 & 	0.41 	 & 	$11.30^{+0.12}_{-0.14}$ 	 & 	$2.72^{+0.17}_{-0.21}$ 	& ...                    &  4.37$^{+0.09}_{-0.13}$ \\ 
\noalign{\vskip 2pt}
\hline
 \noalign{\vskip 2pt}
G254.74\_EUC\_1 	 & 	60.7651 	 & 	$-$47.4641 	 & 	$1.81^{+0.01}_{-0.07}$ 	 & 	12 	 & 	3.1 	 & 	2.73 	 & 	0.43 	 & 	$11.07^{+0.11}_{-0.11}$ 	 & 	$2.56^{+0.18}_{-0.22}$ 	& 2.53$^{+0.28}_{-0.28}$ &  4.31$^{+0.10}_{-0.11}$ \\ 
 \noalign{\vskip 2pt}
G254.74\_EUC\_2 	 & 	60.8219 	 & 	$-$47.4658 	 & 	$2.01^{+0.11}_{-0.09}$ 	 & 	17 	 & 	4.3 	 & 	3.17 	 & 	0.48 	 & 	$11.26^{+0.11}_{-0.11}$ 	 & 	$2.61^{+0.17}_{-0.23}$ 	& 3.20$^{+0.22}_{-0.19}$ &  4.56$^{+0.05}_{-0.07}$ \\ 
\noalign{\vskip 2pt}
\hline
 \noalign{\vskip 2pt}
G257.45\_EUC\_1 	 & 	57.5446 	 & 	$-$48.7012 	 & 	$2.47^{+0.10}_{-0.05}$ 	 & 	7 	 & 	3.3 	 & 	3.59 	 & 	0.34 	 & 	$11.11^{+0.13}_{-0.12}$ 	 & 	$2.36^{+0.27}_{-0.09}$ 	& ...                    &  4.50$^{+0.07}_{-0.12}$ \\ 
\noalign{\vskip 2pt}
\hline
 \noalign{\vskip 2pt}
G257.71\_EUC\_1 	 & 	59.6343 	 & 	$-$49.3654 	 & 	$1.39^{+0.08}_{-0.01}$ 	 & 	7 	 & 	3.4 	 & 	1.35 	 & 	0.40 	 & 	$10.60^{+0.11}_{-0.10}$ 	 & 	$1.85^{+0.17}_{-0.21}$ 	& ...                    &  4.09$^{+0.09}_{-0.13}$ \\ 
 \noalign{\vskip 2pt}
G257.71\_EUC\_2 	 & 	59.7149 	 & 	$-$49.2533 	 & 	$1.79^{+0.06}_{-0.14}$ 	 & 	20 	 & 	4.0 	 & 	2.00 	 & 	0.53 	 & 	$11.34^{+0.11}_{-0.11}$ 	 & 	$2.62^{+0.27}_{-0.15}$ 	& ...                    &  4.09$^{+0.09}_{-0.13}$ \\ 
 \noalign{\vskip 2pt}
G257.71\_EUC\_3 	 & 	59.7054 	 & 	$-$49.3269 	 & 	$2.69^{+0.11}_{-0.08}$ 	 & 	9 	 & 	5.3 	 & 	5.38 	 & 	0.33 	 & 	$10.96^{+0.12}_{-0.15}$ 	 & 	$2.48^{+0.21}_{-0.15}$ 	& ...                    &  4.55$^{+0.05}_{-0.06}$ \\ 
 \noalign{\vskip 2pt}

\noalign{\vskip 2pt}
\hline
\end{tabular}
\tablefoot{
 Column descriptions: \Planck\ + \Euclid overdensity name, RA and Dec in degrees, \texttt{DETECTIFz} photometric redshift, based on \Euclid photometric redshifts, number of galaxies in the protocluster, signal-to-noise ratio (S/N) of the overdensity detection by \texttt{DETECTIFz}, equivalent radius of overdensity in the highest S/N redshift slice, as estimated by \texttt{DETECTIFz} in arcminutes, angular $R_{200}$ of a virialised halo at the corresponding redshift, as estimated in \cite{mo2002}, total stellar mass of the protocluster (the sum of the stellar masses of all member galaxies, coming from \Euclid \texttt{NNPZ}), total SFR of the protocluster (the sum of the SFRs of all member galaxies, coming from \Euclid \texttt{NNPZ}), and the SFR derived with data from \Euclid\ (lower limits), \Planck, and \Herschel\ (see Sect.~\ref{sect:sfrs}). Absence of data is denoted with ellipsis.}
\end{table*}

\section{\label{sect:Algorithms}Protocluster detection algorithms and selection}

\subsection{\label{sect:input_selection}Selection for input}

Uncertainties on photometry and astrometry come from the \Euclid SGS MER PF \citep{Q1-TP004}. Photometric redshifts and derived physical parameters (stellar masses and SFRs) are obtained from \texttt{PHZ} \citep{Q1-TP005} and \texttt{NNPZ} \citep{Q1-SP031}. We thus have two different values for each parameter. Before applying the \texttt{DETECTIFz} algorithm (see Sect.~\ref{sect:detectifz}), we make the following selections: 

\begin{itemize}
    \item $\HE < 23.5$ (Fig.~\ref{fig:histo_magnitudes_I_H}). We take a stronger limit than the \texttt{PHZ} usual limit at $\HE < 24$ \citep{Q1-TP005} to remain on the cautious side for the photometric redshift determination. According to our tests, a higher cut allows for more detections, but these have S/N $<3$ in general.
    \item Sources must be relatively isolated from bright stars and/or known remaining artefacts. To account for this we select MER sources with a spurious probability below 20\%.
    \item Sources have no flags raised in the \texttt{PHZ} catalogue.
    \item The redshift estimated by \texttt{PHZ} is within $1 \,\sigma$ error margin of the redshift estimated by \texttt{NNPZ} (contours in Fig.~\ref{fig:nnpz_vs_phz}). Both pipelines are prone to different systematics: MER and \texttt{PHZ} can suffer from biased flux measurements due to persistences in the infrared bands, and \texttt{NNPZ} can blend the infrared fluxes from two sources if their angular separation lies below \Spitzer 's angular resolution. Ensuring that both methods give consistent results allows us to eliminate part of these problems.
    \item $\delta z_{\rm ph} / (1+z_{\rm ph}) < 0.20$ , with $z_{\rm ph}$ and $\delta z_{\rm ph} $ being the photometric redshift and uncertainty respectively. Above this threshold, the redshift error blends \texttt{DETECTIFz}'s 3D map (Sect.~\ref{sect:detectifz}).
\end{itemize}

\subsection{\label{sect:detectifz} \texttt{DETECTIFz} protocluster-finding algorithm}

With the goal of identifying the counterparts of the \Planck-detected protocluster candidates, we apply the \texttt{DETECTIFz} (DElaunay TEssellation ClusTer IdentiFication with photo-$z$) algorithm \citep{sarron2021} on the galaxies located less than \ang{1;;} around each \Planck\ protocluster candidate. 
The \texttt{DETECTIFz} algorithm uses the Delaunay Tessellation Field Estimator (DTFE) to identify extended galaxy overdensities within redshift slices. This method is entirely empirical and independent of cosmological models, relying exclusively on galaxy sky coordinates and samples drawn from the photometric redshift probability distribution.

Beginning with a galaxy catalogue containing sky coordinates and photometric redshift probabilities, \texttt{DETECTIFz} constructs a 3D overdensity map. Overdensities are estimated in redshift slices spaced by 0.01 and with varying widths of $\sigma_{\rm z}({\rm z})$, derived from the input galaxy catalogue.
For each slice, \texttt{DETECTIFz} generates 100 Monte Carlo realisations of the overdensity map by sampling from the photometric redshift probability distribution and applying DTFE to each sample. The final density map is obtained by averaging over these realisations. 

If the redshift error is too high, galaxies will spread over a large redshift range in the 100 Monte Carlo realisations, resulting in an estimation of the mean density biased towards lower values. This motivated our choice to only use galaxies fulfilling $\delta z_{\textrm{ph}} / (1+z_{\textrm{ph}}) < 0.20$ as an input for \texttt{DETECTIFz}, so the detection is based upon galaxies with accurate enough photometric redshifts.

\begin{figure}[htbp!]
\centering
\includegraphics[angle=0,width=\hsize]{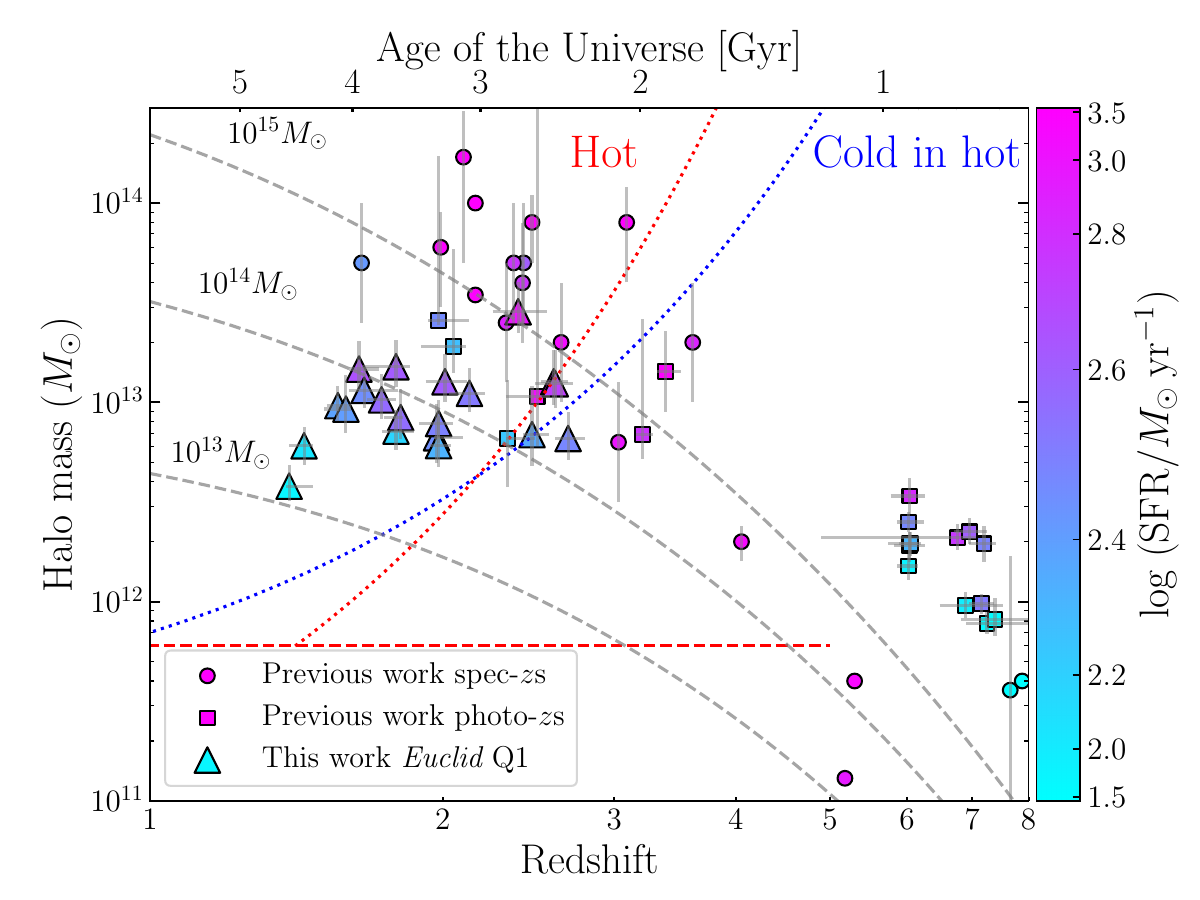}
\caption{Protocluster halo mass ($M_{\rm h}$) versus redshift $z$, colour-coded by star-formation rate (SFR, colour bar). Triangles represent \Euclid protocluster, with the halo mass estimated using Method~B \citep[see Sect.~\ref{sect:halo_mass} and][]{shuntov2022}. Circles and squares are protoclusters from the literature \citep{casey2016,hill2020,polletta2021,laporte2022,morishita2023,morishita2024,sillassen2024,shimakawa2024}. The red and blue lines illustrate the different predicted gas-cooling regimes: the red dotted line comes from \cite{dekel06} and the blue dotted line from \cite{daddi2022} and separate loci of cold gas in a hot medium (right) and hot gas (left); halo masses below the horizontal red dashed line $M_{\rm shock}$, coming from \citet{dekel06}, are predicted to contain only cold flows with no shock heating within halos at $z<5$. The grey lines indicate the average halo mass growth history for halos of present day masses $10^{13}$, $10^{14}$, and $10^{15} \: M_\odot$ as determined from \citet{behroozi2013}.
}
\label{fig:arthur_mh_vs_z}
\end{figure}

Within each slice, overdensities are identified as extended peaks in the density map. For each detection, the algorithm records a bounding box ($\text{RA}_\text{min}$, $\text{RA}_\text{max}$, $\text{Dec}_\text{min}$, $\text{Dec}_\text{max}$) that encloses regions with pixel values above the (S/N)$_{\text{min}}$ threshold. To eliminate multiple detections of the same protocluster across adjacent slices or substructures, the algorithm merges peaks that are contiguous along the line of sight (i.e., in the redshift direction) and have either overlapping bounding boxes or peak separations of less than 2 comoving Mpc. These merged regions constitute the final galaxy protocluster candidates.

This selected sample (Sect.~\ref{sect:input_selection}) is used as an input for \texttt{DETECTIFz}, to which we provide celestial coordinates and the full photometric redshift PDF. We force \texttt{DETECTIFz} to search within the redshift range $1.35$--$3.5$, and to only select structures above an S/N of 3. \texttt{DETECTIFz} produces a list of overdensities and includes S/N, as well as measurements of overdensity and radius \citep{sarron2021}. In \texttt{DETECTIFz}, S/N is computed for each detection in each redshift slice, within a disc of fixed comoving radius of 500 kpc ($R_{\rm det}$)\footnote{This is equivalent to radii of about \ang{;0.4;} at $2.5<z<3.5$, and \ang{;0.5;}--\ang{;0.6;} at $1.35<z<2$. This radius is more than an order of magnitude smaller than the \Planck\ beam (FWHM$\,=\,$\ang{;4.7;}, and similar to the \Herschel\ beam (FWHM$\,\simeq\,$24\arcsec\ at 350 $\mu$m).} around the peak. For each slice, we calculate the mean ($\mu_\delta$) and standard deviation ($\sigma_\delta$) of the logarithm of the entire overdensity map $\delta$, after applying a $3\,\sigma$-clipping to remove outliers. The mean overdensity signal in the region around the peak is then compared to these global values to form the overall S/N:

\vspace{-0.2cm}
\begin{equation}
{\rm S/N} = \frac{\langle \logten(1 + \delta) \rangle_{R_{\rm det}} - \mu_\delta}{\sigma_\delta} \cdot
\end{equation}
\vspace{-0.2cm}

\noindent For each detection, \texttt{DETECTIFz} computes an equivalent angular scale corresponding to the radius of a disc having the same area as the zone inside which S/N $>3$.

\subsection{\label{sect:outputselection}Post detection after \texttt{DETECTIFz}: the \Euclid-\Planck\ protocluster sample}
 
For each overdensity in the output catalogue of \texttt{DETECTIFz}, we build a catalogue of member galaxies. 
We chose a conservative approach and consider a galaxy as a member if its angular separation with the overdensity centre is smaller than the overdensity radius, and its most probable redshift (estimated by \texttt{NNPZ}) falls within the $3\,\sigma$ range of the overdensity redshift.

Among all overdensities, we only select those having at least seven galaxy members. Structures containing at least seven star-forming galaxies at $z=2$ have a 92\% chance of reaching a virial mass $\geq 10^{14} M_\odot$ at $z=0$ \citep{gouin2022}. Although we do not impose an SFR threshold in the input catalogue, we ran tests on simulations to ensure that this criterion alone leads to a higher rate of true detections (Sect.~\ref{sect:purity}). 

We are then left with 20 \Euclid\ counterparts of eight \Planck\ protocluster candidates (Tab.~\ref{tab:selected_protoclusters_top}). Six of the eight \Planck\ protocluster candidates actually have several \Euclid\ counterparts. In total, there are 285 member galaxies in these 20 \Euclid-\Planck\ protoclusters. Cutouts are displayed in Fig.~\ref{fig:cutouts}

\subsection{\label{sect:purity}Estimating the purity and completeness of our detection method with simulations}

\begin{table}[h!]
    \centering
    \caption{Purity of \texttt{DETECTIFz} outputs for different redshift bins and S/N thresholds, as determined from a run on the GAEA light cone.}
    \begin{tabular}{l@{\hskip 2em} cccc}
        \noalign{\vskip 2pt}
        \hline 
        \hline
        \noalign{\vskip 2pt}
       $z$ bin & 1.3--2 & 2--2.5 & 2.5--3 \\
       \hline
       \noalign{\vskip 2pt}
        S/N$>2$ & 0.69 & 0.85 & 0.76 \\
       \noalign{\vskip 2pt}
        S/N$>3$ & 0.81 & 0.97 & 0.93 \\
        \noalign{\vskip 2pt}
        S/N$>3$ , 7+gal & 0.93 & 0.98 & 0.96 \\
        \noalign{\vskip 2pt}
        \hline
        \noalign{\vskip 2pt}
    \end{tabular}
    \tablefoot{The last line shows the purity of detections with S/N $>3$ and at least seven member galaxies, which is the criterion used in this work.}
    \label{tab:purity}
\end{table}

We make use of the simulated GAEA lightcone developed by the galaxy cluster working group of the \Euclid\ consortium \citep{EP-Boehringer}, containing a galaxy catalogue spanning a disc of radius \ang{1;;} and replicating \Euclid's redshift selection function\footnote{\url{https://drive.google.com/drive/folders/14lQ1H_wFmp8kITiKHgL2X5jzttK5KU9A}} as well as a truth table\footnote{\url{https://drive.google.com/drive/folders/1PMEkg039_N0bSiHCvlVhifI9oVXS69b9}}. We run \texttt{DETECTIFz} on the simulated data and compute how many of the resulting detections are true. According to \cite{sarron2021}, the purity of the \texttt{DETECTIFz} output depends on redshift and S/N, therefore it is more convenient to compute it in different redshift bins (cf. Table~\ref{tab:purity}). Each entry in this table is computed from a subsample of at least 114 detections. Higher S/N leads to better purity, and the latter is higher in the redshift bin $2<z<2.5$. The purity inside the $1.3<z<2$ redshift bin is significantly raised by selecting overdensities with at least seven galaxy members, as we do with the \Euclid\ Q1 data (Sect.~\ref{sect:outputselection}). Taking a mean purity of 95\%, we can therefore state with confidence that at least 19 out of our 20 detections are indeed protoclusters.

By comparing the number of true protoclusters detected with the total number of protoclusters inside the lightcone, we obtain the completeness of our output. Using the same cuts as those listed in Sect.~\ref{sect:outputselection}, we detect 804 out of the 934 simulated protoclusters, which corresponds to a completeness of 86\%. 

\begin{table*}[htbp!]
\caption{Halo masses $M_{\rm h}$ for selected \Euclid protoclusters detected with \texttt{DETECTIFz} at the locations of \Planck+\Herschel\ protocluster candidates, obtained from the stellar masses with Methods~A and B (Sect.~\ref{sect:halo_mass}).}
\label{tab:halo_masses}
\centering
\begin{tabular}{l@{\hskip 2em}ccccc}
\hline\hline
\noalign{\vskip 2pt}
\Planck\ + \Euclid    &   $M_{\ast {\rm c}}$  & $M_{\rm h}$ (A)            & $M_{\rm h}$ (\citetalias{behroozi2013})     & $M_{\rm h}$ (\citetalias{legrand2019})     & $M_{\rm h}$ (\citetalias{shuntov2022})  \\ 
overdensity name &   [$\logten(\Mh/M_\odot)$]        &  [$\logten(\Mh/M_\odot)$] & [$\logten(\Mh/M_\odot)$] & [$\logten(\Mh/M_\odot)$] & [$\logten(\Mh/M_\odot)$] \\
 \noalign{\vskip 2pt}
  \hline
  
\noalign{\vskip 2pt}
G221.09\_EUC\_1 	 & 	$11.15^{+0.09}_{-0.10}$ 	 & 	$11.88^{+0.09}_{-0.10}$ 	 & 	$12.35^{+0.10}_{-0.14}$ 	 & 	$12.99^{+0.08}_{-0.06}$ 	 & 	$13.01^{+0.13}_{-0.10}$ 	 \\ 
 \noalign{\vskip 2pt}
G221.09\_EUC\_2 	 & 	$10.95^{+0.12}_{-0.12}$ 	 & 	$11.68^{+0.12}_{-0.12}$ 	 & 	$12.30^{+0.09}_{-0.14}$ 	 & 	$12.76^{+0.11}_{-0.07}$ 	 & 	$12.78^{+0.15}_{-0.11}$ 	 \\ 
 \noalign{\vskip 2pt}
G221.09\_EUC\_3 	 & 	$10.97^{+0.12}_{-0.13}$ 	 & 	$11.70^{+0.12}_{-0.13}$ 	 & 	$12.25^{+0.09}_{-0.17}$ 	 & 	$12.86^{+0.09}_{-0.07}$ 	 & 	$12.89^{+0.12}_{-0.10}$ 	 \\ 
\noalign{\vskip 2pt}
\hline
 \noalign{\vskip 2pt}
G222.05\_EUC\_1 	 & 	$10.92^{+0.06}_{-0.09}$ 	 & 	$11.65^{+0.06}_{-0.09}$ 	 & 	$12.19^{+0.13}_{-0.10}$ 	 & 	$12.69^{+0.07}_{-0.07}$ 	 & 	$12.78^{+0.09}_{-0.09}$ 	 \\ 
 \noalign{\vskip 2pt}
G222.05\_EUC\_2 	 & 	$11.02^{+0.10}_{-0.09}$ 	 & 	$11.75^{+0.10}_{-0.09}$ 	 & 	$12.26^{+0.09}_{-0.13}$ 	 & 	$12.94^{+0.07}_{-0.05}$ 	 & 	$12.98^{+0.10}_{-0.08}$ 	 \\ 
 \noalign{\vskip 2pt}
G222.05\_EUC\_3 	 & 	$10.95^{+0.10}_{-0.11}$ 	 & 	$11.67^{+0.10}_{-0.11}$ 	 & 	$12.24^{+0.09}_{-0.15}$ 	 & 	$12.82^{+0.07}_{-0.06}$ 	 & 	$12.86^{+0.11}_{-0.09}$ 	 \\ 
 \noalign{\vskip 2pt}
G222.05\_EUC\_4 	 & 	$11.29^{+0.11}_{-0.12}$ 	 & 	$12.02^{+0.11}_{-0.12}$ 	 & 	$12.64^{+0.09}_{-0.13}$ 	 & 	$13.08^{+0.10}_{-0.08}$ 	 & 	$13.11^{+0.16}_{-0.12}$ 	 \\ 
\noalign{\vskip 2pt}
\hline
 \noalign{\vskip 2pt}
G222.75\_EUC\_1 	 & 	$11.15^{+0.10}_{-0.12}$ 	 & 	$11.88^{+0.10}_{-0.12}$ 	 & 	$12.25^{+0.10}_{-0.21}$ 	 & 	$13.03^{+0.08}_{-0.06}$ 	 & 	$13.06^{+0.11}_{-0.09}$ 	 \\ 
\noalign{\vskip 2pt}
\hline
 \noalign{\vskip 2pt}
G223.18\_EUC\_1 	 & 	$11.00^{+0.10}_{-0.12}$ 	 & 	$11.73^{+0.10}_{-0.12}$ 	 & 	$12.45^{+0.10}_{-0.15}$ 	 & 	$12.80^{+0.10}_{-0.09}$ 	 & 	$12.82^{+0.17}_{-0.12}$ 	 \\ 
 \noalign{\vskip 2pt}
G223.18\_EUC\_2 	 & 	$11.64^{+0.12}_{-0.12}$ 	 & 	$12.36^{+0.12}_{-0.12}$ 	 & 	$12.36^{+0.12}_{-0.16}$ 	 & 	$13.43^{+0.10}_{-0.07}$ 	 & 	$13.45^{+0.15}_{-0.11}$ 	 \\ 
\noalign{\vskip 2pt}
\hline
 \noalign{\vskip 2pt}
G254.49\_EUC\_1 	 & 	$11.12^{+0.11}_{-0.11}$ 	 & 	$11.85^{+0.11}_{-0.11}$ 	 & 	$12.47^{+0.15}_{-0.14}$ 	 & 	$12.94^{+0.10}_{-0.08}$ 	 & 	$12.97^{+0.17}_{-0.12}$ 	 \\ 
 \noalign{\vskip 2pt}
G254.49\_EUC\_2 	 & 	$11.32^{+0.09}_{-0.11}$ 	 & 	$12.05^{+0.09}_{-0.11}$ 	 & 	$12.45^{+0.12}_{-0.17}$ 	 & 	$13.14^{+0.09}_{-0.07}$ 	 & 	$13.17^{+0.14}_{-0.10}$ 	 \\ 
 \noalign{\vskip 2pt}
G254.49\_EUC\_3 	 & 	$11.16^{+0.12}_{-0.11}$ 	 & 	$11.89^{+0.12}_{-0.11}$ 	 & 	$12.29^{+0.10}_{-0.13}$ 	 & 	$13.02^{+0.09}_{-0.07}$ 	 & 	$13.05^{+0.13}_{-0.09}$ 	 \\ 
 \noalign{\vskip 2pt}
G254.49\_EUC\_4 	 & 	$11.30^{+0.12}_{-0.14}$ 	 & 	$12.03^{+0.12}_{-0.14}$ 	 & 	$12.57^{+0.13}_{-0.07}$ 	 & 	$13.07^{+0.11}_{-0.09}$ 	 & 	$13.09^{+0.17}_{-0.12}$ 	 \\ 
\noalign{\vskip 2pt}
\hline
 \noalign{\vskip 2pt}
G254.74\_EUC\_1 	 & 	$11.07^{+0.11}_{-0.11}$ 	 & 	$11.80^{+0.11}_{-0.11}$ 	 & 	$12.39^{+0.09}_{-0.15}$ 	 & 	$12.90^{+0.09}_{-0.07}$ 	 & 	$12.92^{+0.14}_{-0.11}$ 	 \\ 
 \noalign{\vskip 2pt}
G254.74\_EUC\_2 	 & 	$11.26^{+0.11}_{-0.11}$ 	 & 	$11.99^{+0.11}_{-0.11}$ 	 & 	$12.36^{+0.12}_{-0.14}$ 	 & 	$13.08^{+0.09}_{-0.07}$ 	 & 	$13.10^{+0.14}_{-0.10}$ 	 \\ 
\noalign{\vskip 2pt}
\hline
 \noalign{\vskip 2pt}
G257.45\_EUC\_1 	 & 	$11.11^{+0.13}_{-0.12}$ 	 & 	$11.84^{+0.13}_{-0.12}$ 	 & 	$12.49^{+0.13}_{-0.15}$ 	 & 	$12.83^{+0.15}_{-0.10}$ 	 & 	$12.84^{+0.25}_{-0.16}$ 	 \\ 
\noalign{\vskip 2pt}
\hline
 \noalign{\vskip 2pt}
G257.71\_EUC\_1 	 & 	$10.60^{+0.11}_{-0.10}$ 	 & 	$11.32^{+0.11}_{-0.10}$ 	 & 	$11.85^{+0.14}_{-0.12}$ 	 & 	$12.50^{+0.08}_{-0.05}$ 	 & 	$12.58^{+0.10}_{-0.08}$ 	 \\ 
 \noalign{\vskip 2pt}
G257.71\_EUC\_2 	 & 	$11.34^{+0.11}_{-0.11}$ 	 & 	$12.07^{+0.11}_{-0.11}$ 	 & 	$12.33^{+0.10}_{-0.15}$ 	 & 	$13.15^{+0.09}_{-0.07}$ 	 & 	$13.18^{+0.14}_{-0.10}$ 	 \\ 
 \noalign{\vskip 2pt}
G257.71\_EUC\_3 	 & 	$10.96^{+0.12}_{-0.15}$ 	 & 	$11.68^{+0.12}_{-0.15}$ 	 & 	$12.50^{+0.14}_{-0.00}$ 	 & 	$12.79^{+0.09}_{-0.08}$ 	 & 	$12.82^{+0.13}_{-0.11}$ 	 \\ 
 \noalign{\vskip 2pt}

\noalign{\vskip 2pt}
\hline
\end{tabular}
\tablefoot{ $M_{\rm h}$ (A) is the mass derived from method~A using the $\Omega_{\rm b}/\Omega_{\rm m}$ ratio and it is expected to give lower values. Method~B uses the stellar to halo mass relations \citetalias{behroozi2013}, \citetalias{legrand2019}, and \citetalias{shuntov2022}. The first one is applied to the most massive galaxy stellar mass of the protocluster (column $M_{\ast {\rm c}}$), whereas the second and third ones take dark matter halos of each member galaxy into account. As there could be more member galaxies than those we detected, these estimates can be taken as lower limits. The overdensity name refers to  Table~\ref{tab:selected_protoclusters_top}.
We plot in Fig.~\ref{fig:arthur_mh_vs_z} $M_{\rm h}$ (B \citetalias{shuntov2022}) versus redshift. Uncertainties come from both the measured stellar masses and the models linking stellar masses to halo masses (see Sect.~\ref{sect:halo_mass}).
}
\end{table*}

\section{\label{sect:DetectionsPlanck}Results}

\begin{figure*}[htbp!]
    \centering
    \includegraphics[width=\hsize]{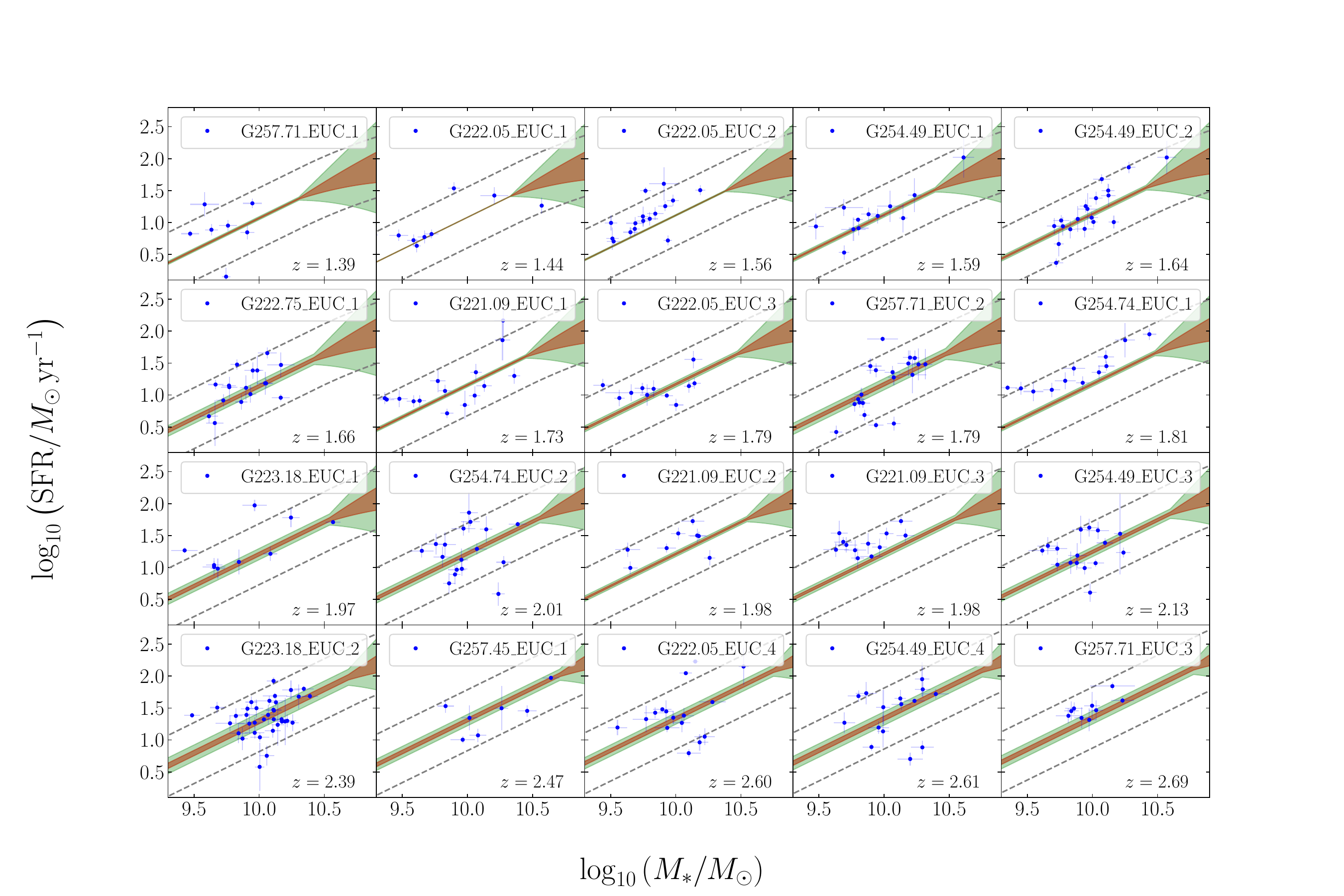}
    \caption{
    SFR$-M_\star$ of member galaxies of each overdensity. Blue dots show the position of each member galaxy of the 20 \Euclid\ protocluster candidates. Shaded region is the main sequence from \cite{schreiber2015} plotted with errors at 1 $\sigma$ (orange) and 3 $\sigma$ (green) computed from uncertainties on the model's parameters. Dashed lines indicate a factor of 3 around the main sequence. Protoclusters are ordered by increasing redshift, from left to right and from top to bottom. Most galaxies remain within a factor of 3 around the main sequence.
    }
    \label{fig:main_sequence}
\end{figure*}

\subsection{\label{sect:protocluster_candidates}Properties of the \Euclid\ protocluster candidates}

Throughout the rest of this paper, we use the term ‘counterpart' to refer to one of the 20 \Euclid-\Planck\ protocluster candidates (Sect.~\ref{sect:outputselection}). The physical parameters of the 20 \Euclid counterparts are listed in Tables~\ref{tab:selected_protoclusters_top} and~\ref{tab:halo_masses}.
These \Euclid\ counterparts are concentrated inside circles of \ang{;5;} radius centred on the coordinates of each \Planck\ high-redshift star-forming field. Each one of the eight fields has at least one \Euclid\ counterpart. We verified that the submillimetric fluxes measured by the \Planck\ HFI instrument for these eight candidates fall within the normal range of the \Planck\ protocluster candidate sample, and are, in that sense, representative of the whole catalogue.

\subsubsection{Position and photometric redshift distribution of the 20 \Euclid\ counterparts}

As mentioned above, among the eight \Planck\ protocluster candidates in \Euclid Q1, six have multiple counterparts. Two \Euclid\ protocluster candidates were found in G223.18$-$54.86 and G254.74$-$47.64, three in G221.09$-$54.59 and G257.71$-$47.99, and four in G254.49$-$47.73 and  G222.05$-$54.24. The combined submm emission of each group of detections fits within \Planck's large beam. Some of these fields could therefore be composed of multiple protoclusters, as previously expected \citep{negrello2017,gouin2022} and already observed \citep{flores-cacho2016,kneissl2019,polletta2021,polletta2022,hill2024}. G222.75$-$55.98 and G257.45$-$49.50 only have one \Euclid\ counterpart each. The EDF-F and EDF-S fields together with \Planck\ and \Euclid\ detections are shown in Fig.~\ref{fig:edf}. Two pairs of \Euclid\ detections, (G221.09\_EUC\_2; G221.09\_EUC\_3) and (G254.49\_EUC\_1; G254.49\_EUC\_2), share similar redshifts ($z_{\rm ph} \simeq 1.98$ and $z_{\rm ph} \simeq 1.6$, respectively, cf. Table~\ref{tab:selected_protoclusters_top}), while being also close to each other on the sky (see Appendix~\ref{sect:merging} for a discussion).  

The photometric redshift distribution for the 20 counterparts can be seen in Fig.~\ref{fig:histos}. There are notably less detections at $z\geq 2$; this should not be regarded as an absence of protoclusters but rather as a consequence of \Euclid's selection function, since source density in the MER catalogue drops at these redshifts, mainly due to the redshifted H $\alpha$ line leaving the \HE filter.

\subsubsection{\label{sect:sfrs}Stellar masses and SFRs}

The total \Euclid\ stellar mass and star-formation rate \citep{Q1-SP031} distributions of the detected structures are shown in Fig.~\ref{fig:histos}. Their respective mean values are $(1.5 \pm 0.8) \expo{11} M_\odot$ and $350 \pm 200 \: M_\odot \: {\rm yr}^{-1}$. Both maxima are reached by the same candidate (G223.18\_EUC\_2), which logically also has the highest number of galaxy members (36, see Table~\ref{tab:selected_protoclusters_top}). 

Neither \Planck\ nor \Herschel\ observations provided estimates of stellar and total masses. They are conducted in the submm and millimetre range, a spectral window suitable for SFR estimates but not to constrain stellar or total masses \citep{michalowski2012,michalowski14,lower2020}. We derive \Planck\ and \Herschel\ SFRs for each protocluster candidate using ${\rm SFR} /M_\odot \, {\rm yr}^{-1}  = 1.49 \times 10^{-10}\, L_{\rm FIR} / L_\odot$, where $L_{\rm FIR}$ is the integrated flux between 300 GHZ and 37.5 THz \citep{kennicutt12}. Several values are possible for \Planck\ SFRs depending on the assumed dust emission temperature, so we keep the estimate yielding the same redshift as the corresponding \Euclid\ counterpart. Concerning \Herschel\ sources, we only consider those for which the Spectral Energy Distribution (SED) best fit yields dust temperatures between 15\,K and 50\,K.

In general, multiple \Herschel\ sources are located inside the sky area of one \Euclid\ candidate. As we cannot estimate their redshift with high enough accuracy, it is not possible to identify which sources actually correspond to the \Euclid\ detection. We therefore choose a conservative approach and sum the SFR contributions of all \Herschel\ sources within the delimited area, thus obtaining a higher limit.

Our SFRs are lower than those estimated by \Planck. We could be missing galaxies inside each structure, as the selection function of \Euclid\ drops around $z=2$. This corresponds to a lower limit of $\logten (M_*/M_\odot) \simeq 9.5$ in the stellar mass of the member galaxies of our protocluster candidates. We do not add a cut at high stellar mass since according to \cite{mitra2024}, \Euclid\ should detect the same dusty star-forming submillimetre galaxies as \Herschel\ at $z>1.5$, even in the absence of far-infrared photometry. We first computed the expected missing stellar mass using the stellar mass function from \cite{ilbert2013}. Then, assuming the majority of galaxies lies on the star-forming main sequence from \cite{schreiber2015}, we computed the ratio between the detected and total SFR. We found that with a $10^{9.5} M_\odot$ cut in stellar mass, we recover 80\% of the total SFR, meaning that the missing low-mass galaxies would raise the estimate by about 25\%. The discrepancy can also partially be explained by an underestimation of the SFR due to \texttt{NNPZ}'s architecture (see Sect.~\ref{sect:nnpz}). Indeed, the SED fitting has been noticed to diverge if the priors allow high SFR values, such as the ones that could be reached by starbursting galaxies. Furthermore, the scaling relation from \cite{kennicutt12} between SFR and low infrared luminosity used to compute \Planck\ SFRs yields higher values for SFR than the SED fitting used for the \Euclid\ catalogue. The difference between both estimations is expected to be of a factor at most 2 \citep[Fig.~3 and 11 of][respectively]{Wuyts2011,figueira2022}. In addition, \Planck\ SFRs are overestimated due to line-of-sight accumulation effects \citep[][]{negrello2017,gouin2022}. Specifically, using the TNG300 simulation of the IllustrisTNG project, \citet[][their Fig.~5]{gouin2022} demonstrated that the SFR derived from the most star-forming halos at the redshift of the protocluster candidates is insufficient to account for the \Planck-derived SFR. Conversely, the \Planck\ SFRs are recovered by summing the former with the contribution of fainter foreground and background sources projected along the line of sight within a region similar to the \Planck\ beam. We can expect up to 1.5 dex overestimation of SFR by \Planck\ due to line-of-sight and beam effects. This is by far the most important effect, and together with the ones listed above, it can account for the SFR discrepancies in Tab.~\ref{tab:selected_protoclusters_top}.

In the case of \Herschel\ SFRs, the overestimation effect seen in \Planck\ by \cite{gouin2022} should also be present, but with a smaller amplitude due to the smaller \Herschel\ beam size compared to \Planck. Even in the absence of far-infrared fluxes measured by \Euclid, the resulting SFR of single galaxies are statistically fairly consistent with the derived values from \Herschel\ \citep{Q1-SP071}. Therefore, the \Herschel\ SFR excess in Tab.~\ref{tab:selected_protoclusters_top} might indicate that several selected \Herschel\ sources do not actually correspond to \Euclid\ candidates.

\subsubsection{\label{sect:halo_mass}Halo masses}

Typical methods used to weigh galaxy clusters, such as measuring the virial mass from the velocity dispersion of galaxies or weak gravitational lensing, cannot work properly for protoclusters. Indeed, the first one relies on the virial hypothesis whereas protoclusters are by definition out of dynamical equilibrium. As for the second, the search for gravitational arcs is challenging, owing to the rapidly declining sky density of bright background sources for $z$ > 2 \citep{madau14}. We describe below two different methods to estimate the mass of each of our 20 \Euclid\ counterparts.

We choose first to perform an estimation of the dark matter mass of the halo by multiplying the stellar mass by the universal cosmological baryonic fraction  $\Omega_{\rm m} / \Omega_{\rm b} -1  = 5.35$ \citep{planck2020_I,planck2020_VI}; we call this `Method~A'. We note that this method is extremely conservative, since it is equivalent to not taking into account the baryonic gas between the galaxies. It is then expected to yield significantly lower halo masses than the other method we choose to use.

We adopt a compromise and make the assumption that central regions of protoclusters may have undergone virialisation. This hypothesis makes sense since according to the hierarchical structure formation, dynamical equilibrium is reached by smaller structures first. Therefore, the central regions should virialise sooner than the outskirts of the protocluster. This method to estimate the mass of protoclusters is based on stellar mass to halo mass scaling relations. We call it `Method~B'. Uncertainties are computed by propagating the errors on stellar mass from the \Euclid \texttt{NNPZ} physical parameter catalogues \citep{Q1-SP031} and also using each model parameter uncertainty, as provided by \citet[][hereafter \citetalias{behroozi2013}]{behroozi2013}, \citet[][hereafter \citetalias{legrand2019}]{legrand2019}, and \citet[][hereafter \citetalias{shuntov2022}]{shuntov2022}. \citetalias{behroozi2013} is a scaling relation from the stellarmass of the central galaxy, whereas \citetalias{legrand2019} and \citetalias{shuntov2022} are scaling relation between the stellar mass and halo mass of each member galaxy, which we need to sum at the end. The estimated halo masses are reported in Table~\ref{tab:halo_masses} for each method and they are shown in Fig.~\ref{fig:arthur_mh_vs_z} for Method~B \citetalias{shuntov2022}. The mean values for method A, \citetalias{behroozi2013}, \citetalias{legrand2019} and \citetalias{shuntov2022} are $(7.9\pm 4.3) \expo{11}\,M_\odot$, $(2.4\pm 0.8) \expo{12}\,M_\odot$, $(9.7\pm 4.9) \expo{12}\,M_\odot$ and $(10.4\pm 5.1) \expo{12}\,M_\odot$ respectively. This is the first time that a halo mass has been estimated for these objects.

\subsection{\label{sect:galaxy_members}Properties of the member galaxies}

Member galaxies all have \HE magnitude below 23.5, as stated in Sect.~\ref{sect:input_selection} (see Fig.~\ref{fig:histo_magnitudes_I_H}). Most of them have magnitude close to this threshold, since we are searching for very distant objects. In general, \IE magnitudes are lower than \HE due to the redshifting of the galaxy's spectrum.

\texttt{NNPZ} provides estimates for the stellar mass and SFR of each galaxy member in our sample (Sect.~\ref{sect:nnpz} and Fig.~\ref{fig:histos}). Figure \ref{fig:main_sequence} shows the SFR--$M_\star$ plane, and includes the main sequence of galaxies \citep{brinchmann2004,daddi2007,elbaz2011,rodighiero11} of \cite{schreiber2015}. We can therefore visualise the star-formation activity inside each protocluster candidates. For our member galaxies, SFR remains within a factor 3 of the one expected for their stellar mass on the main sequence. This means that they are presenting at most a slightly enhanced star-formation, and none of them is clearly starbursting or quenching. We note however that these SFR values should be regarded as lower limits because the \texttt{NNPZ} pipeline might underestimate high SFRs. This pipeline has been noticed to have strong divergence problems in estimating parameters if the priors on SFRs are not capped. It is then possible that higher star-formation events are occurring in these structures. 

We can also use the \IE$-$\YE vs \JE$-$\HE colour-colour diagram \citep{bisigello2020} from the four \Euclid\ photometric bands (Fig.~\ref{fig:colour_colour}) as a complementary way to assess the activity of galaxies. All identified galaxies at photometric redshifts $z_{\rm ph}<2.5$ lie in the active region of the colour-colour plots. We cannot assess the activity of galaxies at $z_{\rm ph}>2.5$ since this diagnostic does not hold anymore at such redshifts. 

\begin{figure*}[tbp!]
    \centering
    \includegraphics[angle=0,width=0.48\hsize]{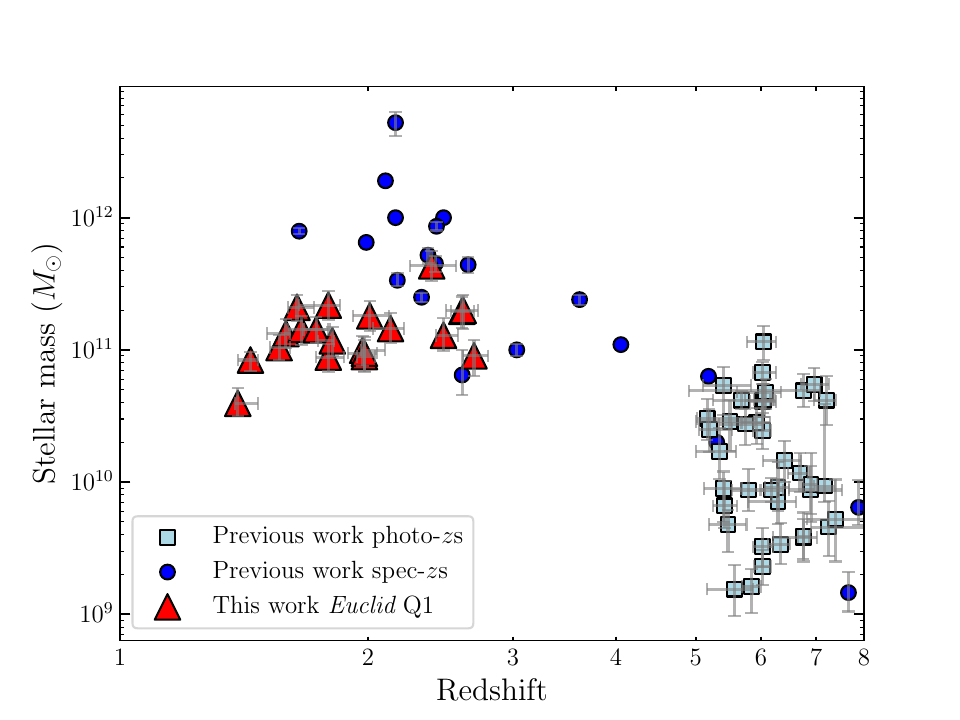}
    \includegraphics[width=0.48\hsize]{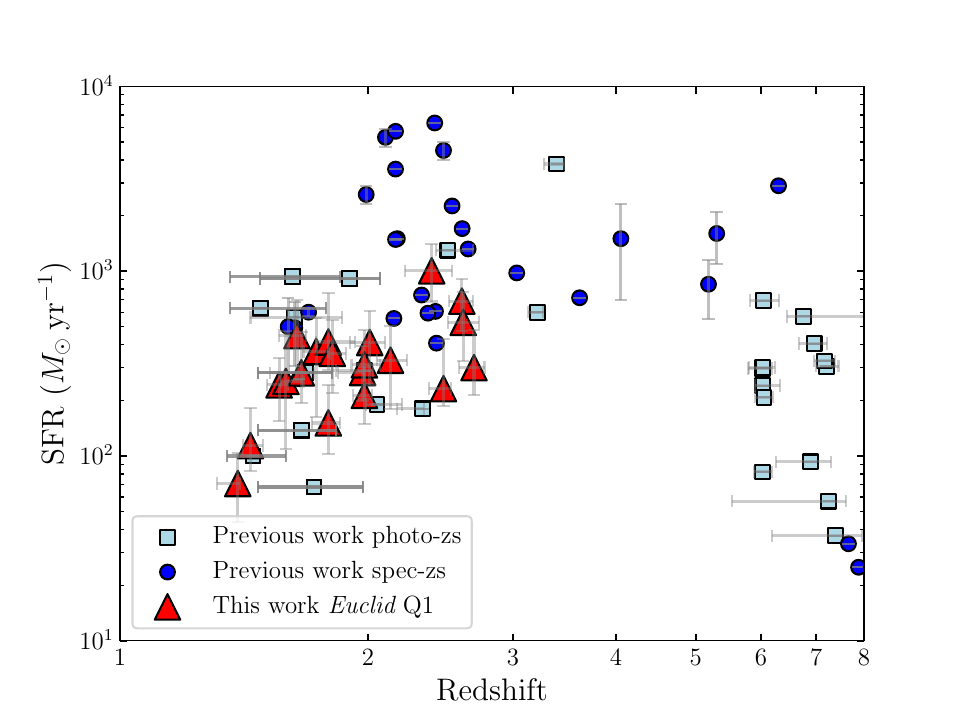}
    \caption{Stellar mass (resp. SFR) as a function of redshift for the \Euclid\ protoclusters (triangles) and from the literature \citep[circles and squares;][]{casey2016,gomez-guijarro2019,polletta2021,laporte2022,morishita2023,morishita2024,sillassen2024,li2024,shimakawa2024}. Dark blue circles have spectroscopic redshifts and light blue squares have photometric redshifts. The peak in star-formation around $z=$ 2 -- 3 is clearly visible. SFR values for \Euclid\ candidates may have been underestimated (see Sect.~\ref{sect:nnpz}).}
\label{fig:Mstar_vs_redshift_arthur}
\end{figure*}

\section{\label{sect:Discussion}Discussion}

\subsection{Comparison with other detection algorithms}

Resulting matches for \texttt{PPM} and \texttt{DIANAS} are presented in Table~\ref{tab:dtfe_ppm}. We find that both alternative algorithms detected most (17 out of 20) structures at the same sky positions and photometric redshifts as those given by \texttt{DETECTIFz}.
All detections are retrieved by \texttt{PPM} within 2 $\sigma$ of the photometric redshift, with similar errors. It is worth noticing that \texttt{PPM} detects several other overdensities along the line of sight (Fig.~\ref{fig:ppm}), in agreement with the hypothesis and ancillary observations that these fields contain several galaxy groups and protoclusters at different redshifts. 

In addition to the two algorithms above, we study the \Spitzer\ flux in IRAC 1 and IRAC 2 bands. We compute the projected surface galaxy overdensity for each of our \Euclid\ protocluster candidates using two density estimators: the $N$th–nearest-neighbour method and fixed-aperture densities. We find that all of our candidates correspond to significant galaxy overdensities when assessed with both density estimation methods. In particular, our highest S/N detection with \Euclid\ (G222.05\_EUC\_3) is also the highest S/N detection with \Spitzer\ among our sample. 

\subsection{Are these \Euclid-\Planck\ objects actual protoclusters ?}

\subsubsection{Comparison with literature : star formation rates and stellar masses}

Figure~\ref{fig:Mstar_vs_redshift_arthur} showcases the relation of $M_*$ and SFR with redshift for our detections and from the literature at $1.7 < z< 8$
\citep{casey2016,gomez-guijarro2019,polletta2021,laporte2022,morishita2023,morishita2024,sillassen2024,li2024,shimakawa2024}.
The 20 \Euclid-\Planck\ protocluster candidates show stellar masses (Fig.~\ref{fig:Mstar_vs_redshift_arthur} left) comparable to that of protoclusters from the literature at $z>2.2$, and lower than the few other known protoclusters at $z<2.2$.
About SFR (Fig.~\ref{fig:Mstar_vs_redshift_arthur} right), our protocluster candidates and the litterature are comparable, despite the fact that the effects of using samples drawn from different selection techniques may have introduced larger differences, not seen here.

All the structures seem to follow a similar trend: stellar mass grows with time (and thus decreases with redshift) and star-formation reaches a peak around $z\sim $ 2 -- 3, consistent with the Cosmic Noon era \citep{madau14,chiang2017}. These points support the idea that we are detecting protoclusters.

Many galaxies in \Euclid\ protocluster candidates lie slightly above the main sequence (Fig.~\ref{fig:main_sequence}), which suggests that they are active (Fig.~\ref{fig:colour_colour}), experiencing star-formation events but not starbursts. This is in line with previous studies of the \Planck\ protocluster candidates \citep{kneissl2019,polletta2021,polletta2022,hill2024,polletta2024}, which have shown that the majority of the protocluster members are on the star-forming main sequence, despite the high SFRs measured by \Planck\ and \Herschel\ (Table~\ref{tab:selected_protoclusters_top} and Sect.~\ref{sect:sfrs}). Other protoclusters like the Spiderweb \citep{perez-martinez2024b} also show galaxy members that are mainly on the main sequence, although counter-examples exist \citep[e.g., USS 1558$-$003, ][]{perez-martinez2024a}. We also note that contrary to protoclusters described in \cite{hill2020,kubo2021,perez_martinez23}, our \Euclid\ counterparts do not seem to host massive galaxies with $M_*>10^{11}M_\odot$. Finally, the presence of (slightly above) main sequence galaxies is also in line with the findings of \cite{zhou24,zhou2025}.

\subsubsection{Non-virialised structures}

For each protocluster, we compute from \cite{mo2002} the radius $R_{200}$ of a virialised halo with the same mass at the protocluster redshift and the corresponding angular size (see Table~\ref{tab:selected_protoclusters_top}). We see that for every protocluster, this calculated angular size is significantly smaller than the angular radius outputed by \texttt{DETECTIFz}, by about one order of magnitude. This result clearly illustrates that the structures we detect are not virialised. This makes a strong argument in favour of our detections being actual protoclusters, since it means their halos have not fully formed yet.

Consistent with this, we found no X-ray or SZ clusters within a radius of 10 arcmin around the \Euclid\ protoclusters listed above. The catalogues used for this search, as compiled \cite{EP-Melin}, are the X-ray Clusters II catalogue MCXC-II \citep{Sadibekova_2024}, the eROSITA catalogue \citep{Bulbul2024,Kluge2024a}, the meta-catalogue of SZ clusters MCSZ (Tarr{\'\i}o et al., in prep), and the joint X-ray and SZ catalogue ComPRASS \citep{Tarrio2019}.

\subsection{Total mass and accretion regime}

The four methods to estimate the halo mass listed in Sect.~\ref{sect:halo_mass} show different results. Method~A yields masses about an order of magnitude lower than those computed using \citetalias{legrand2019} and \citetalias{shuntov2022}. This is expected since we do not take the mass of the intergalactic baryonic gas into account. The three estimates using Method~B are consistent within at most 0.5\,dex differences. Since \citetalias{behroozi2013} scales from the stellar mass of the central galaxy only, the values yielded by this method remain lower than that of \citetalias{legrand2019} and \citetalias{shuntov2022}, although they are higher than values from Method~A by at least 0.6\,dex. Finally, the halo masses derived using the prescriptions from \citetalias{legrand2019} and \citetalias{shuntov2022} usually agree to better than 0.1\,dex, as they consist in the same analytic formulae the parameters of which differ slightly. 

We use the mass estimated by method~B \citetalias{shuntov2022} -- as it is the most recent and fine-tuned relation -- to compare our detections with protoclusters from the literature at $1.7 < z < 8$ identified by photometric or spectroscopic redshifts \citep{casey2016,hill2020,polletta2021,laporte2022,morishita2023,morishita2024,sillassen2024,shimakawa2024} in Fig.~\ref{fig:arthur_mh_vs_z}.
At $z>2.2$ our protocluster candidates exhibit halo masses comparable to those reported in the litterature. At $z<2.2$, our protocluster candidates show lower masses than known protoclusters, highlighting the sensitivity of Euclid to this window of lower halo mass protoclusters.

Several protocluster candidates occupy the zone delimited by $\Mh \simeq 10^{13} M_\odot$ and 2 $<z<$ 3, where a transition between cold streaming in hot media and accretion of hot gas is expected \citep[][our Fig.~\ref{fig:arthur_mh_vs_z}]{dekel06,daddi2022}. This observation is in line with the SFR values of our candidates, the member galaxies of which are active yet not starbursting. Other protoclusters with similar mass and redshift also show the same order in magnitude of SFR. 

According to the relation for halo mass growth in \cite{behroozi2013}, our detections are expected to form structures of masses between $10^{13}$ and $10^{15} \: M_\odot$ at $z=0$, making the most massive ones solid candidates as progenitors of today's galaxy clusters.

\section{Conclusions}
\label{sect:conclusion}

We have searched in the \Euclid Q1 data for galaxy protocluster counterparts of eight \Planck\ protocluster candidates, using an improved \Euclid\ catalogue containing additional \Spitzer\ photometry. We found 20 \Euclid\ counterparts to eight \Planck\ protocluster candidates for which we estimated the photometric redshifts as well as lower limits for total stellar masses, star-formation rates, and halo masses.
Our detections span photometric redshifts $1.4 < z_{\rm ph} < 2.7$ and halo masses $12.6 <\logten (\Mh/M_\odot) < 13.4$.

The photometric redshift distribution of the 20 \Euclid\ counterparts matches the estimation by \Planck\ and previous follow-up observations of \Planck\ protocluster candidates. Several detections lie, however, at redshifts slightly lower than anticipated ($z<2$). Several \Euclid counterparts of the \Planck\ candidates occupy, in the $\Mh$--$z$ plane, a location close to the expected transition between cold flows in hot media to the accretion of hot material. A first search for counterparts of these \Euclid protoclusters in the X-ray and SZ  cluster catalogues shows no association; this suggests that the halos do not contain gas that is hot and dense enough to emit in the X-ray or to cast shadows in the cosmic microwave background via the Sunyaev--Zeldovich effect.

The estimated lower limits of $\Mh$, together with the photometric redshift range probed (part of our sample lies towards the end of Cosmic Noon), and the star-formation rates of the protoclusters, allow us to hypothesise that the \Planck\ catalogue of high-redshift star-forming fields mainly contains protoclusters in their maturing phase. In particular, the SFR is typical of galaxies that are active, but not starbursting. According to the position of our detections in the $\Mh$--$z$ plane (Fig.~\ref{fig:arthur_mh_vs_z}), the accreted matter is hot and diffuse rather than channeled along cold streams inside the halo. We may therefore be witnessing the last moments of star-formation activity in these structures, marking the onset of quenching driven by the stabilisation of the virial shock and the subsequent inability of the halo to retain cold inflows. 

Future observations and analyses in the millimetre with SPT, in the X-rays with eROSITA or the ‘XMM Fornax heritage programme'\footnote{\url{https://fornax.cosmostat.org}}, and optical/NIR/mm/radio spectroscopy (including \Euclid NIR spectra) will help to obtain a clearer view of these structures and on the thousands of protoclusters that will be detected in the forthcoming \Euclid DR1 data set. Indeed, in total, we can expect of the order of 40\,000 protoclusters with a mass limit of $M \ge 8\times10^{13} M_{\odot}$ in the complete EWS in the redshift range 1.5--4 \citep{EP-Boehringer}.

\begin{acknowledgements}
We thank L. Legrand for providing us with the stellar mass versus halo mass relations electronically and P. Behroozi and M. Shuntov for providing their relations online. TD acknowledges financial support from the Centre national d’études spatiales (CNES), France (ROR: https://ror.org/04h1h0y33) within the framework of the \Euclid\ mission. TD acknowledges support from the France 2030 programme ANR-11-IDEX-0003, attributed through the
Astrophysical Axis of the Graduate School of Physics of the Universit{\'e}
Paris-Saclay.
  MP acknowledges financial support from INAF mini-grant 2023 `Galaxy growth and fuelling in high-$z$ structures'.
  OC acknowledges financial support from the PRC grant ANR-21-CE33-0022 (InterPlay) and the PIA grant ANR-21-ESRE-0030 (CONTINUUM).
  HD and JMPM acknowledge support from the Agencia Estatal de Investigaci{\'o}n del Ministerio de Ciencia, Innovaci{\'o}n y Universidades (MCIU/AEI) under grant (Construcci{\'o}n de c{\'u}mulos de galaxias en formaci{\'o}n a trav{\'e}s de la formaci{\'o}n estelar oscurecida por el polvo) and the European Regional Development Fund (ERDF) with reference (PID2022-143243NB-I00/10.13039/501100011033).
\AckEC 
\AckQone
The authors acknowledge support from the ELSA project. "ELSA: Euclid Legacy Science Advanced analysis tools" (Grant Agreement no. 101135203) is funded by the European Union. Views and opinions expressed are however those of the author(s) only and do not necessarily reflect those of the European Union or Innovate UK. Neither the European Union nor the granting authority can be held responsible for them. UK participation is funded through the UK Horizon guarantee scheme under Innovate UK grant 10093177
 This paper made use of the \Planck\ data publicly available at \url{https://www.esa.int/Planck}. \Planck\ is a project of the European Space
Agency (ESA) with instruments provided by two scientific consortia funded by ESA member states and led by Principal Investigators from France and Italy, telescope reflectors provided through a collaboration between ESA and a scientific consortium led and funded by Denmark, and additional contributions from NASA (USA).
The \Herschel\ spacecraft was designed, built, tested, and launched under a contract to ESA managed by the \Herschel/\Planck\ Project team by an industrial consortium under the overall responsibility of the prime contractor Thales Alenia Space (Cannes), and including Astrium (Friedrichshafen) responsible for the payload module and for system testing at spacecraft level, Thales Alenia Space (Turin) responsible for the service module, and Astrium (Toulouse) responsible for the telescope, with in excess of a hundred subcontractors.  SPIRE has been developed by a consortium of institutes led by Cardiff University (UK) and including Univ. Lethbridge (Canada), NAOC (China), CEA, LAM (France), IFSI, Univ. Padua (Italy), IAC (Spain), Stockholm Observatory (Sweden), Imperial College London, RAL, UCL-MSSL, UKATC, Univ. Sussex (UK), and Caltech, JPL, NHSC, Univ. Colorado (USA). This development has been supported by national funding agencies: CSA (Canada), NAOC (China), CEA, CNES, CNRS (France), ASI (Italy), MCINN (Spain), SNSB (Sweden), STFC, UKSA (UK), and NASA (USA).

\end{acknowledgements}

\bibliography{Euclid,biblio_aa60169-26}

\begin{appendix}
\onecolumn

\section{\Euclid VIS images and sky positions of the Q1 EDF-S and EDF-F with the \Planck\ protocluster candidates\label{appendix:VIS_images_planck}}

\begin{figure*}[htbp!]
\centering
\includegraphics[angle=0,height=0.45\hsize]{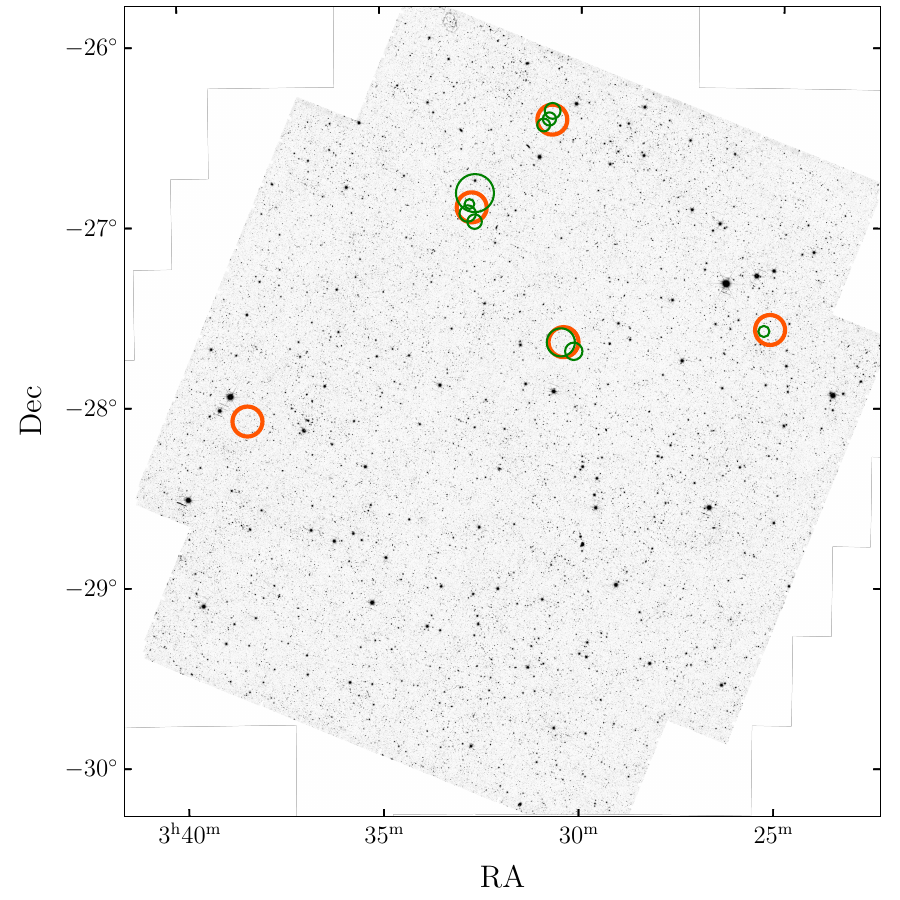}~
\includegraphics[angle=0,height=0.45\hsize]{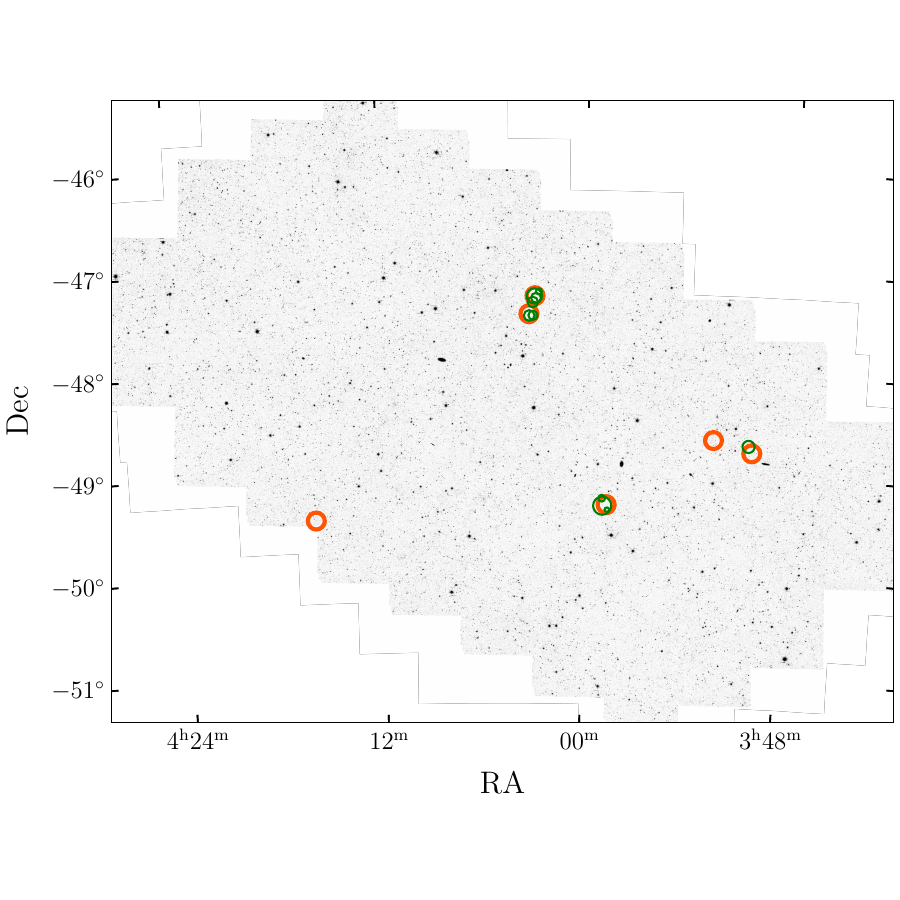}
\caption{\Euclid{} EDF-F (left) and \Euclid{} EDF-S (right) VIS images, with the \Planck{} protocluster candidates shown in orange (\ang{;5;} diameter circles) and the \Euclid{} counterparts in green (circle sizes coming from Table~\ref{tab:selected_protoclusters_top}). }
\label{fig:edf}
\end{figure*}

In this section, we show in Fig.~\ref{fig:edf} the \Euclid\ VIS images of the EDF-F and EDF-S fields. Sky locations of the \Planck\ protocluster candidates and the \Euclid\ counterparts are shown in the figure, and are reported in Table~\ref{tab:planck_fields}.

\begin{table*}[htbp!]
\caption{The eight \Planck\ protocluster candidates located in the \Euclid Q1 footprint \citep{Q1-TP001,Q1cite}, in seven tiles. }
\label{tab:planck_fields}
\centering
\begin{tabular}{l@{\hskip 3em}ccccc}
\hline\hline
\noalign{\vskip 2pt}
\Euclid field     & \Planck\ source   & \Planck\ ID & RA [deg] & {\pd}Dec [deg] & MER tile \\ 
\hline
\noalign{\vskip 2pt}
EDF-F & G221.09$-$54.59\tablefootmark{a}& 1904& 52.6798& $-$26.4142& 102\,046\,112  \\
EDF-F & G222.05$-$54.24\tablefootmark{a}& 1399& 53.1806& $-$26.9014& 102\,045\,467 \\
EDF-F & G222.75$-$55.98\tablefootmark{a}& \pd108& 51.3123& $-$27.5725& 102\,044\,821  \\
EDF-F & G223.18$-$54.86\tablefootmark{a}& 2151& 52.6060& $-$27.6495& 102\,044\,824   \\
\hline
\noalign{\vskip 2pt}
EDF-S & G257.45$-$49.50\tablefootmark{\pa}& \pd320 & 57.4918& $-$48.7670& 102\,020\,530 \\
EDF-S & G254.74$-$47.64\tablefootmark{a}& 1308& 60.8251& $-$47.4485& 102\,021\,984 \\
EDF-S & G254.49$-$47.73\tablefootmark{\pa}& 1494 & 60.7380& $-$47.2715& 102\,021\,984 \\
EDF-S & G257.71$-$47.99\tablefootmark{\pa}& 1926 & 59.6462& $-$49.3164& 102\,020\,059   \\
\hline

\end{tabular}
\tablefoot{
\tablefoottext{a}{\small \Herschel/SPIRE observations are available and come from \cite{oliver2012} in EDF-F and \cite{planck15} in EDF-S.}
Columns are: \Euclid field name, \Planck\ field name, \Planck\ ID \citep[in the high-$z$ source candidate catalogue of][and available on the \Planck\ Legacy Archive at ESA]{planck16}, RA and Dec in degrees for the \Planck\ source, and the \Euclid\ MER tile number \citep{Q1-TP004}. Notice that two of the \Planck\ sources fall in the same \Euclid\ tile.
}
\end{table*}

\section{\label{sect:cutouts}Cutouts of the 20 \Euclid-\Planck\ protocluster candidates}

We showcase here the cutouts around each of the 20 protoclusters (Figs.~\ref{fig:cutouts} \& \ref{fig:cutouts_negative}). For a better visualisation, they are accessible in high resolution together with close cutouts of each of the 285 member galaxies {\href{https://zenodo.org/records/23015936?token=eyJhbGciOiJIUzUxMiJ9.eyJpZCI6Ijk0N2I2ZDNkLTBiY2MtNGViYy05OGQ3LTg4MjZmMjNmOGI1OSIsImRhdGEiOnt9LCJyYW5kb20iOiIxYTkwMjJjMTY3OTZiYzYxYWI0MTc4NDczNmQyMmUyMiJ9.x0dhJDiE2CKAoookC9psANSpgdkUOMicJsrZbc77E-3PKiyGBfSDAWjVwh-5Ksd1NREYoUv2D_kVHoAWynn_ZA}{here}}.

\def\cutoutsize{0.24}
\def\protowidth{0.24}
\begin{figure*}[htbp!]
    \centering
    \includegraphics[width=\protowidth\textwidth]{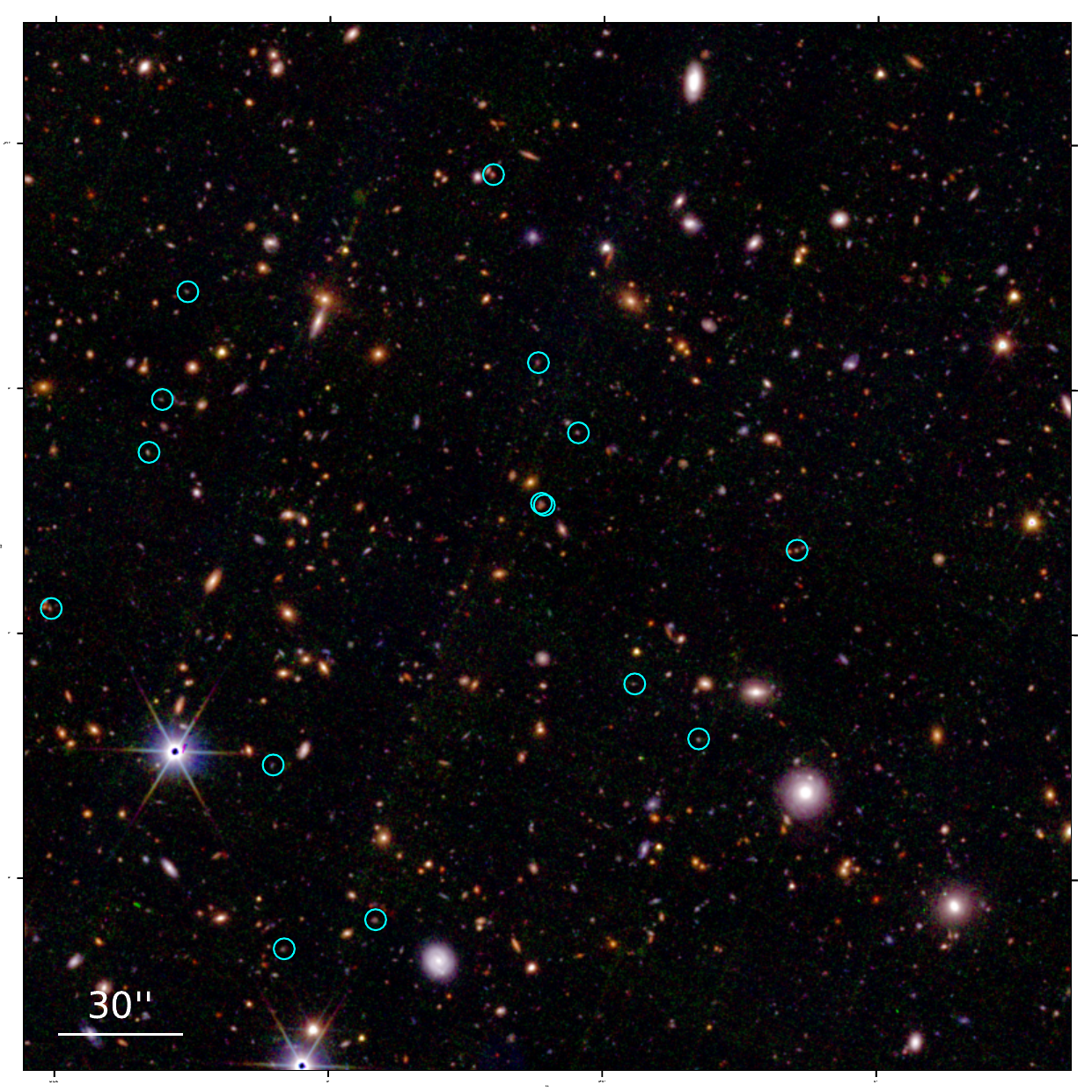}
    \includegraphics[width=\protowidth\textwidth]{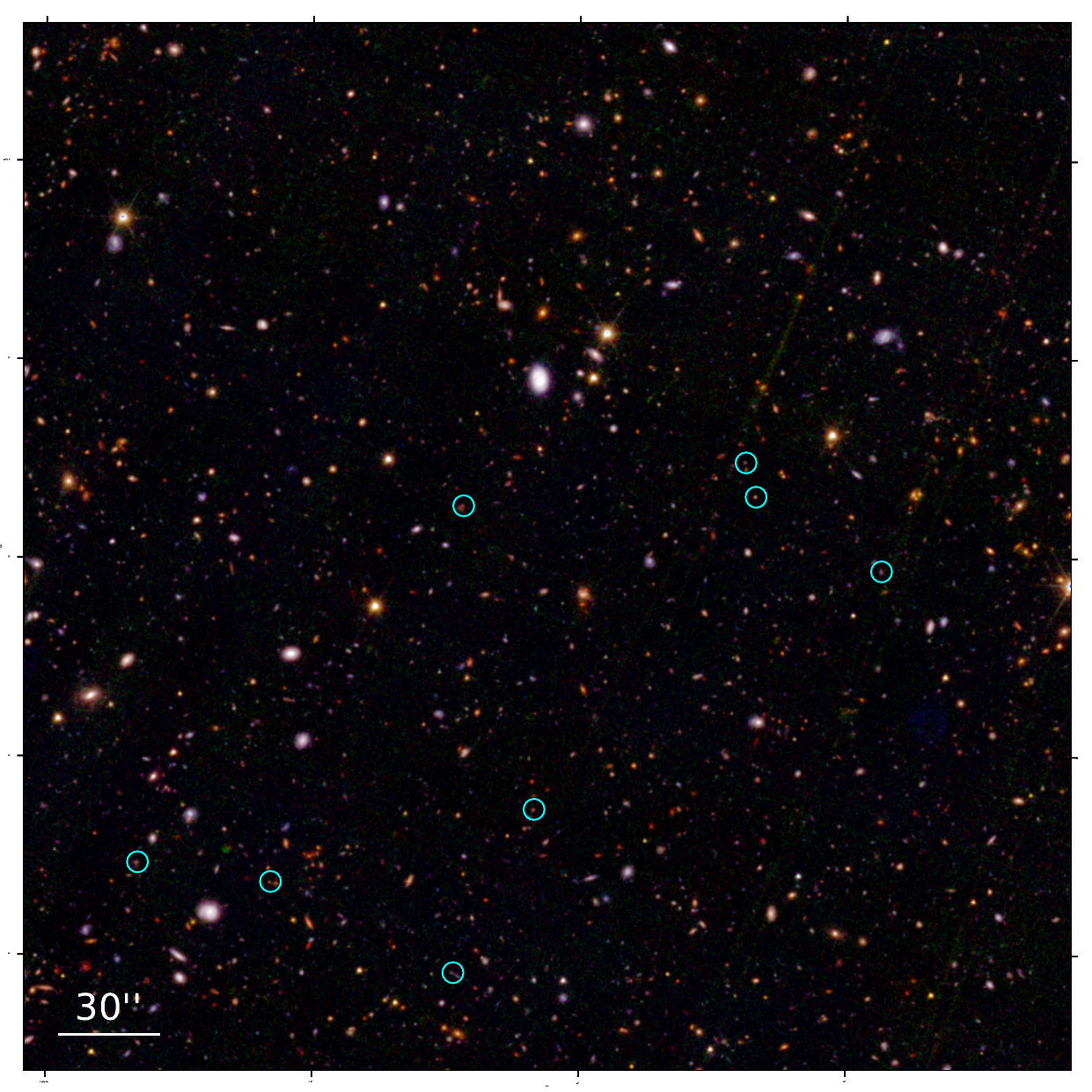}
    \includegraphics[width=\protowidth\textwidth]{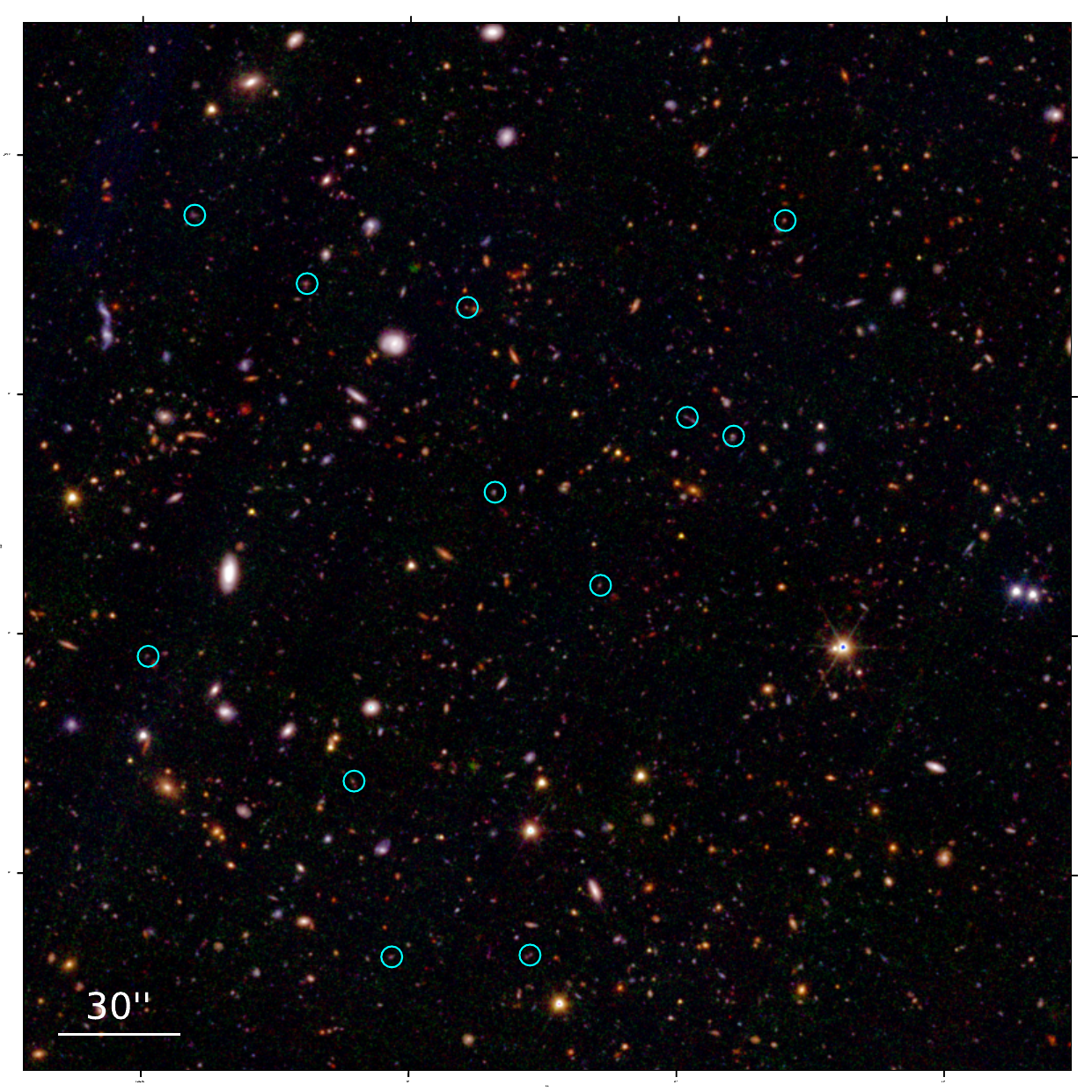}
    \includegraphics[width=\protowidth\textwidth]{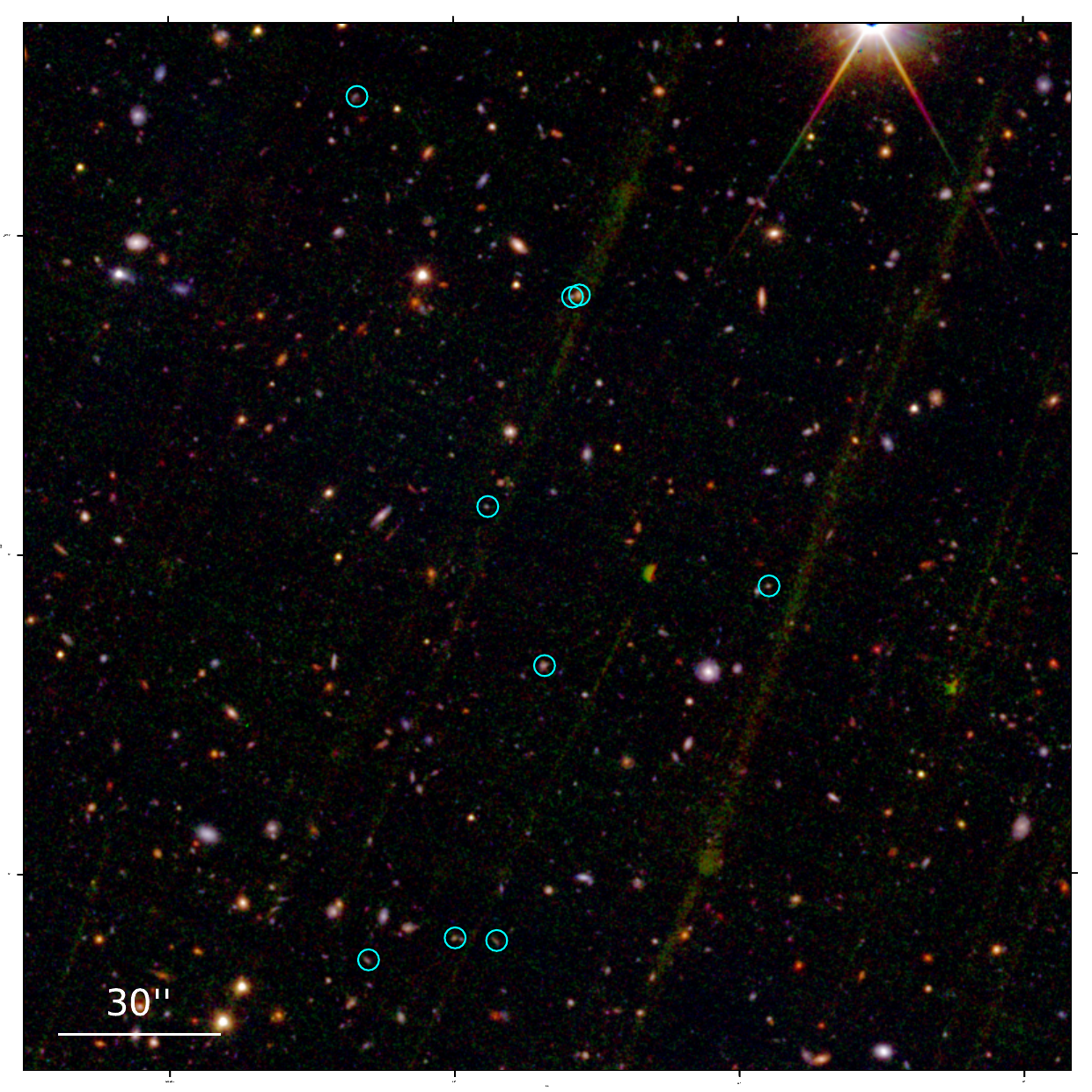}
    \includegraphics[width=\protowidth\textwidth]{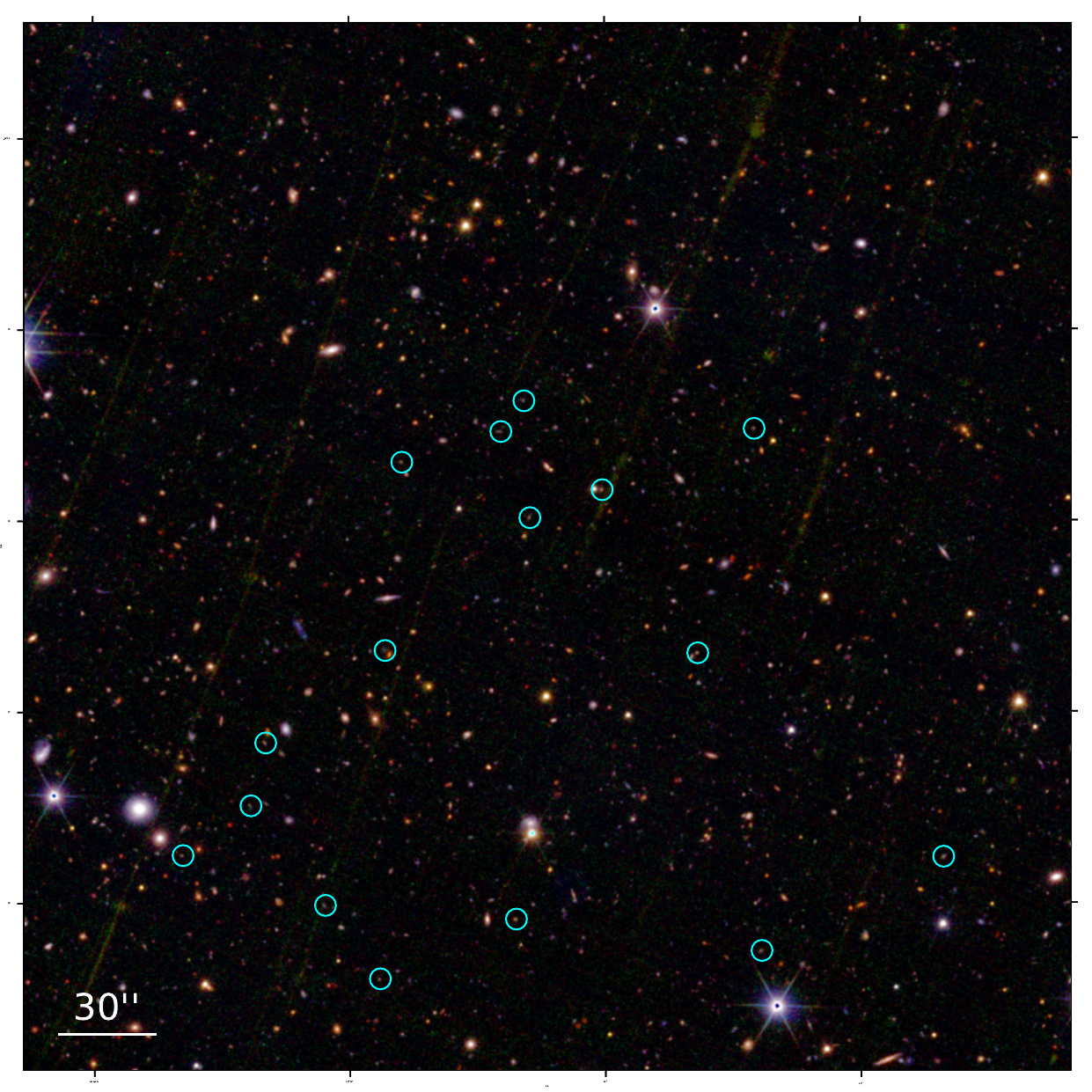}
    \includegraphics[width=\protowidth\textwidth]{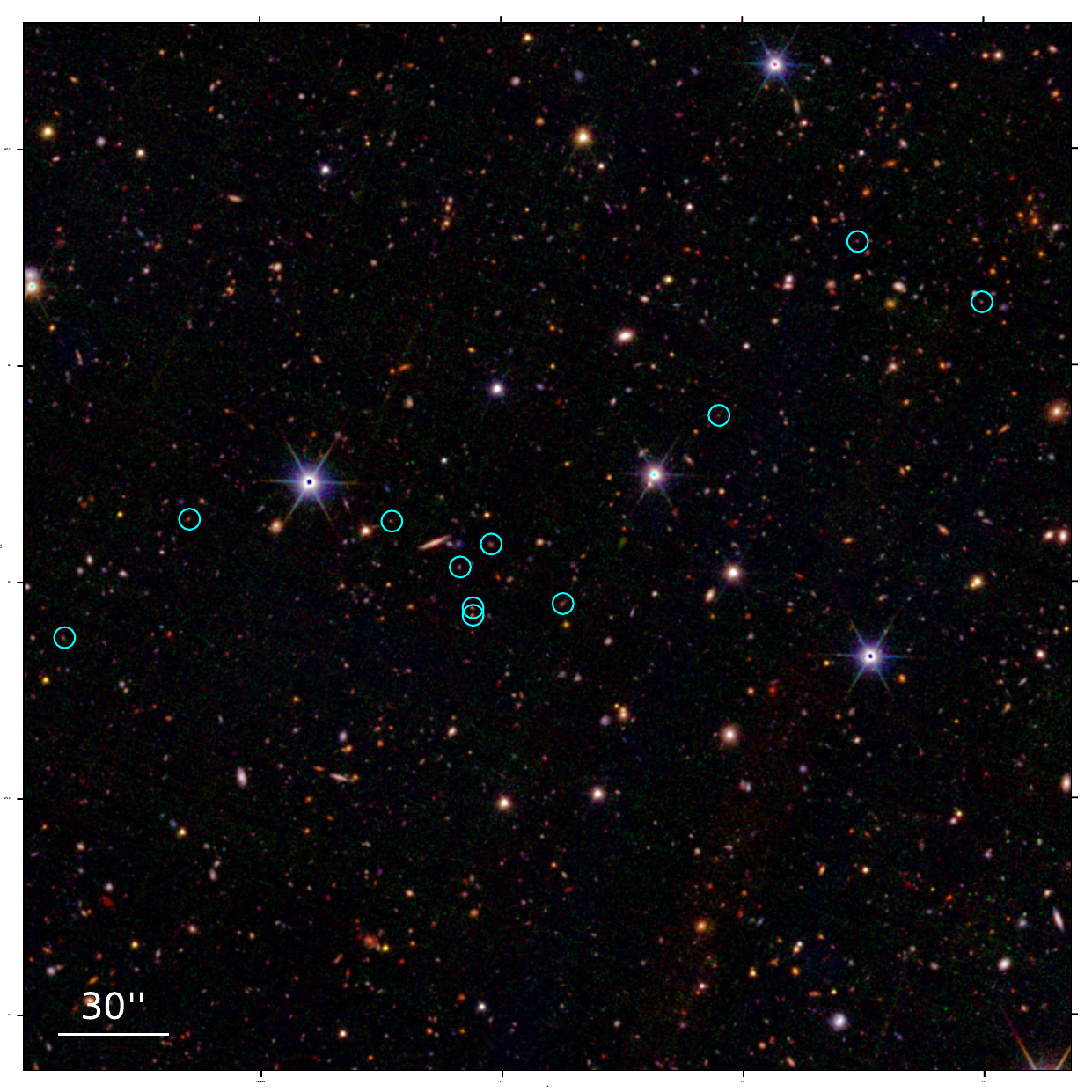}
    \includegraphics[width=\protowidth\textwidth]{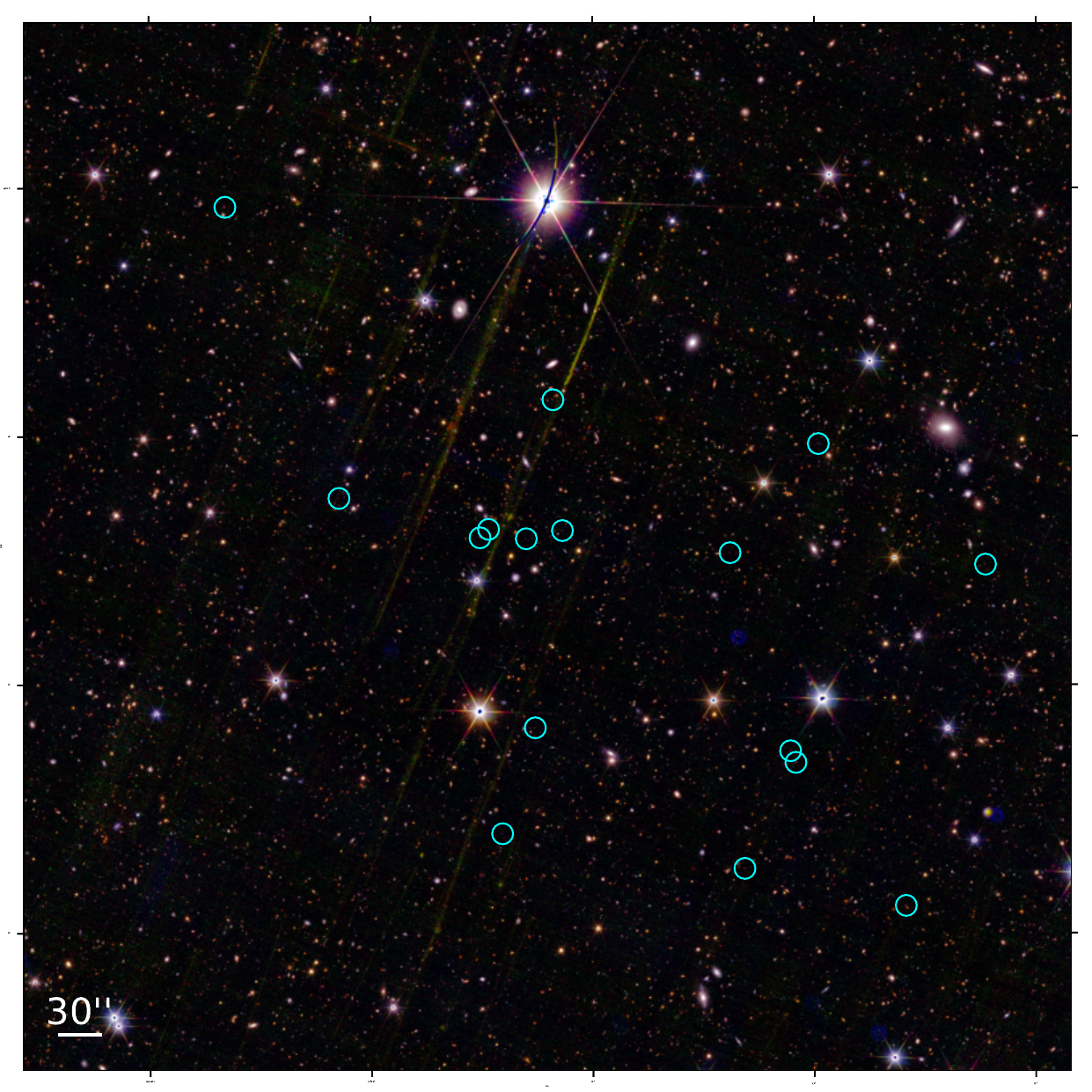}
    \includegraphics[width=\protowidth\textwidth]{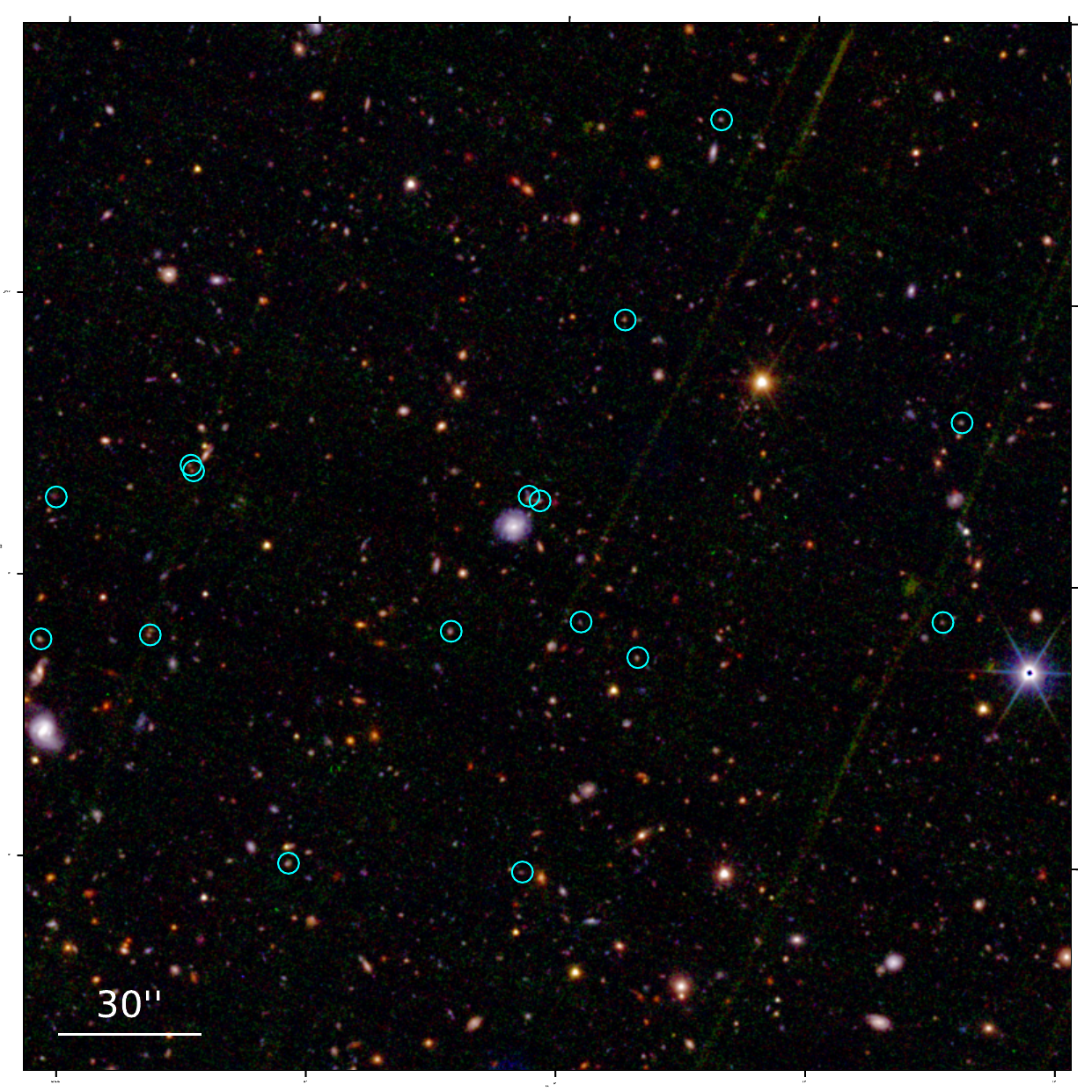}
    \includegraphics[width=\protowidth\textwidth]{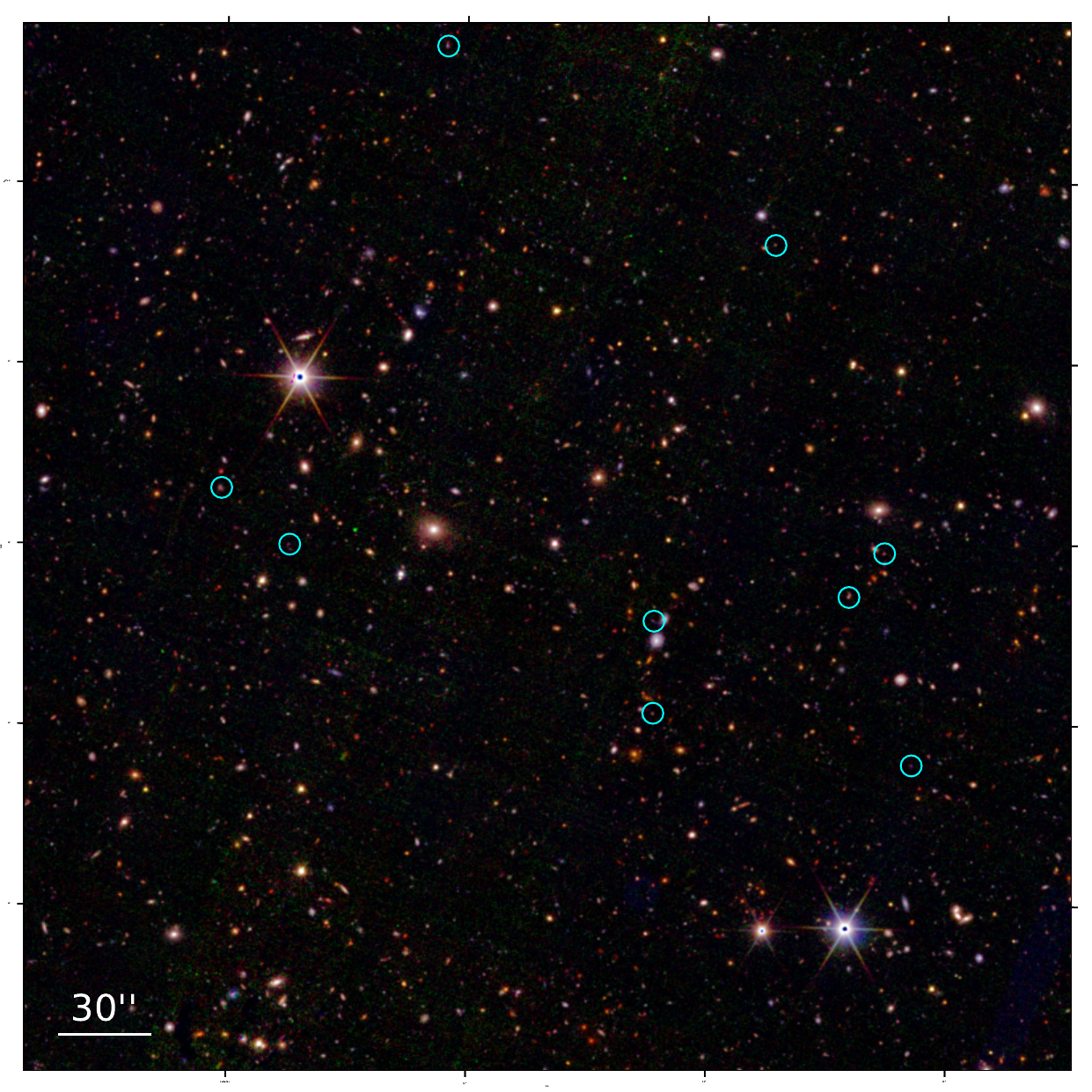}
    \includegraphics[width=\protowidth\textwidth]{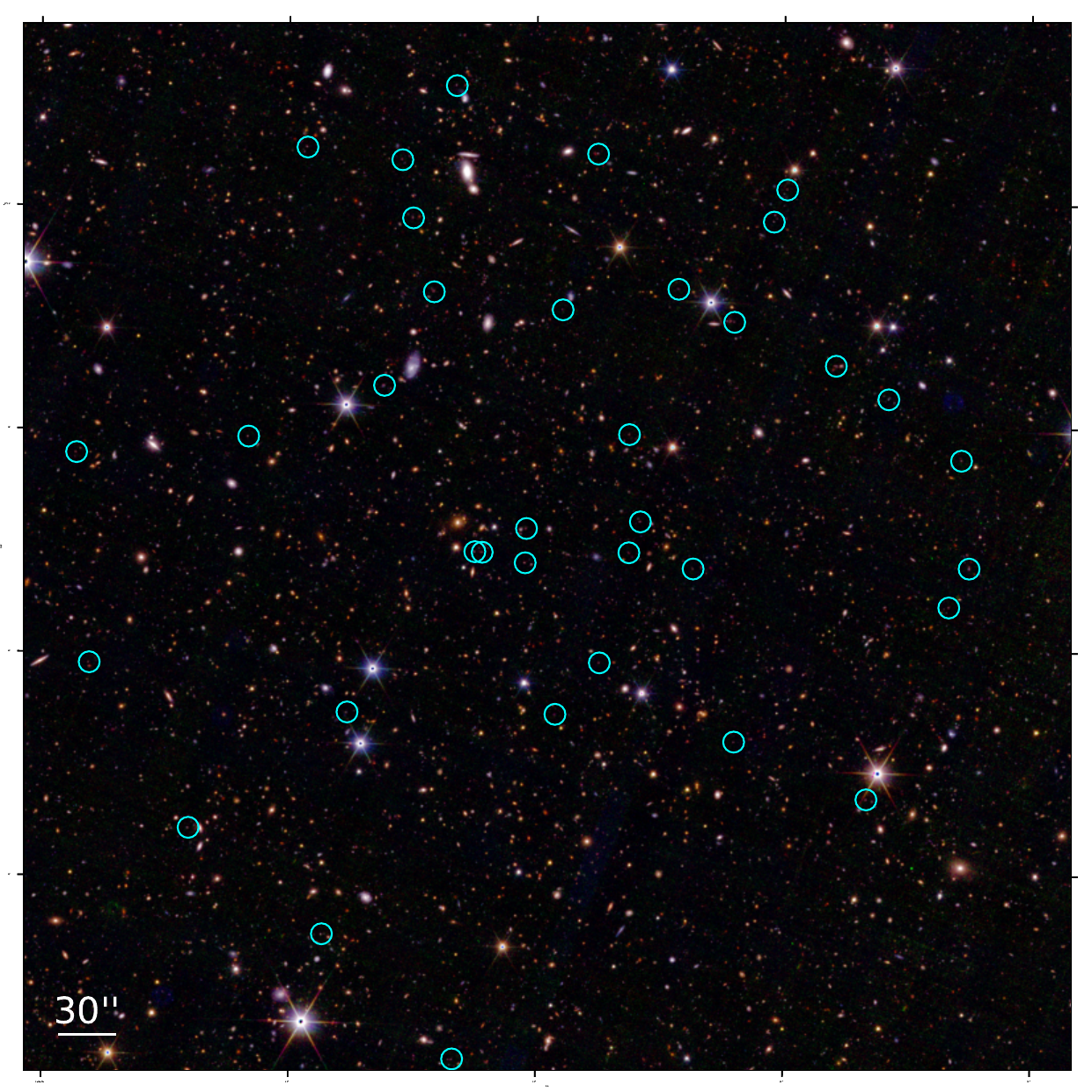}
    \includegraphics[width=\protowidth\textwidth]{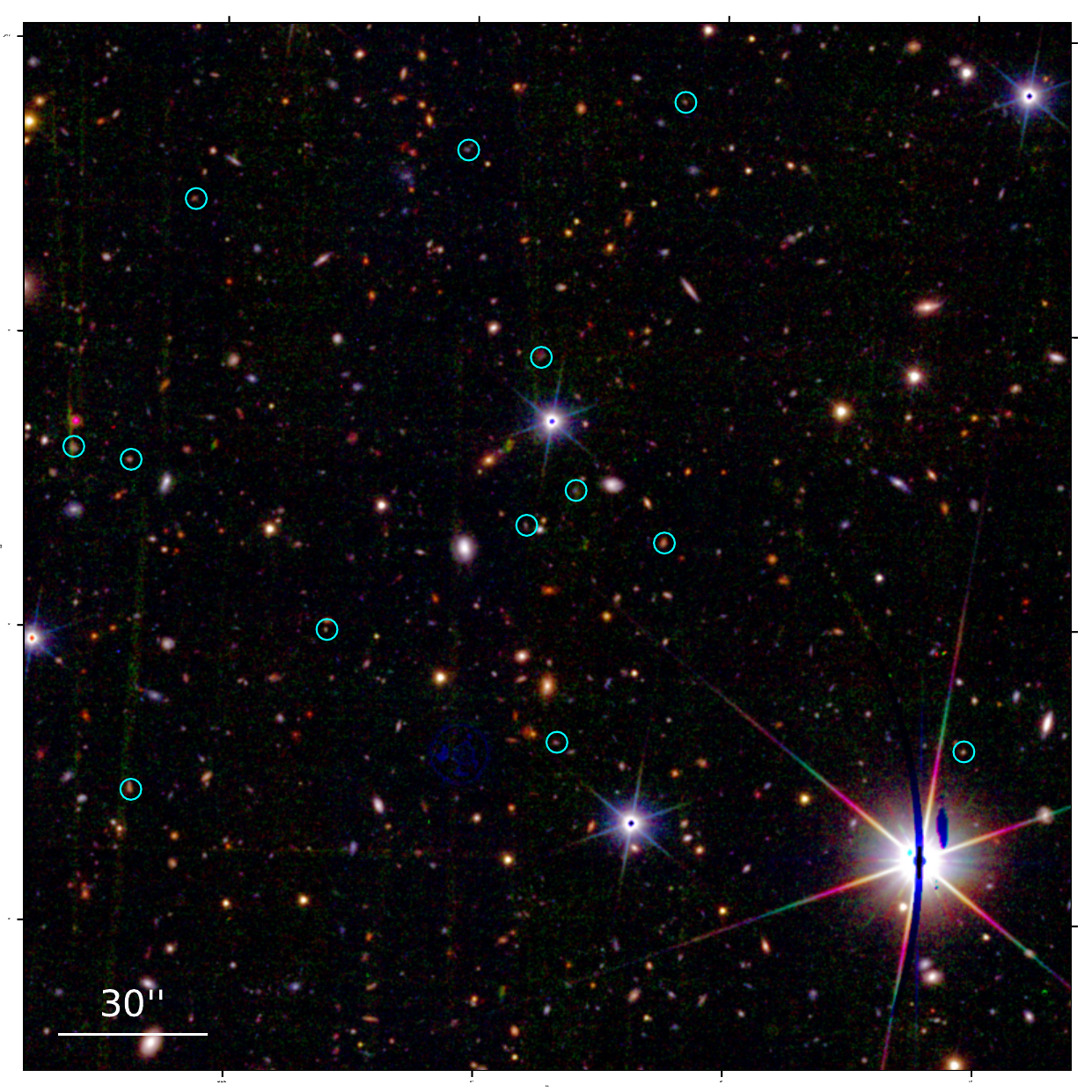}
    \includegraphics[width=\protowidth\textwidth]{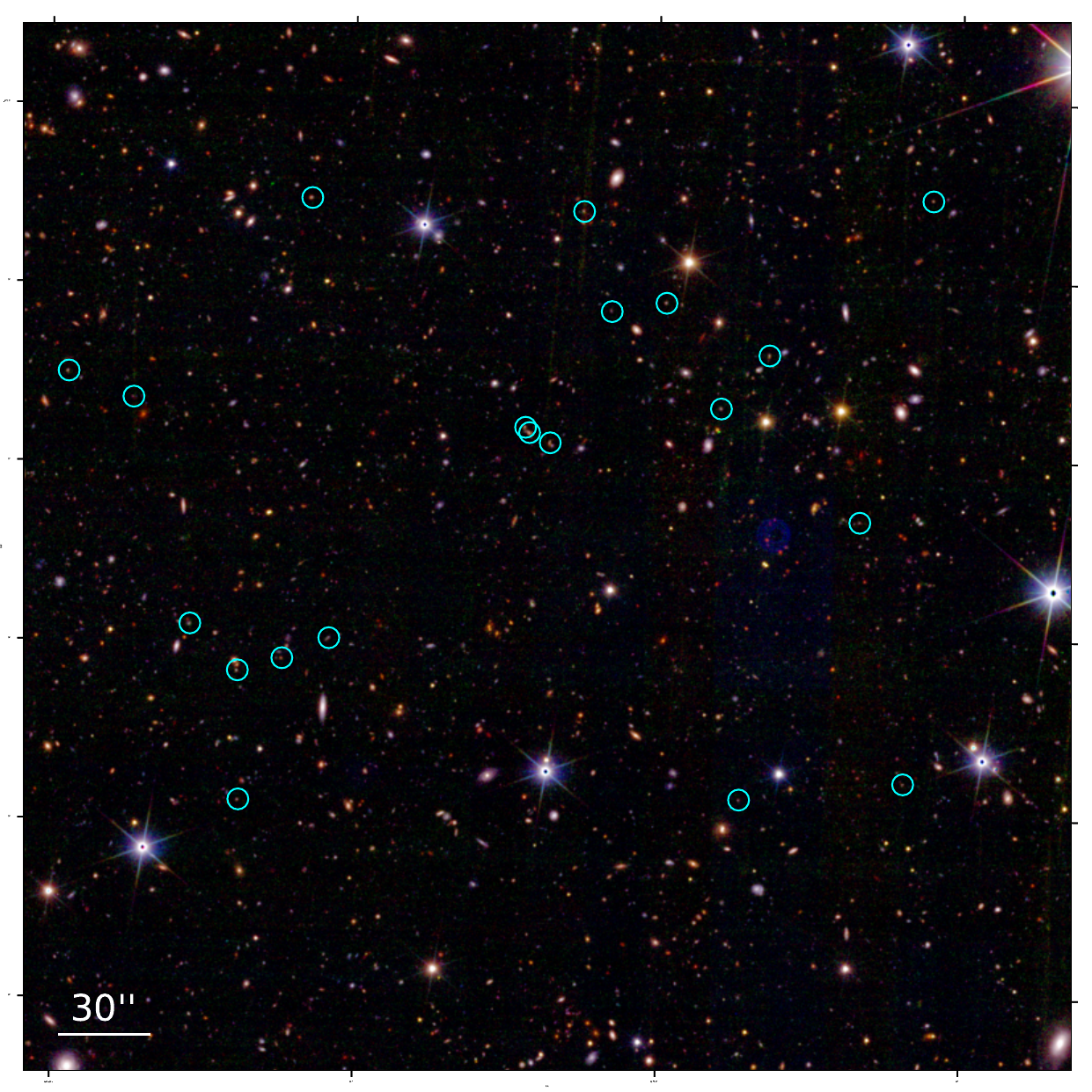}
    \includegraphics[width=\protowidth\textwidth]{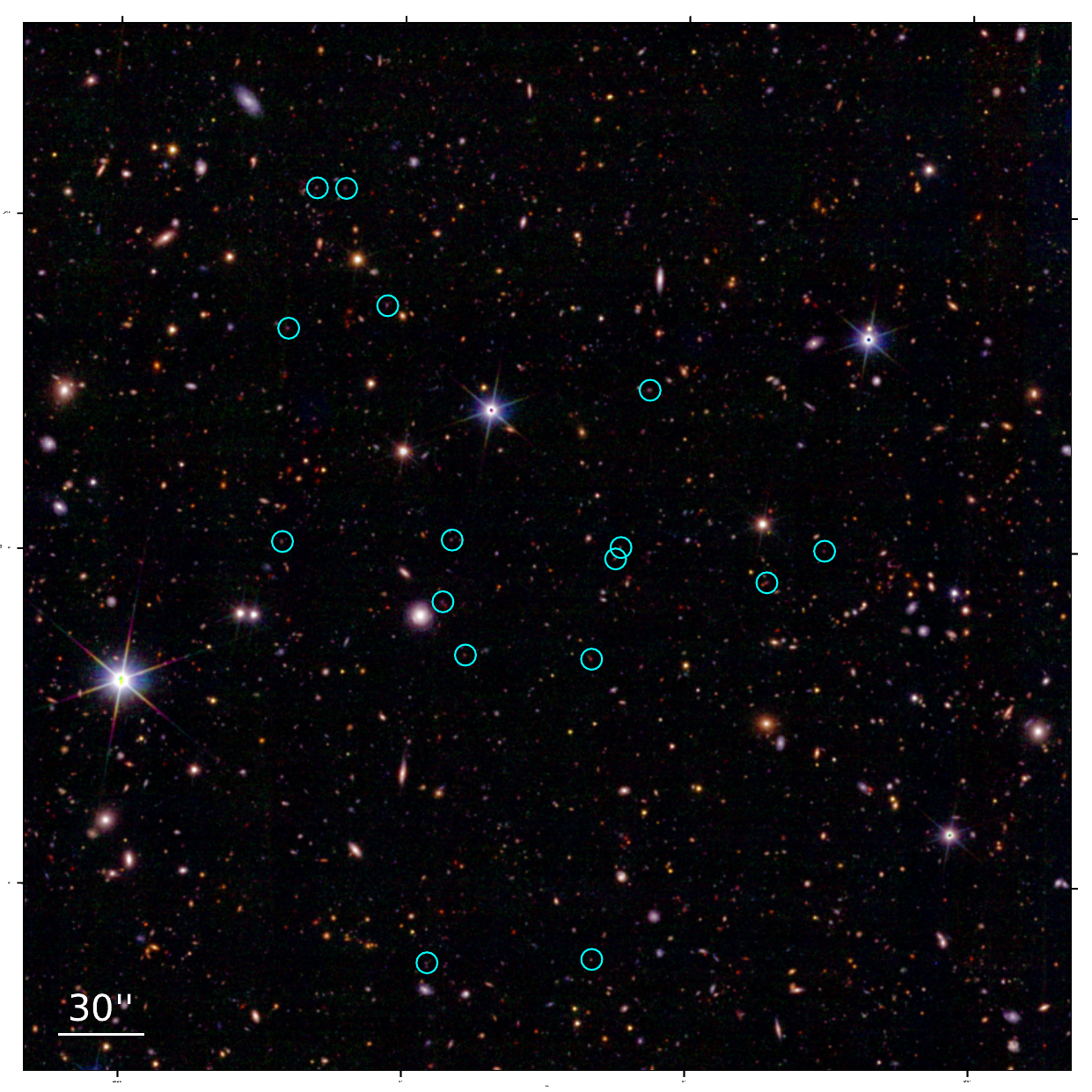}
    \includegraphics[width=\protowidth\textwidth]{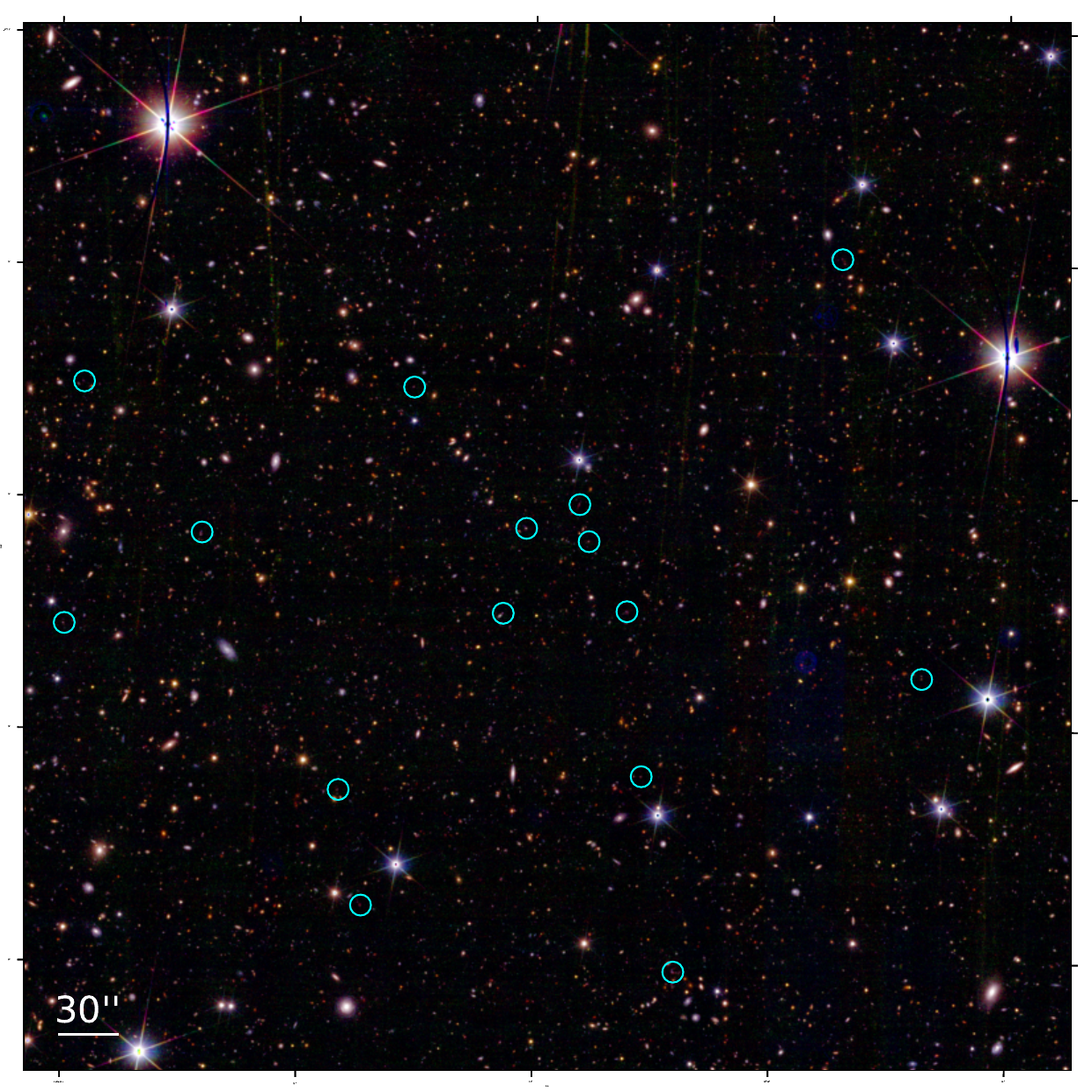}
    \includegraphics[width=\protowidth\textwidth]{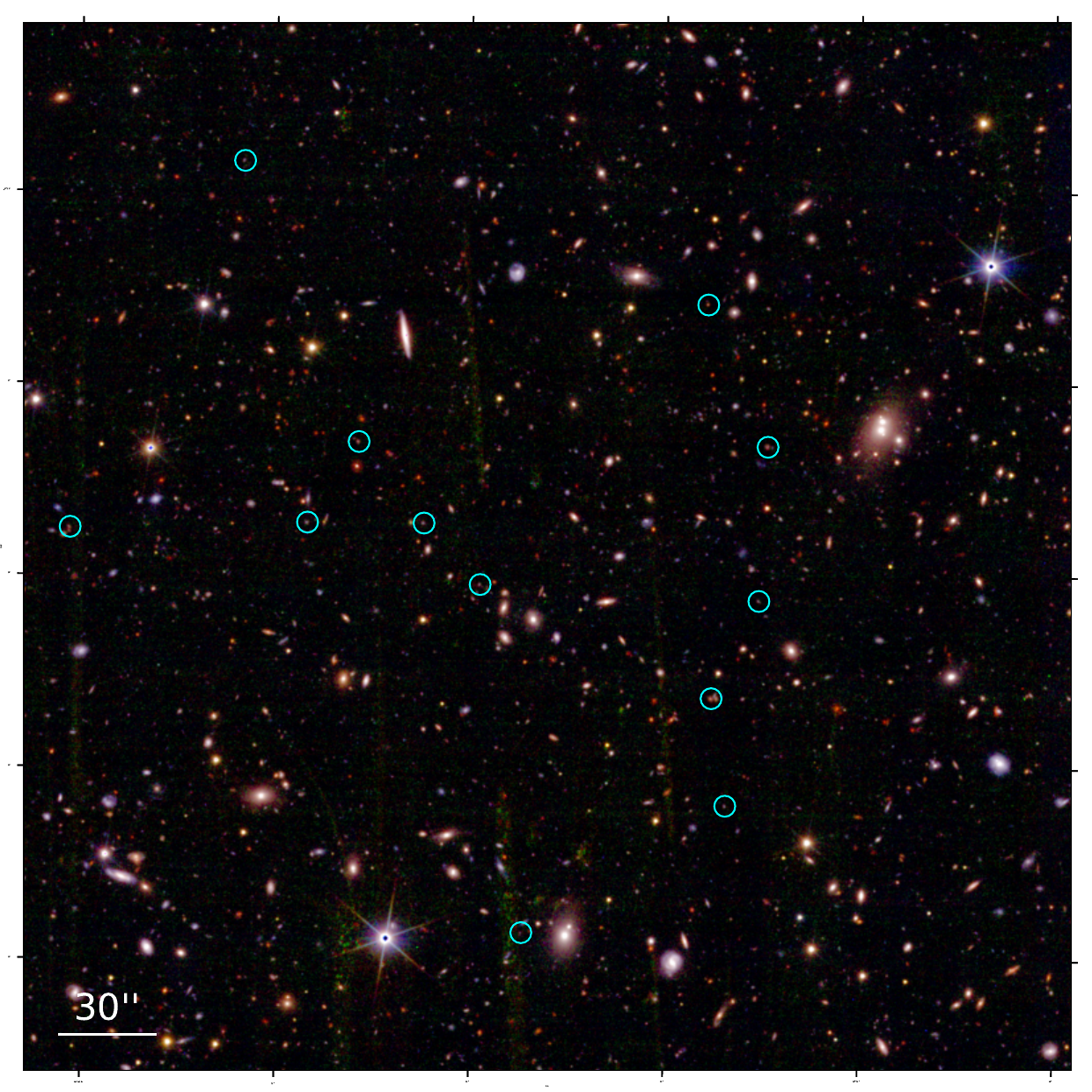}
    \includegraphics[width=\protowidth\textwidth]{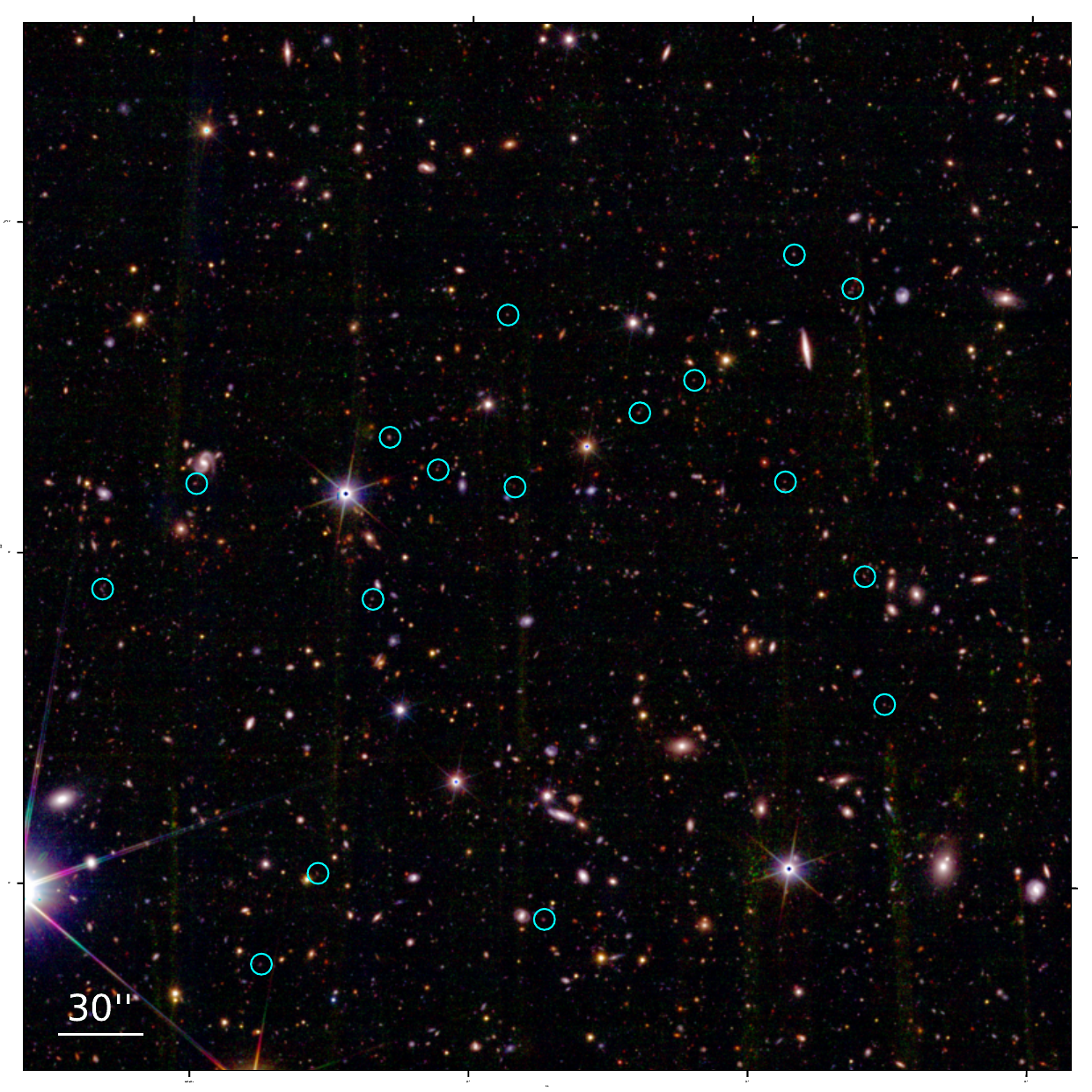}
    \includegraphics[width=\protowidth\textwidth]{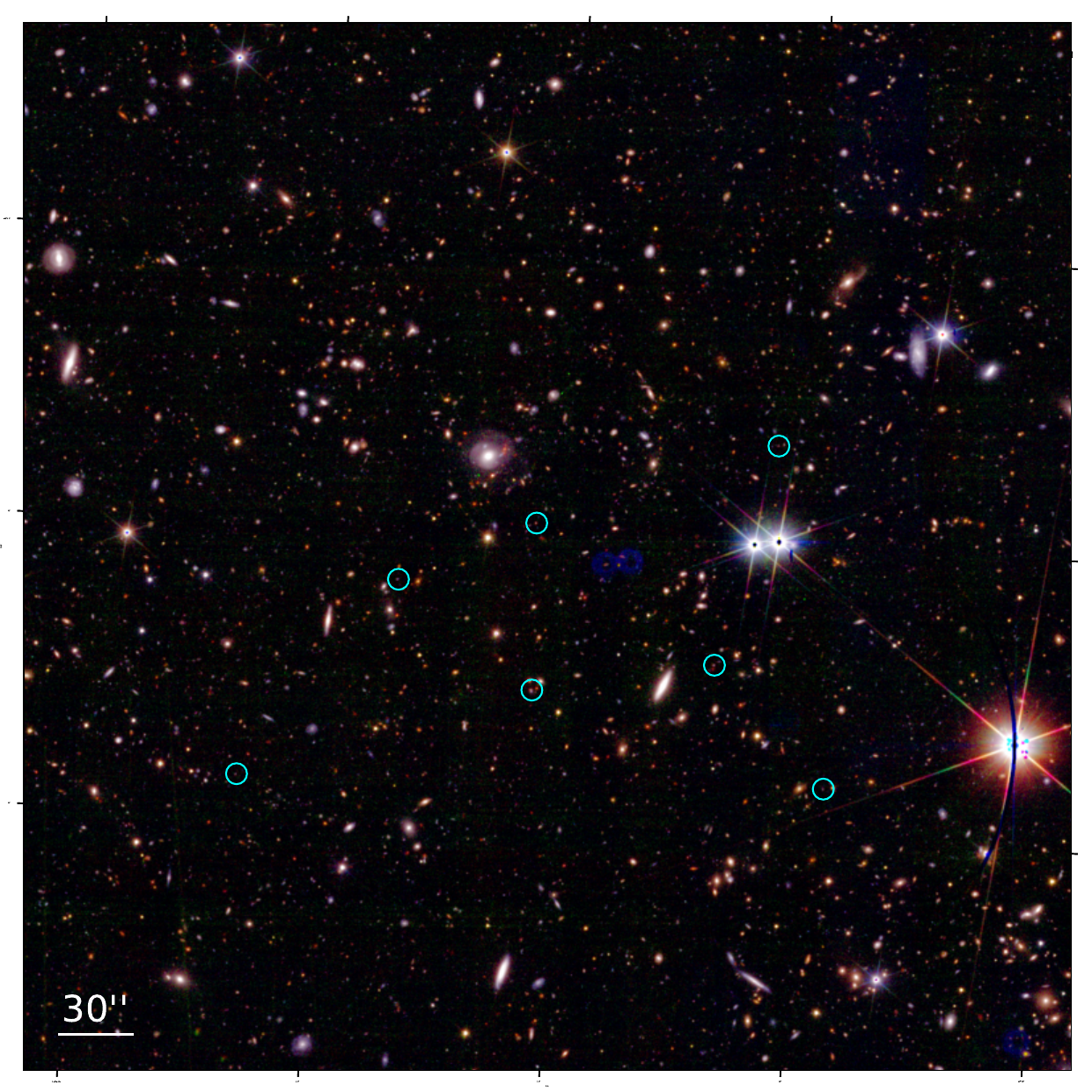}
    \includegraphics[width=\protowidth\textwidth]{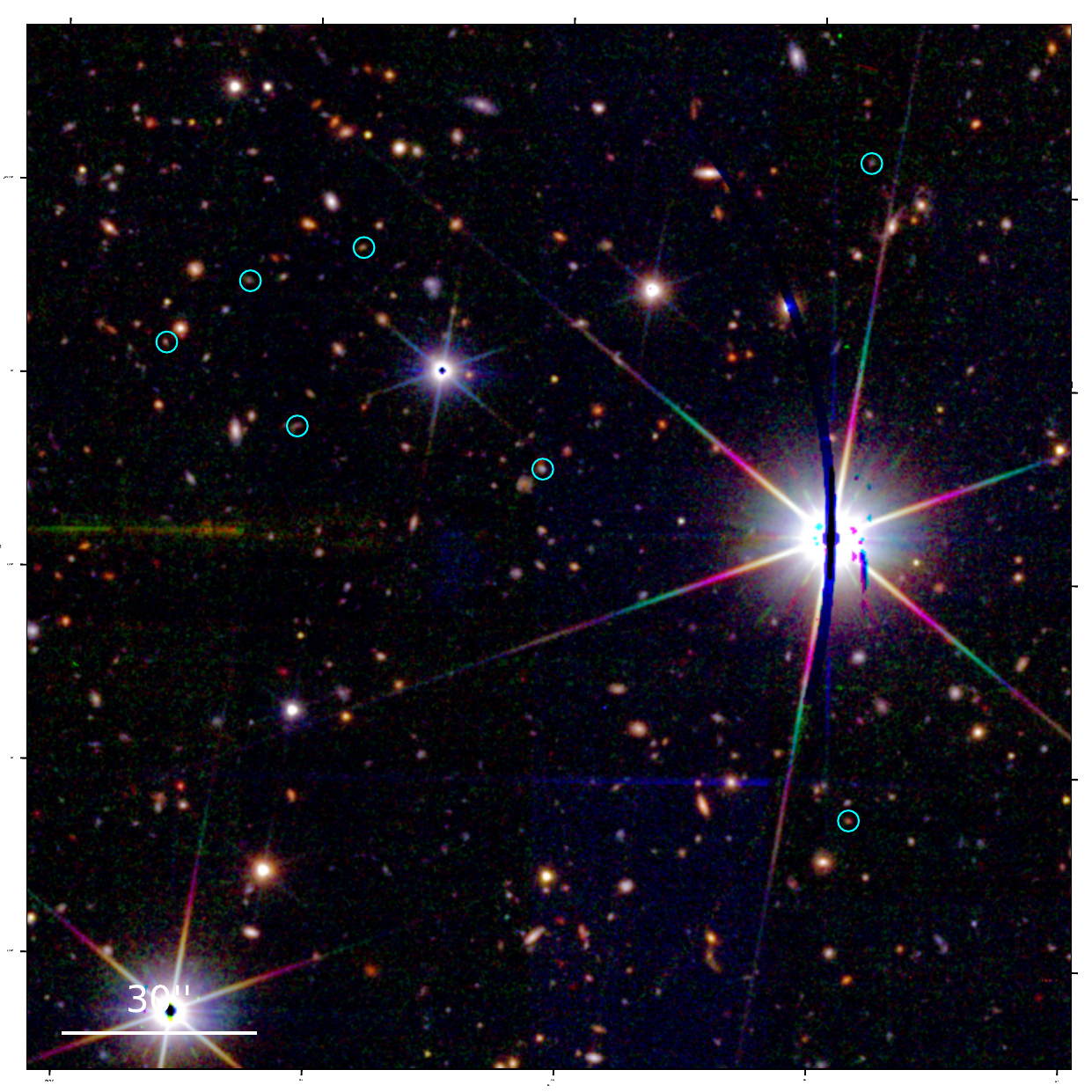}
    \includegraphics[width=\protowidth\textwidth]{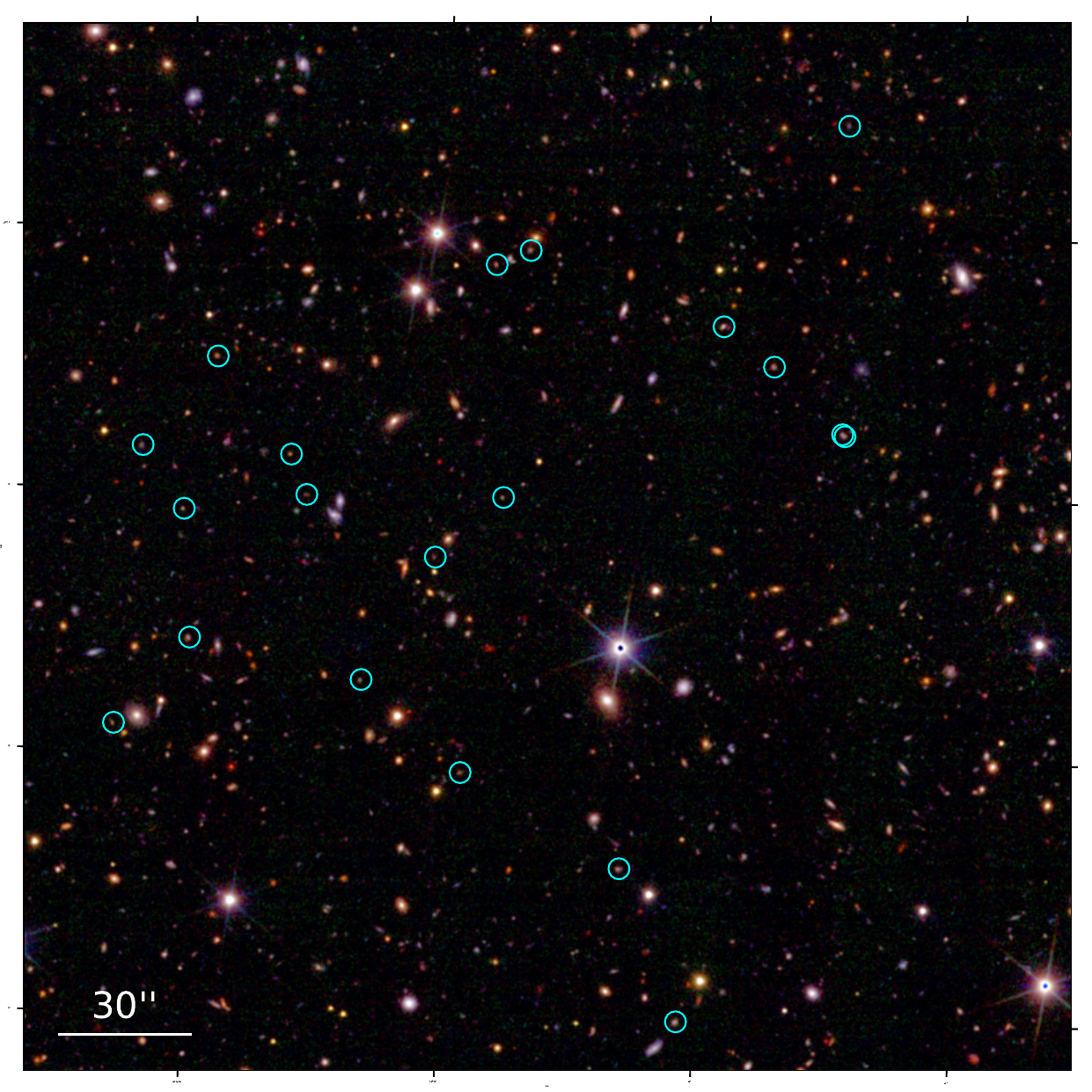}
    \includegraphics[width=\protowidth\textwidth]{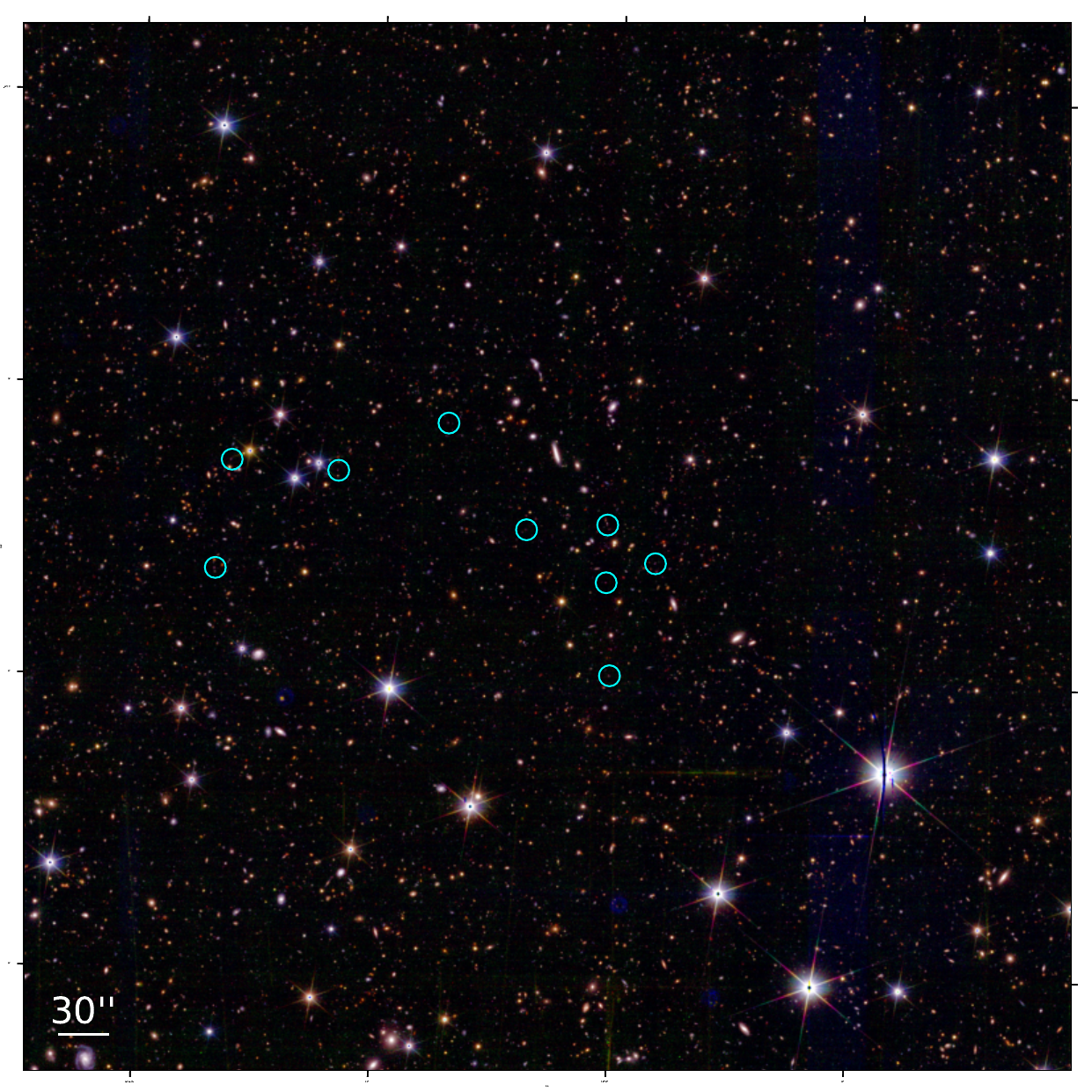}
    \caption{\Euclid Q1 images cutouts around the 20 protocluster candidates (blue=VIS \IE; green=NISP \YE; red = NISP \HE). From left to right and top to bottom, they are presented in the same order as in Tab.~\ref{tab:selected_protoclusters_top}. Cyan circles denote the member galaxies. The side of each image is equal to the angular diameter of the structure estimated by \texttt{DETECTIFz} (Tab.~\ref{tab:selected_protoclusters_top}, column "Radius [arcmin]").}
    \label{fig:cutouts}
\end{figure*}
\begin{figure*}[htbp!]
    \centering
    \includegraphics[width=\protowidth\textwidth]{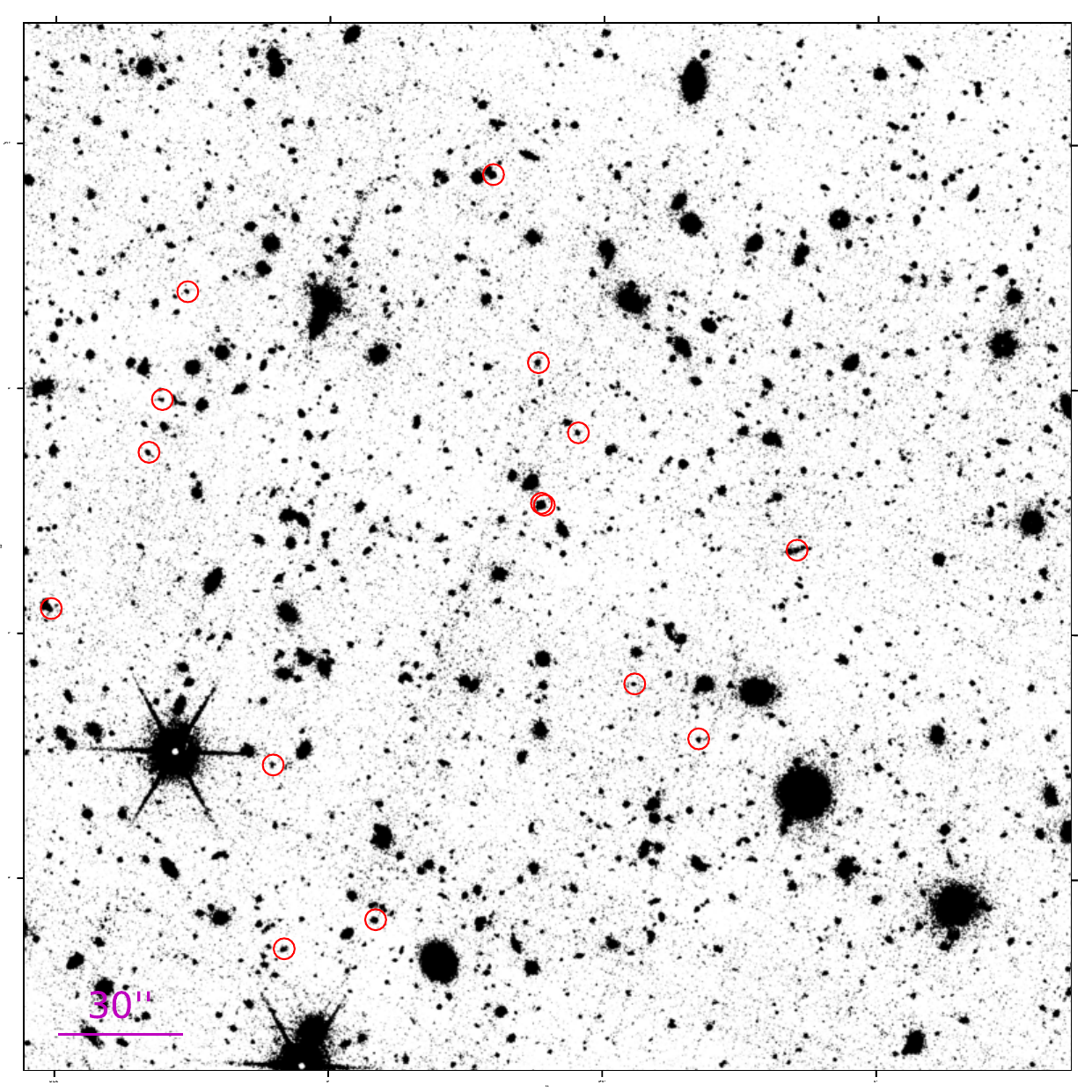}
    \includegraphics[width=\protowidth\textwidth]{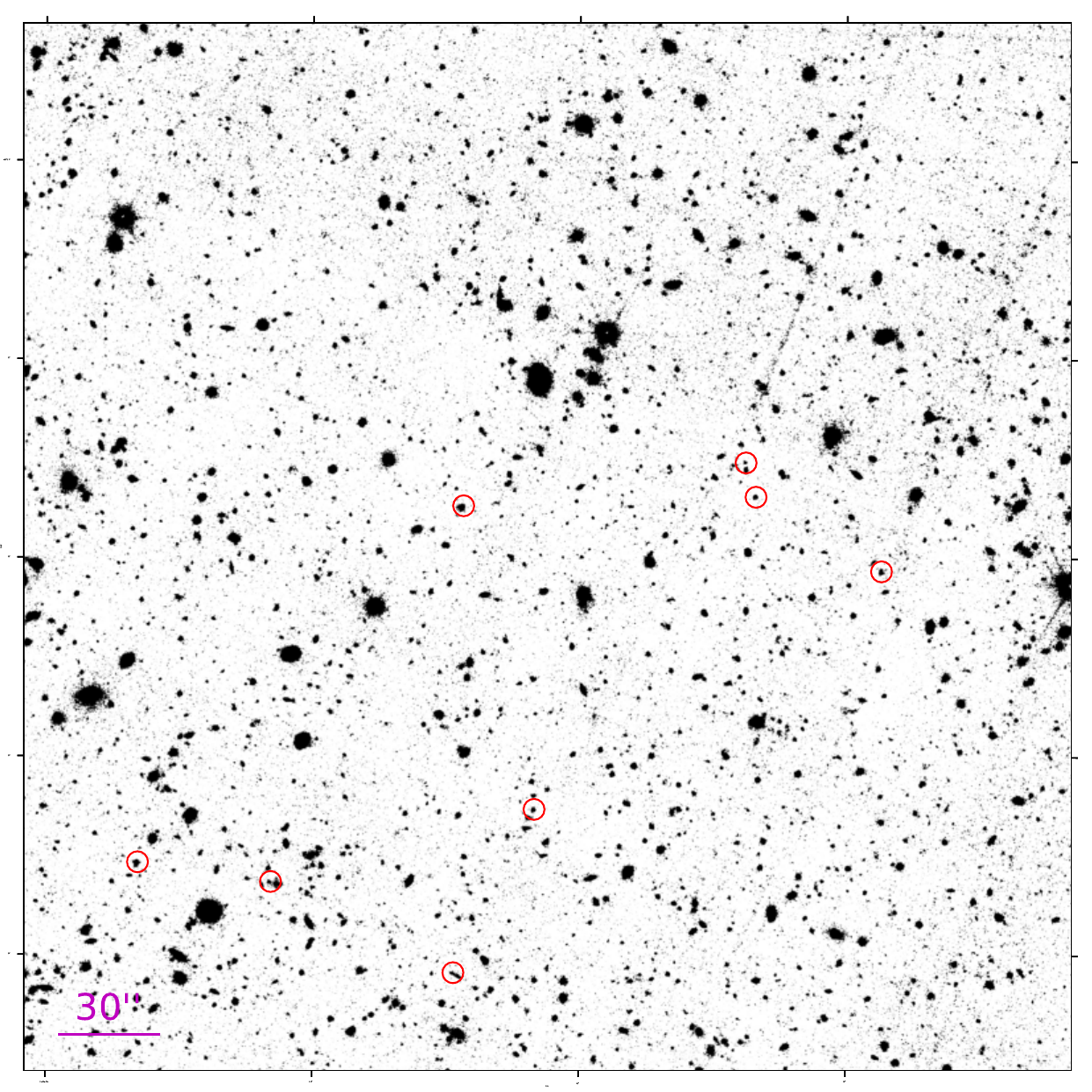}
    \includegraphics[width=\protowidth\textwidth]{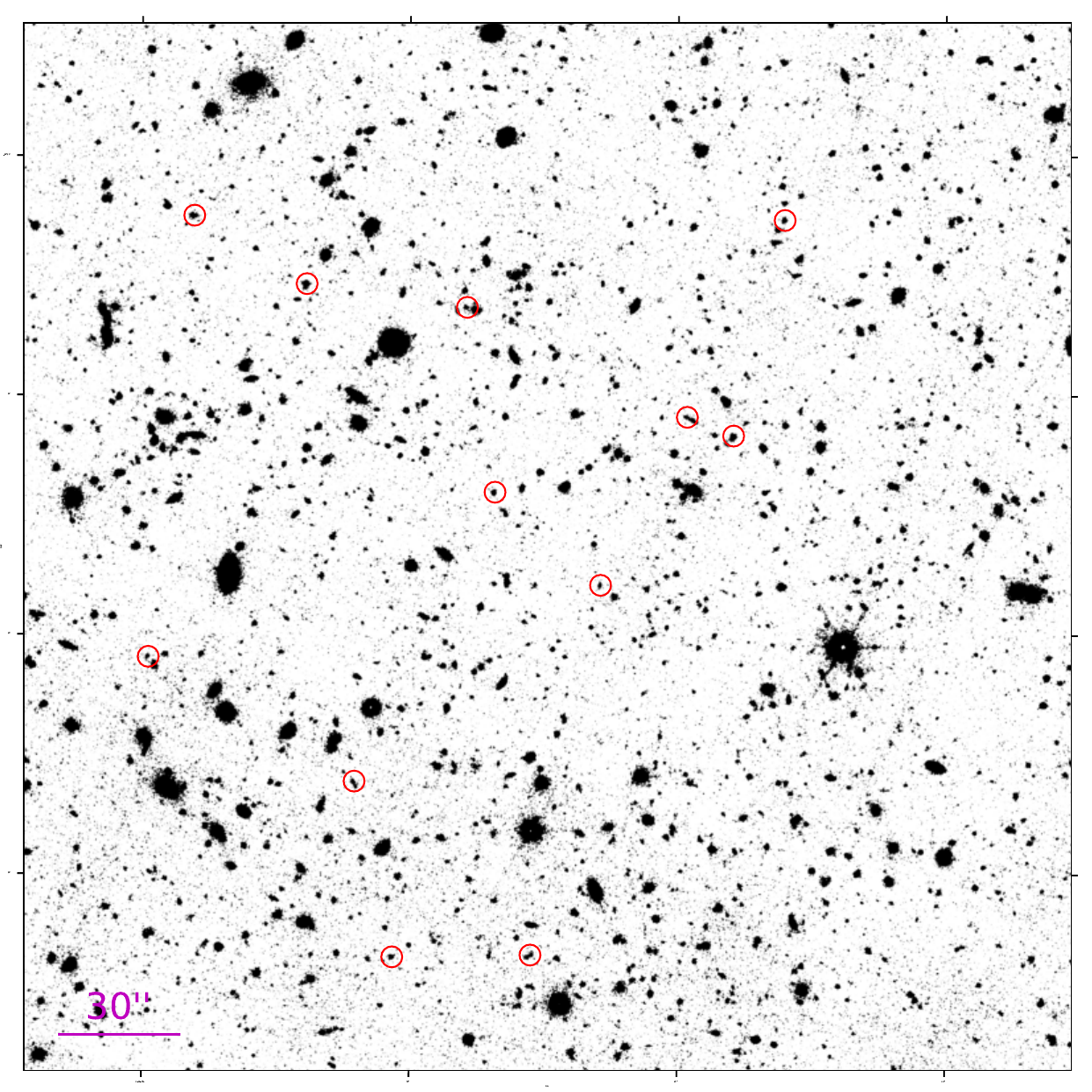}
    \includegraphics[width=\protowidth\textwidth]{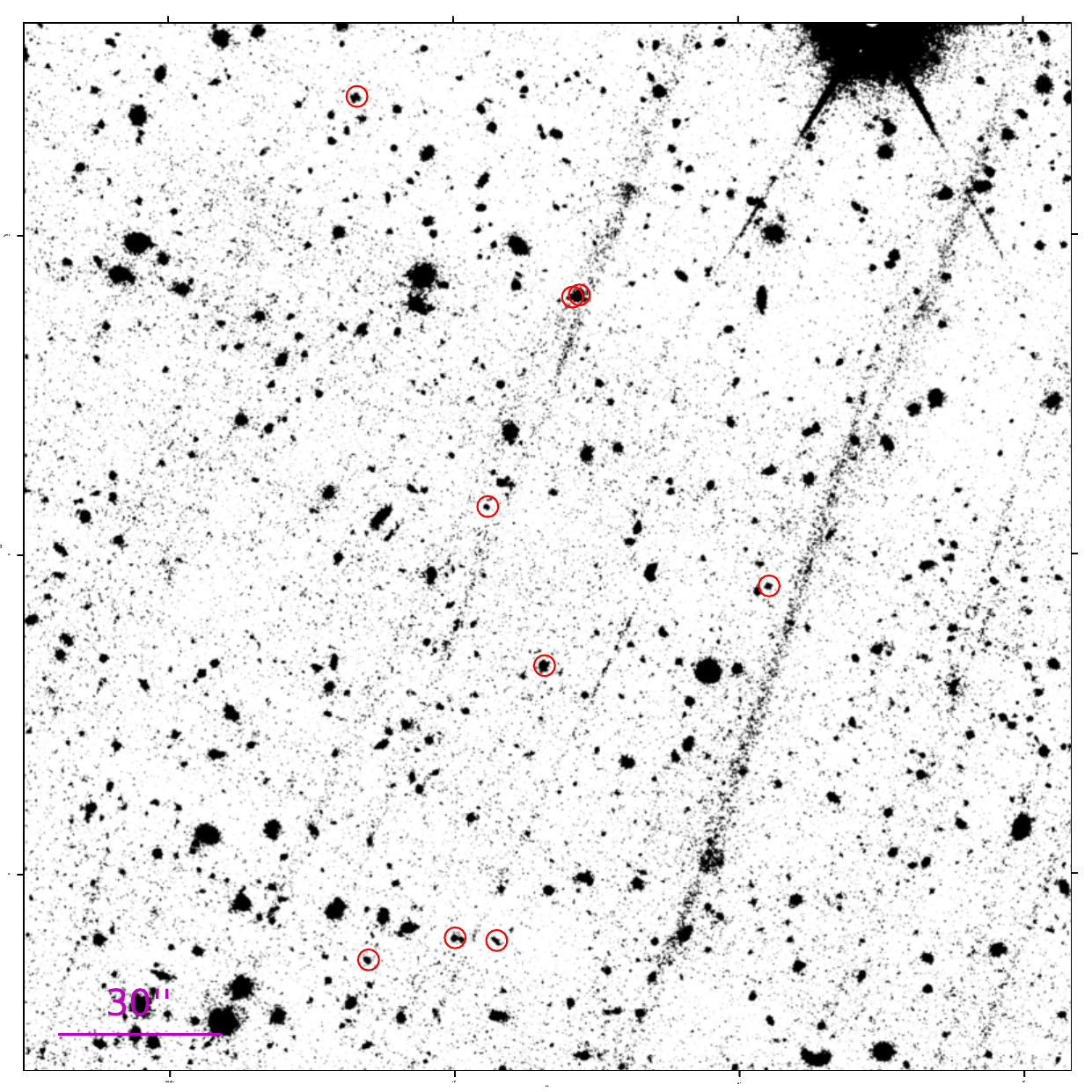}
    \includegraphics[width=\protowidth\textwidth]{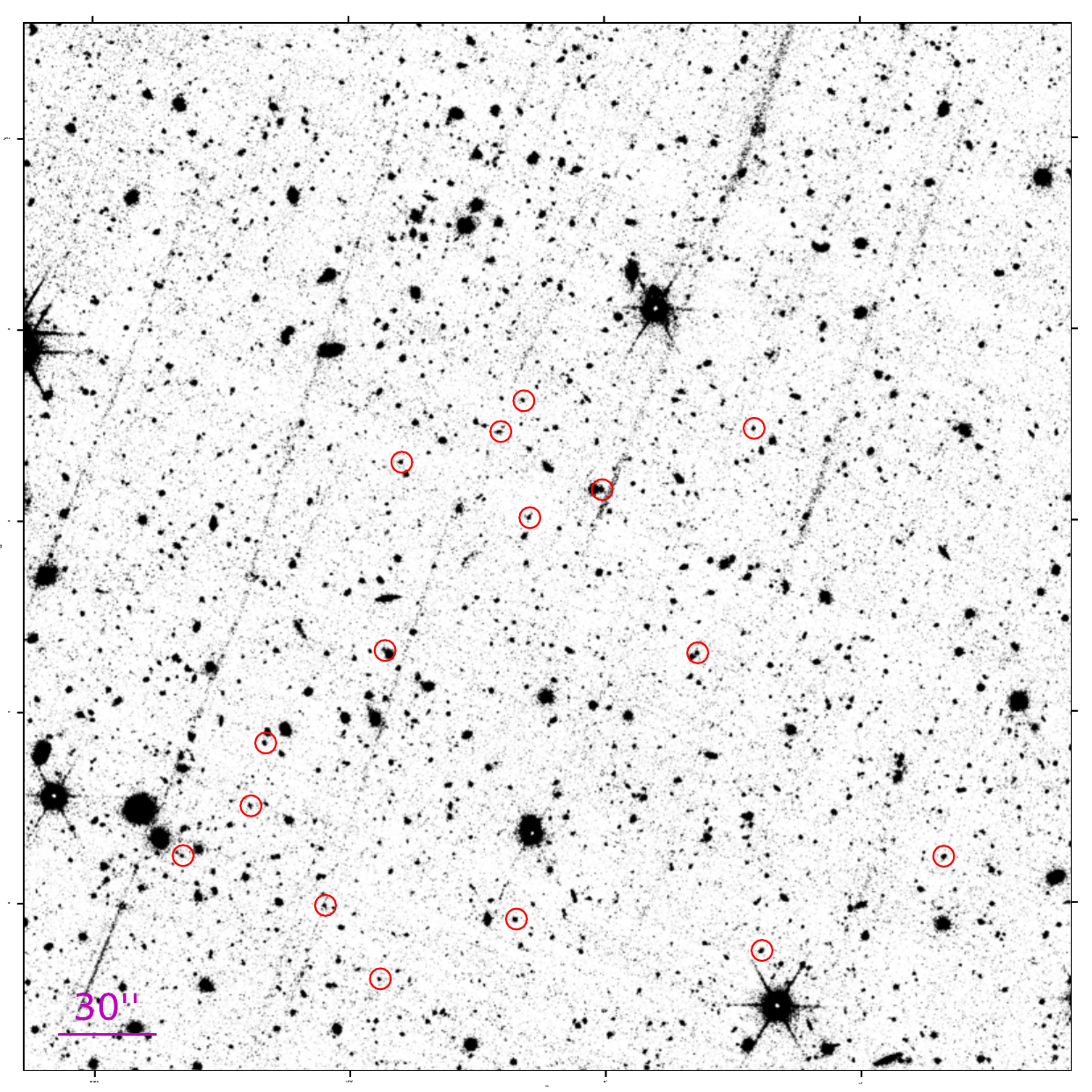}
    \includegraphics[width=\protowidth\textwidth]{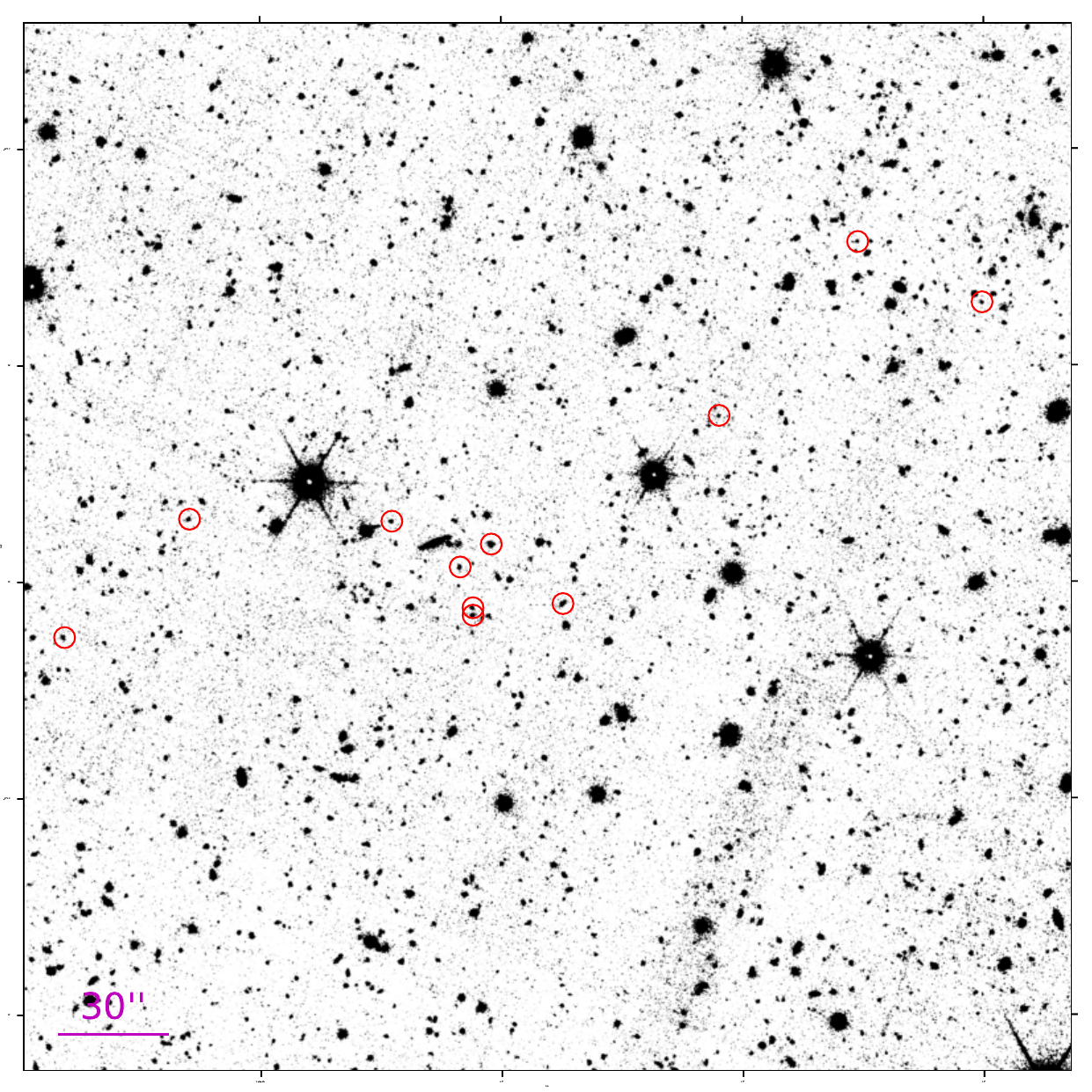}
    \includegraphics[width=\protowidth\textwidth]{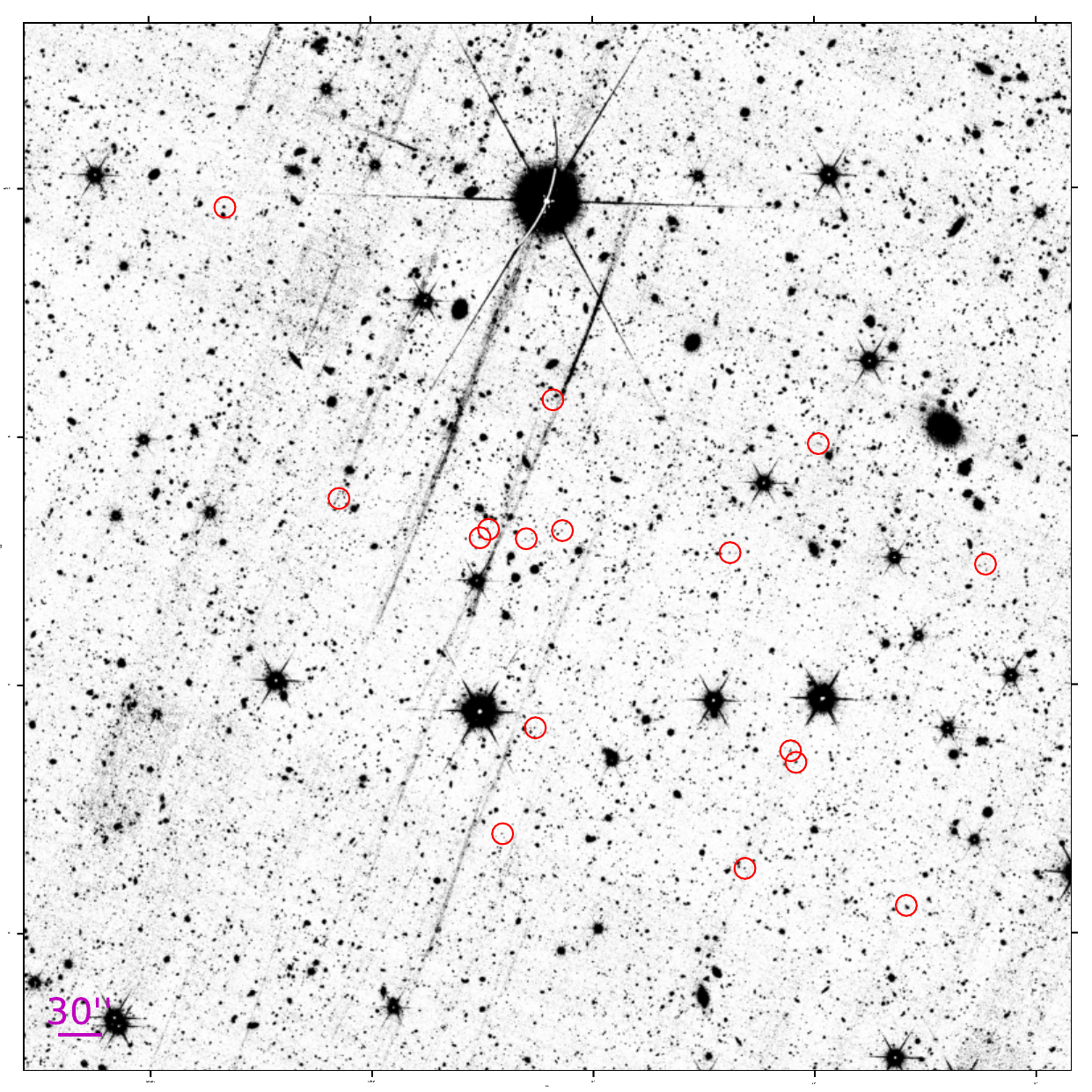}
    \includegraphics[width=\protowidth\textwidth]{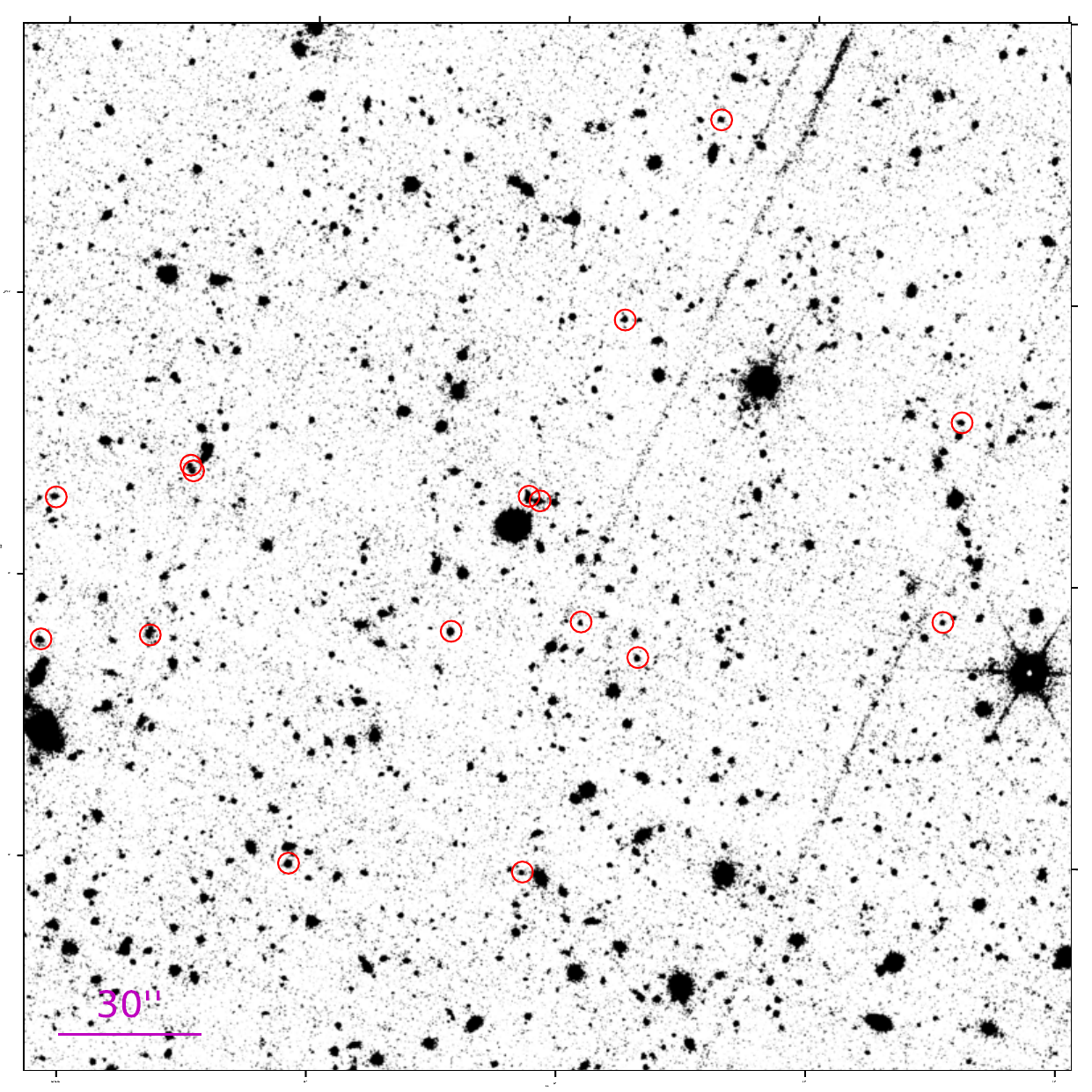}
    \includegraphics[width=\protowidth\textwidth]{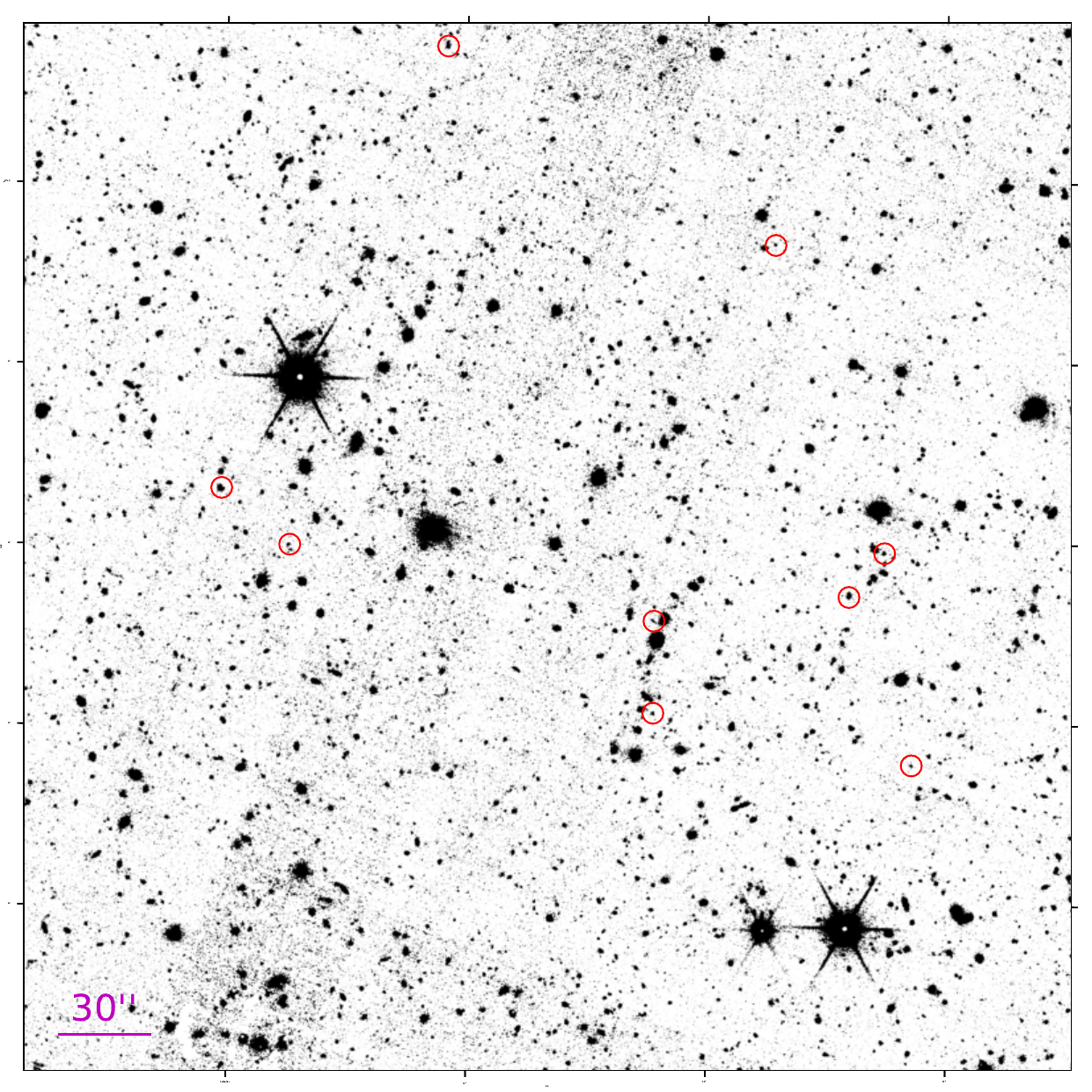}
    \includegraphics[width=\protowidth\textwidth]{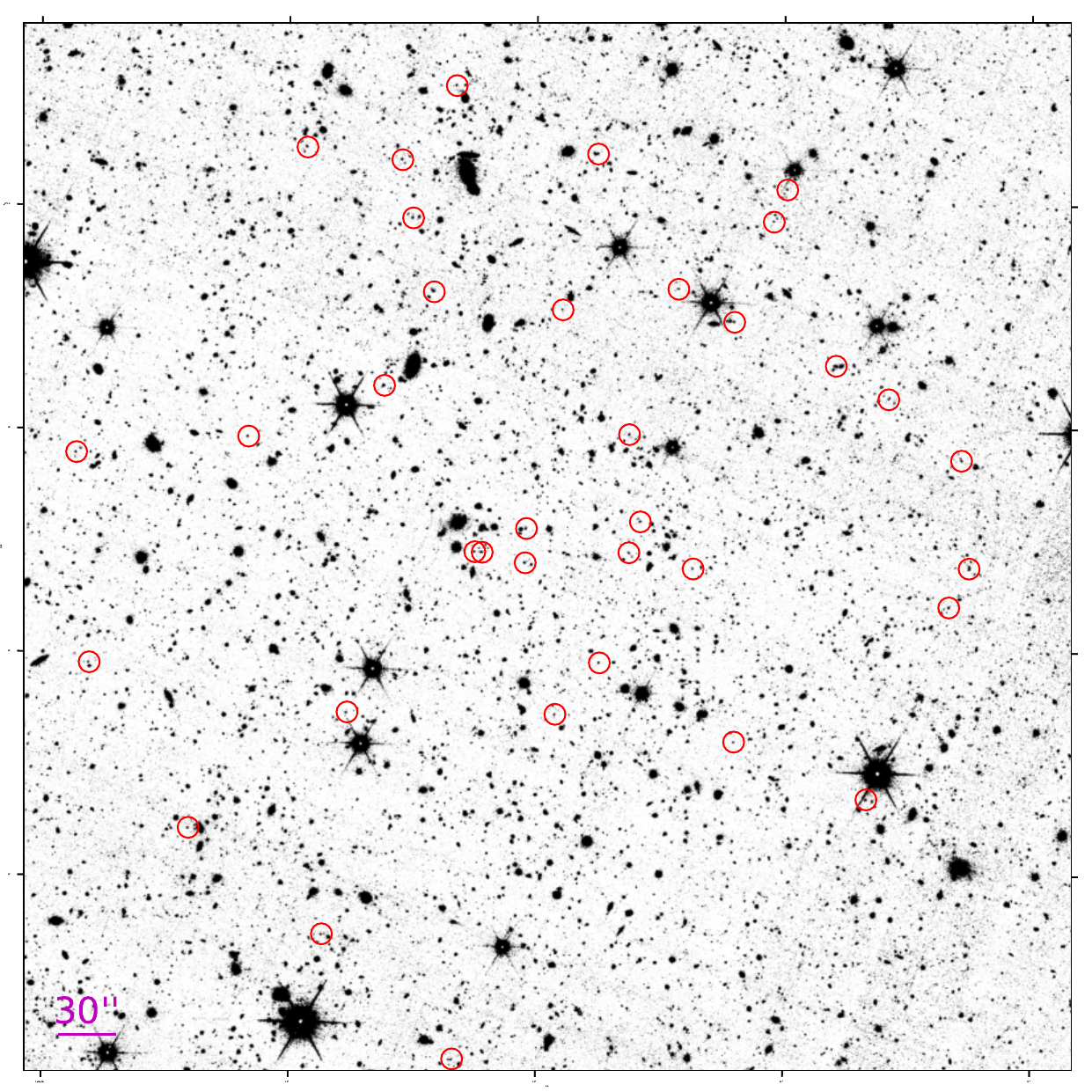}
    \includegraphics[width=\protowidth\textwidth]{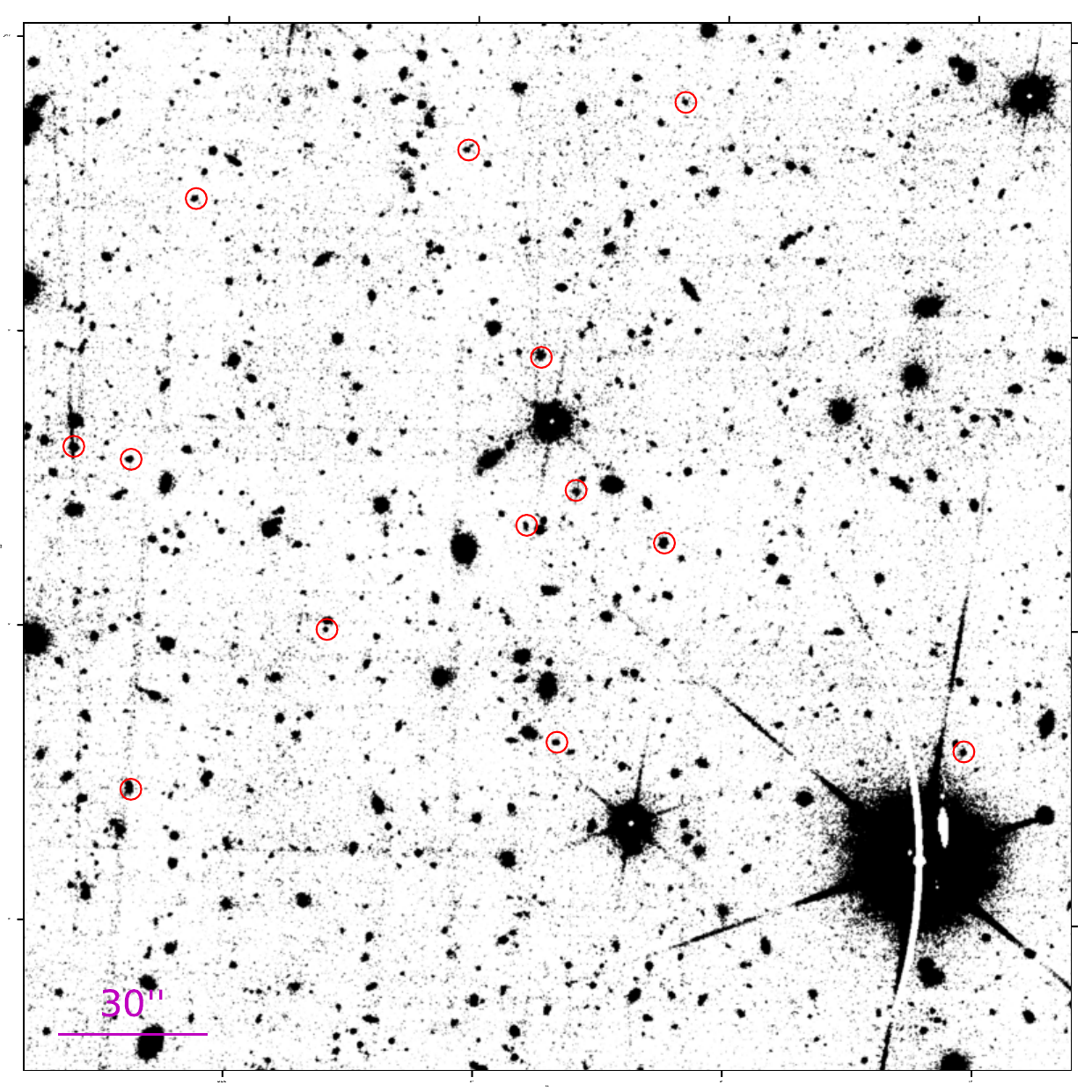}
    \includegraphics[width=\protowidth\textwidth]{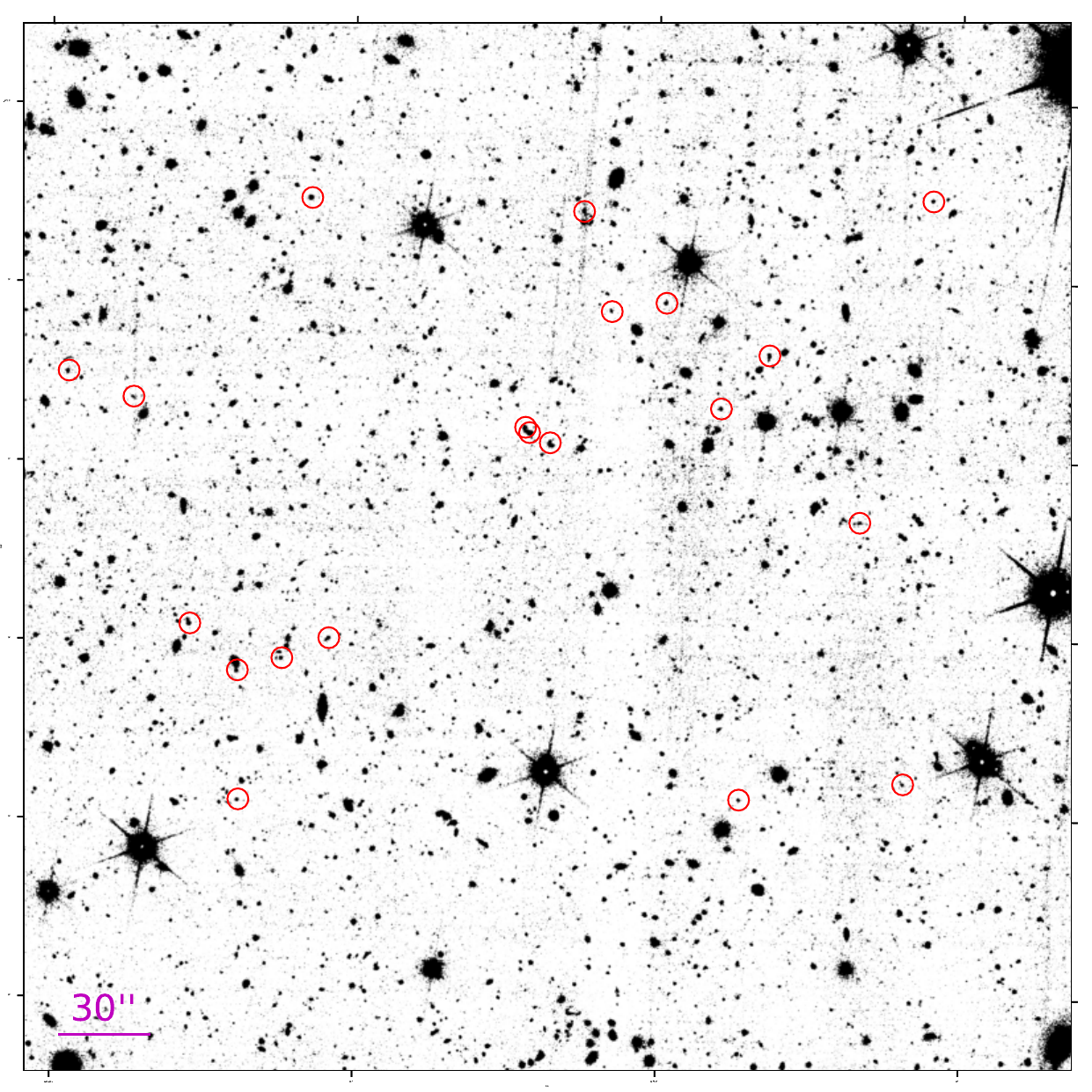}
    \includegraphics[width=\protowidth\textwidth]{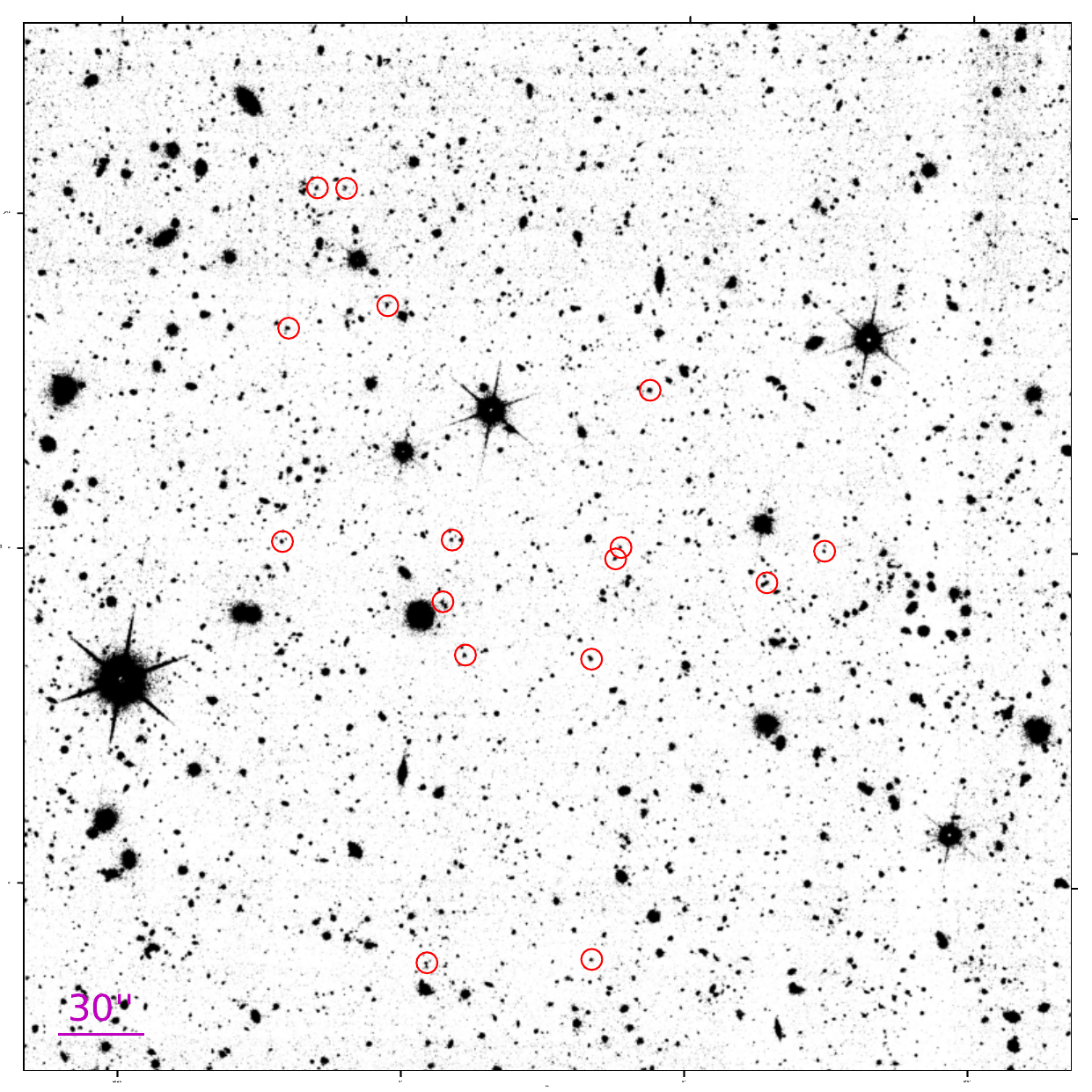}
    \includegraphics[width=\protowidth\textwidth]{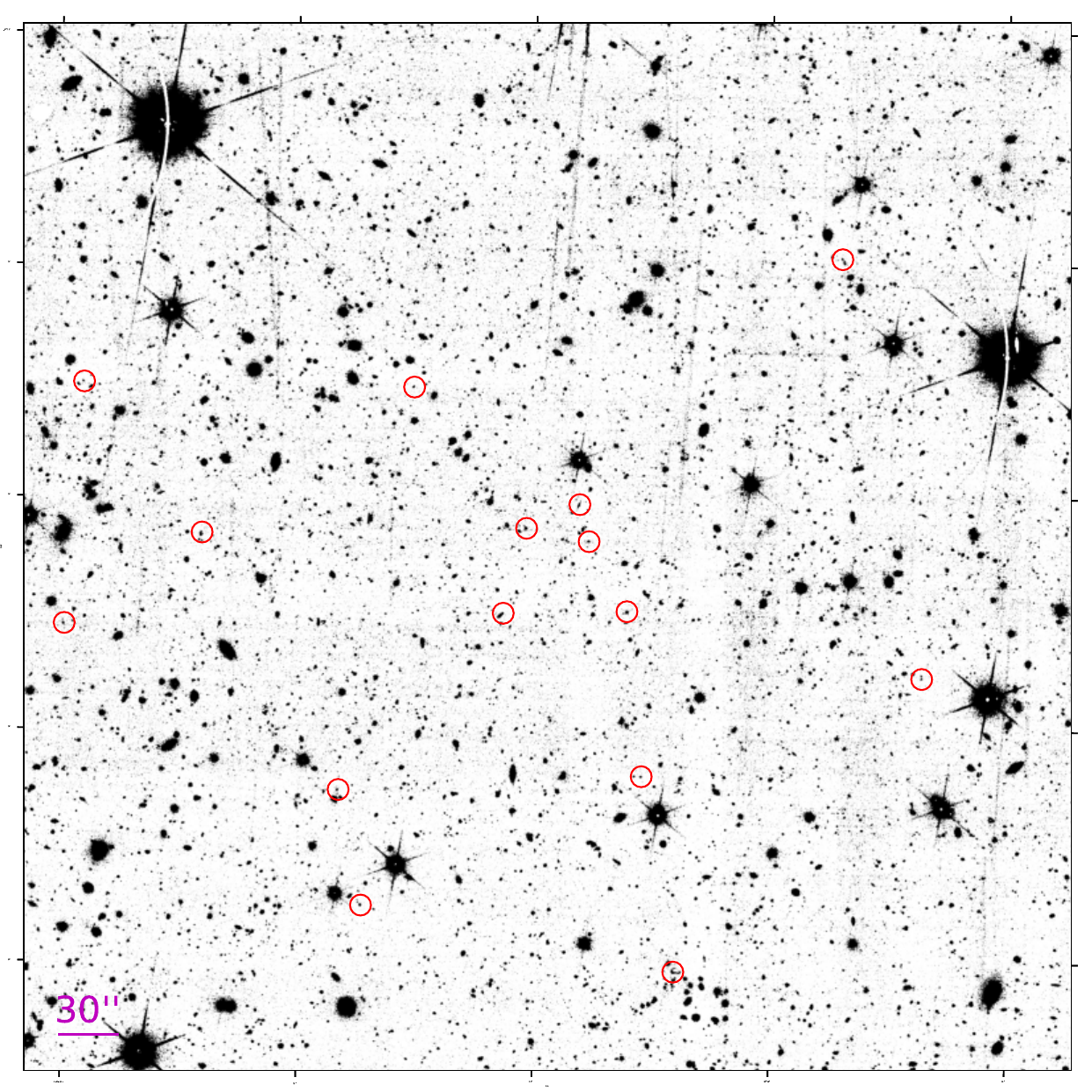}
    \includegraphics[width=\protowidth\textwidth]{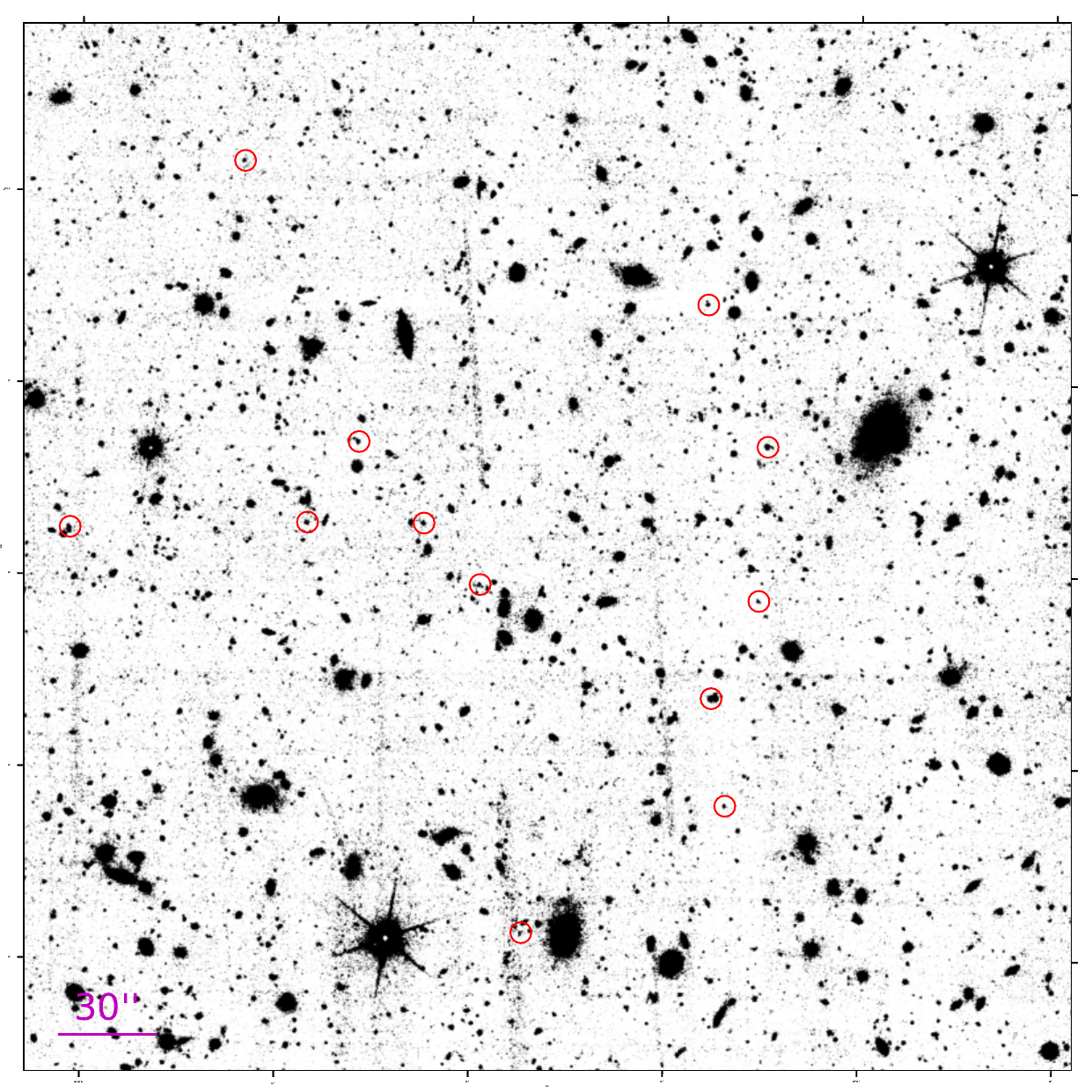}
    \includegraphics[width=\protowidth\textwidth]{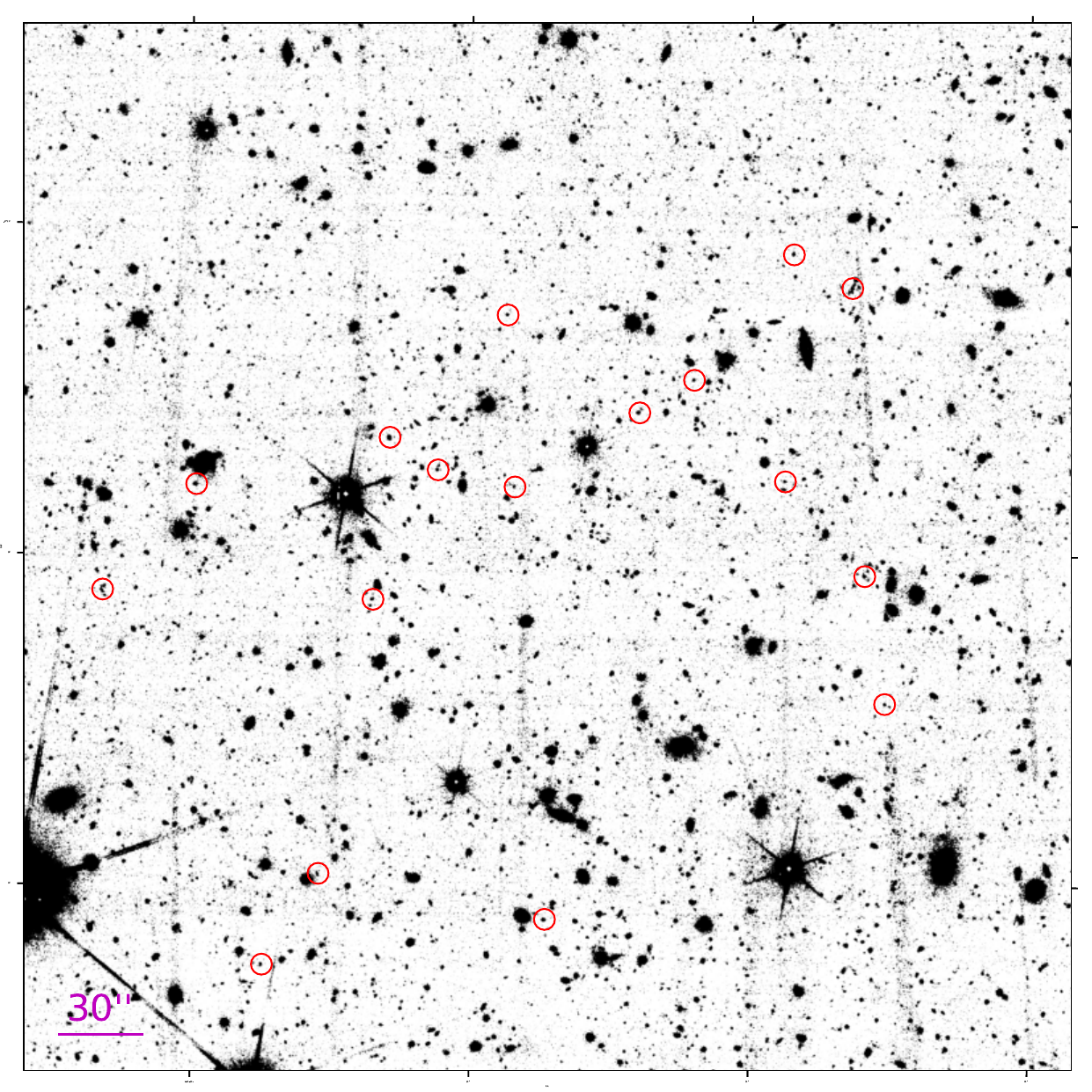}
    \includegraphics[width=\protowidth\textwidth]{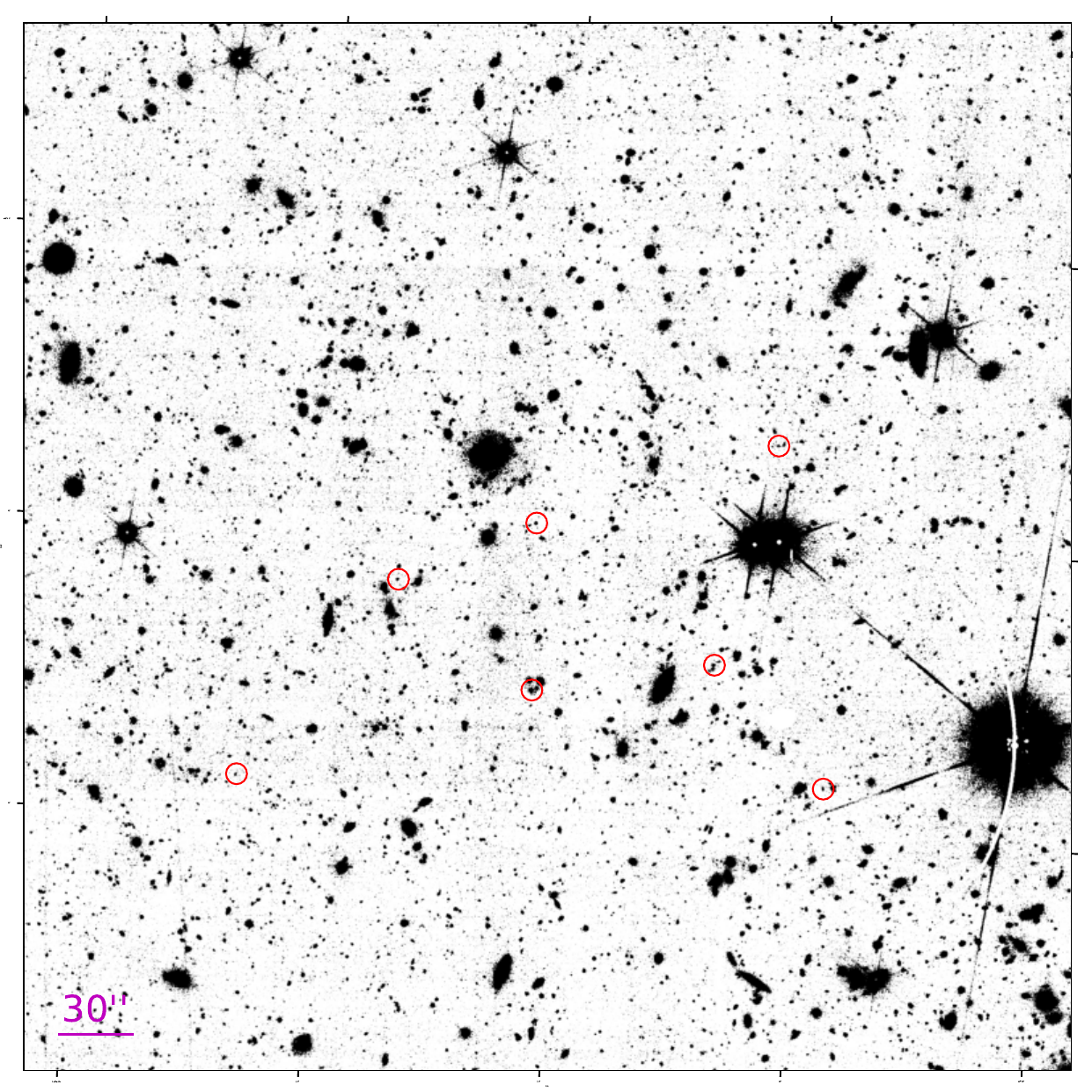}
    \includegraphics[width=\protowidth\textwidth]{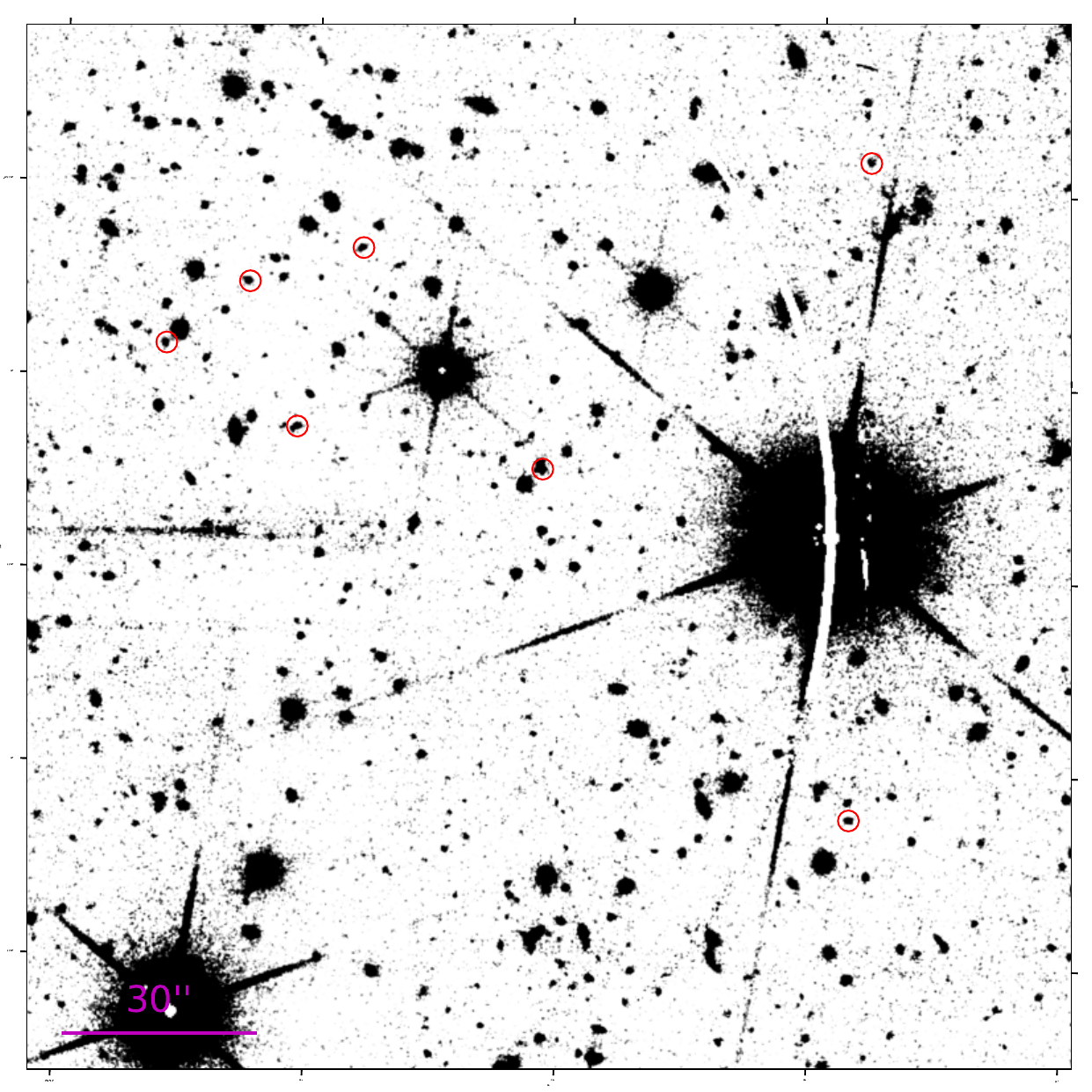}
    \includegraphics[width=\protowidth\textwidth]{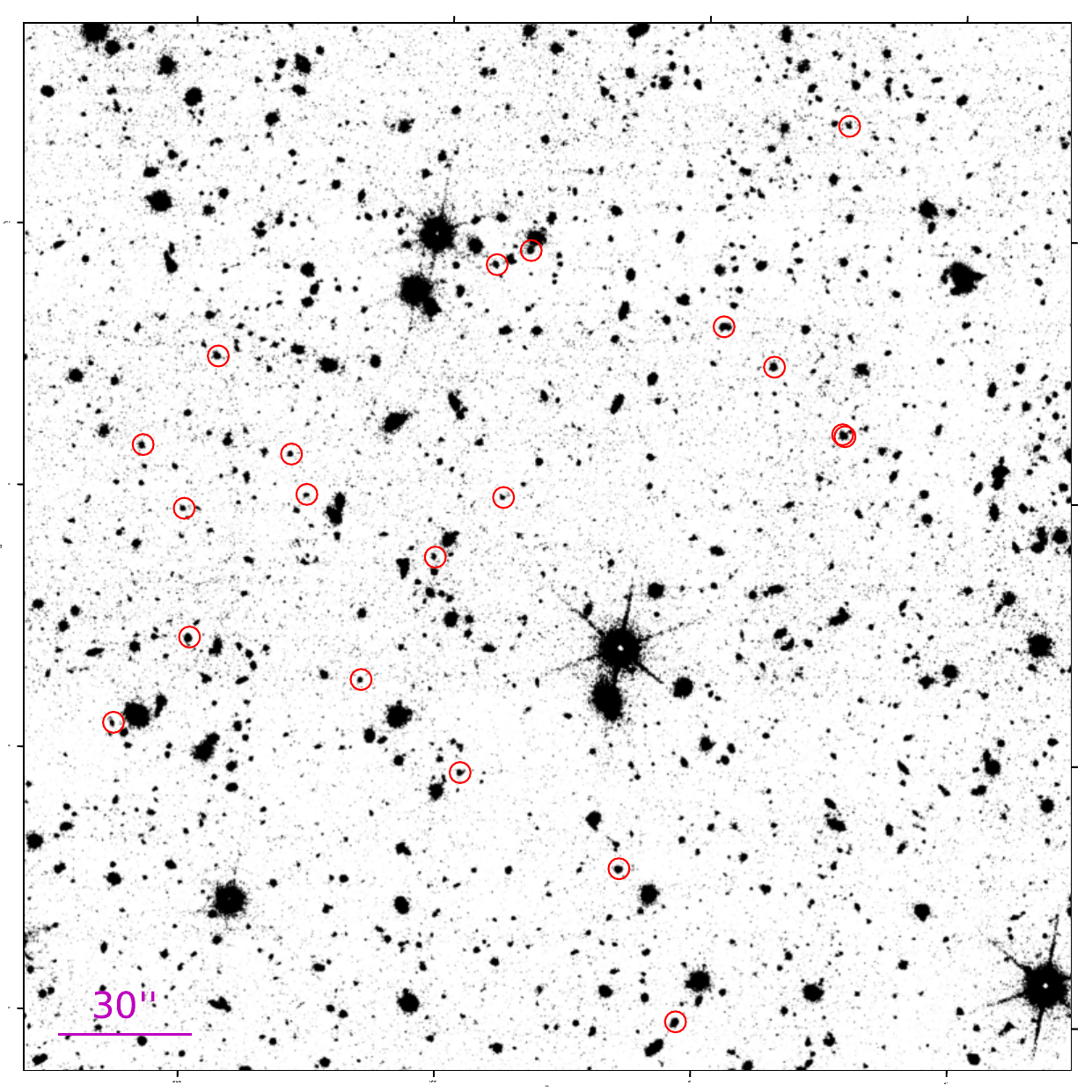}
    \includegraphics[width=\protowidth\textwidth]{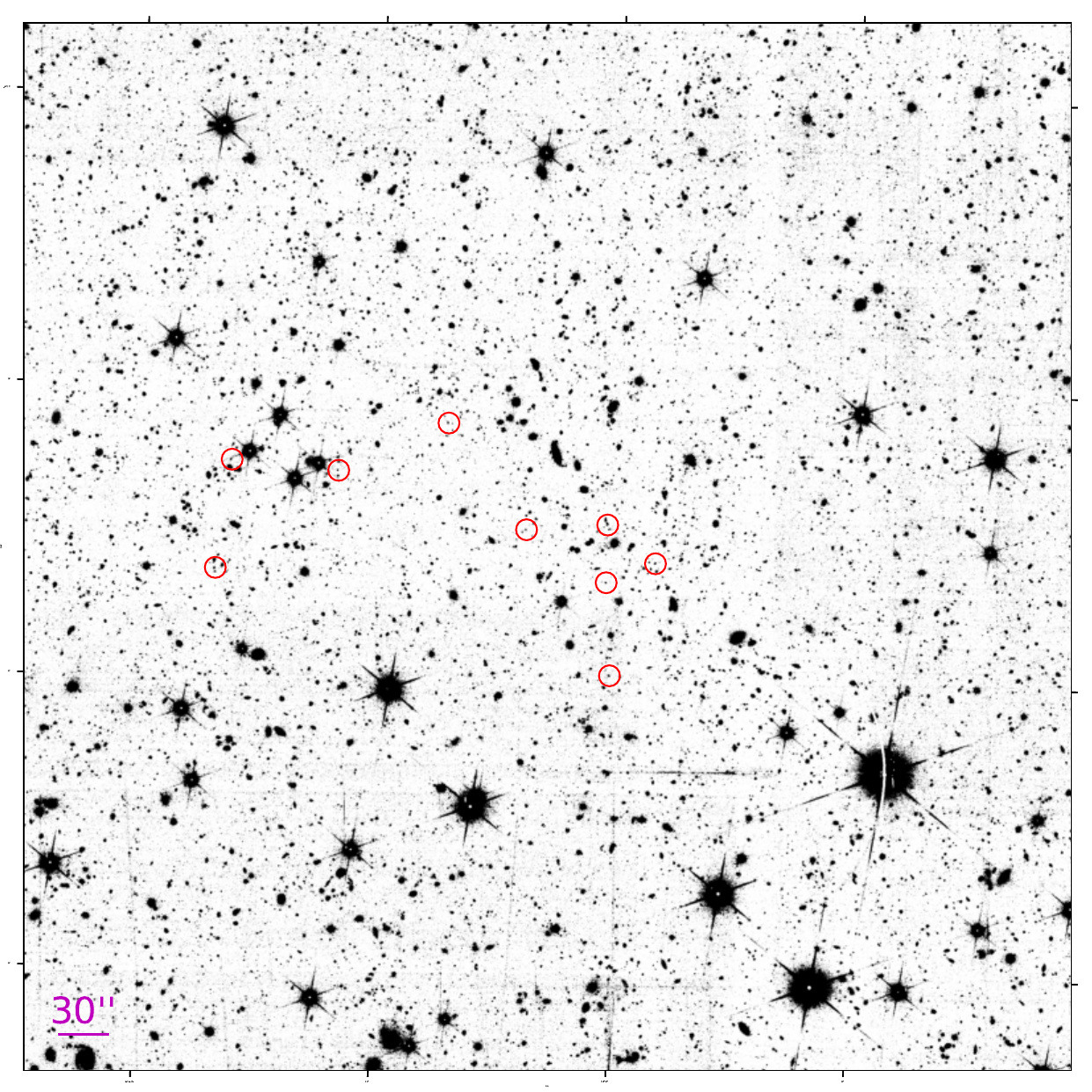}
    \caption{\Euclid Q1 images cutouts around the 20 protocluster candidates (NISP \HE). From left to right and top to bottom, they are presented in the same order as in Tab.~\ref{tab:selected_protoclusters_top}. Cyan circles denote the member galaxies. The side of each image is equal to the angular diameter of the structure estimated by \texttt{DETECTIFz} (Tab.~\ref{tab:selected_protoclusters_top}, column "Radius [arcmin]").}
    \label{fig:cutouts_negative}
\end{figure*}

\clearpage

\section{\label{sect:color-color}Colour-Colour diagram for the member galaxies of the 20 \Euclid-\Planck\ protoclusters}

\begin{figure*}[htbp!]
    \centering
    \includegraphics[width=0.85\hsize]{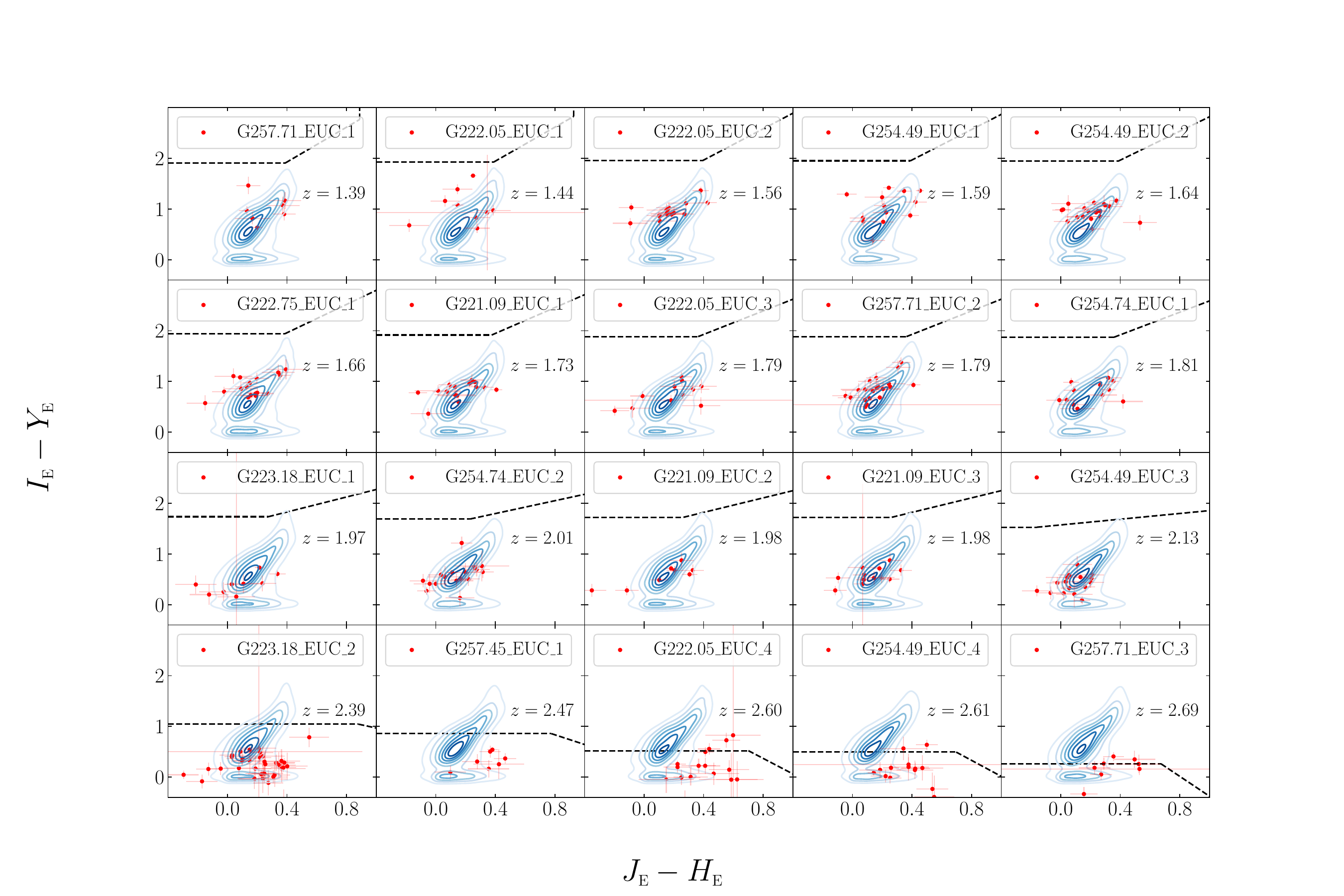}
    \caption{
    \Euclid colour-colour diagram: \IE $-$ \YE versus \JE $-$ \HE. 
    Red dots show the position of each member galaxy of the 20 \Euclid\ protocluster candidates. 
    Dashed lines indicate the active zone from \cite{bisigello2020} in the lower part of the graphs. Contours are the average density on the whole zone used for our detection. Protoclusters are ordered by increasing redshift, from left to right and from top to bottom. Galaxies present redder colours as redshift increases.
    }
    \label{fig:colour_colour}
\end{figure*}

We provide one colour-colour diagram for each of our 20 protoclusters. The member galaxies are plotted in the $\IE-\YE$ versus $\JE-\HE$ plane of the four \Euclid\ photometric bands (Fig.~\ref{fig:colour_colour}).

\section{\label{sect:crossvalidation}\texttt{DIANAS} and \texttt{PPM} detections}

We provide here the results of \texttt{PPM} and \texttt{DIANAS} detections along with the localisation and S/N of our 20 \texttt{DETECTIFz} detections. We also add a brief description of both methods.

\subsection{\texttt{PPM}}

The Poisson Probability Method (\texttt{PPM}) searches for high-$z$ megaparsec-scale overdensities of galaxies around a given target. It operates by statistically analyzing the distribution of photometric redshift probability estimates across all galaxies in the survey field. Through the use of a solid positional prior and accurate photometric redshift sampling, \texttt{PPM} partially overcomes the limitations deriving from low number-count statistics and shot-noise fluctuations, which are particularly relevant in the high-$z$ Universe, such as in the case of protoclusters. More specifically, \texttt{PPM} method uses photometric redshifts of galaxies to search for overdensities around each target along the line of sight. In this work, we used the projected space coordinates of \texttt{DETECTIFz} as input. To search for associated overdensities, \texttt{PPM} adopts an accurate sampling of the photometric redshift information although it has a less sophisticated tessellation of the projected space, which is performed in terms of concentric annuli centred around each target. We refer to previous studies for a detailed description of the method \citep{castignani2014,castignani2014b}, its wavelet-based extension (\texttt{wPPM} \citealt{castignani2019}), and the applications \citep{castignani2014b,castignani2019,calvi2023}. 

\subsection{\texttt{DIANAS}}

We also crossmatch our results with the protocluster validation (Q1) catalogue\footnote{The Q1 \texttt{DIANAS} Validation Protocluster Catalogue is a work in preparation \citep{NRCH2024eas} using the Detection Algorithm for NAscent Structures.}. \texttt{DIANAS} detection strategy was specifically designed and fine-tuned for galaxy protoclusters in EWS. 

Given that the \Spitzer\ data will not be available on the whole EWS footprint, to mimic the protocluster detectability in EWS, we considered the original MER and \texttt{PHZ} catalogues from the Q1 data release. The galaxy selection sources in the MER and \texttt{PHZ} catalogues satisfy the same criteria as the Q1 galaxy cluster catalogue by \cite{Q1-SP050}, except that we followed a stricter \HE limit and photometric quality cut: $\HE < 23.5$, $\delta z/(1 + \zphz) < 0.1$, and no \texttt{PHZ} flags. No VIS\_DET flag was applied, as we targeted structures at $z>1.4$. The 3D Gaussian filters were specifically calibrated with GAEA and MAMBO simulations, to mimic the protocluster properties in \Euclid-like survey as described in \cite{EP-Boehringer}. The \Euclid\ sources in the MER catalogue have photometric quality comparable to GAEA and MAMBO if the same cuts as above are applied.

 The \texttt{DIANAS} detection code was fine-tuned to detect rich protoclusters ($N_{\rm gal} \geq 10$) at $z>1.4$, with photometric uncertainty\footnote{The nominal photometric uncertainty for EWS correspond to $ \sigma_{\mathrm{NMAD}} = 0.05$ \citep{Scaramella-EP1}.} $\sigma_{\mathrm{NMAD}} = 0.06$, as estimated for the faint end of \IE magnitude \citep[see Fig.9 in ][]{Q1-TP005}. Meaning, we use the photometric redshifts from the Q1 release (\zphz) and not those estimated by \texttt{NNPZ} (\znnpz). This results in a sensitivity loss, explaining why several of our protoclusters are not retrieved by this method.

\begin{table*}[htbp!]
\setlength{\tabcolsep}{3pt}
\renewcommand{\arraystretch}{1.4}
\centering

\caption{\Euclid protoclusters detected with \texttt{DETECTIFz} at the location of \Planck\ protocluster candidates.}
\label{tab:dtfe_ppm}
\begin{tabular}{l@{\hskip 2em}cccc@{\hskip 2em}cccc@{\hskip 2em}cc}
\hline\hline

\Planck\ + \Euclid & \multicolumn{4}{c}{\texttt{DETECTIFz}} & \multicolumn{4}{c}{\texttt{DIANAS}} & \multicolumn{2}{c}{\texttt{PPM}}\\
overdensity name  & RA & Dec  & $z_{\rm ph}$ &  S/N & RA   & Dec   & $z_{\rm ph}$  & S/N & $z_{\rm ph}$ & S/N  \\ 
   & [deg]   &  [deg]  &   &      & [deg] & [deg] & & &  &\\
   
\hline

G221.09\_EUC\_1 	 & 	52.7334 	 & 	$-$26.4440 	 & 	$1.73^{+0.06}_{-0.06}$  & 	3.6 & $\cdots$ & $\cdots$ & $\cdots$ & $\cdots$ & $1.76^{+0.07}_{-0.07}$ & 5.0 \\ 
 
G221.09\_EUC\_2 	 & 	52.6780 	 & 	$-$26.3657 	 & 	$1.98^{+0.06}_{-0.04}$ 	& 	3.1 & $\cdots$ & $\cdots$ & $\cdots$ & $\cdots$ & $1.97^{+0.09}_{-0.09}$ & 3.0 \\ 
 
G221.09\_EUC\_3 	 & 	52.6976 	 & 	$-$26.4105 	 & 	$1.98^{+0.07}_{-0.09}$ 	& 	3.1 & 52.7105 	 & 	$-$26.4153 	 & 	$2.04^{+0.28}_{-0.28}$ 	 & 	5.4 & $1.85^{+0.09}_{-0.09}$ & 5.9 \\ 
 
 \hline
G222.05\_EUC\_1 	 & 	53.1945 	 & 	$-$26.8829 	 & 	$1.44^{+0.03^*}_{-0.05}$ 	& 	3.1 & 	53.1503 	 & 	$-$26.8293 	 & 	$1.44^{+0.24}_{-0.24}$ 	 & 	3.9 & $1.53^{+0.08^*}_{-0.08}$ & 4.5 \\ 
 
G222.05\_EUC\_2 	 & 	53.2056 	 & 	$-$26.9356 	 & 	$1.56^{+0.05}_{-0.04}$ 	& 	3.8 & $\cdots$ & $\cdots$ & $\cdots$ & $\cdots$ & $1.63^{+0.08}_{-0.08}$ & 2.8 \\ 
 
G222.05\_EUC\_3 	 & 	53.1627 	 & 	$-$26.9806 	 & 	$1.79^{+0.08}_{-0.06}$ 	& 	5.5 & $\cdots$ & $\cdots$ & $\cdots$ & $\cdots$ & $1.95^{+0.09^*}_{-0.09}$ & 2.3 \\
& & & & & & & & & $1.61^{+0.08^*}_{-0.08}$ & 4.7 \\
 
G222.05\_EUC\_4 	 & 	53.1602 	 & 	$-$26.8222 	 & 	$2.60^{+0.09}_{-0.08}$ 	& 	5.1 & 	53.1450 	 & 	$-$26.8503 	 & 	$2.44^{+0.24}_{-0.24}$ 	 & 	2.0 & $2.61^{+0.06}_{-0.06}$ & 2.1 \\ 
 
 \hline
G222.75\_EUC\_1 	 & 	51.3510 	 & 	$-$27.5813 	 & 	$1.66^{+0.14}_{-0.06}$ 	& 	3.4 & 	51.3599 	 & 	$-$27.5841 	 & 	$1.64^{+0.24}_{-0.24}$ 	 & 	3.2 & $1.76^{+0.07}_{-0.07}$ & 2.7 \\ 
 
 \hline
G223.18\_EUC\_1 	 & 	52.5416 	 & 	$-$27.7002 	 & 	$1.97^{+0.13}_{-0.01}$ 	& 	3.2 & $\cdots$ & $\cdots$ & $\cdots$ & $\cdots$ & $2.03^{+0.09}_{-0.09}$ & 2.5 \\ 
 
G223.18\_EUC\_2 	 & 	52.6231 	 & 	$-$27.6509 	 & 	$2.39^{+0.17}_{-0.14}$ 	& 	5.5 & 	52.5954 	 & 	$-$27.6208 	 & 	$2.36^{+0.24}_{-0.24}$ 	 & 	5.2 & $2.47^{+0.08}_{-0.08}$ & 2.7 \\ 
 
 \hline
G254.49\_EUC\_1 	 & 	60.6815 	 & 	$-$47.2287 	 & 	$1.59^{+0.02}_{-0.08}$ 	& 	3.6 & $\cdots$ & $\cdots$ & $\cdots$ & $\cdots$ & $1.62^{+0.09}_{-0.09}$ & 2.7 \\ 
 
G254.49\_EUC\_2 	 & 	60.7235 	 & 	$-$47.2912 	 & 	$1.64^{+0.08}_{-0.04}$ 	& 	4.8 & $\cdots$ & $\cdots$ & $\cdots$ & $\cdots$ & $1.65^{+0.09}_{-0.09}$ & 3.0 \\ 
 
G254.49\_EUC\_3 	 & 	60.7705 	 & 	$-$47.3329 	 & 	$2.13^{+0.08}_{-0.10}$ 	& 	4.7  & 	60.7710 	 & 	$-$47.2990 	 & 	$2.00^{+0.28}_{-0.28}$ 	 & 	3.2 & $2.23^{+0.09^*}_{-0.09}$ & 3.5 \\ 
 
G254.49\_EUC\_4 	 & 	60.7472 	 & 	$-$47.2737 	 & 	$2.61^{+0.11}_{-0.12}$ 	& 	4.2 & $\cdots$ & $\cdots$ & $\cdots$ & $\cdots$ & $2.59^{+0.08}_{-0.08}$ & 2.8 \\ 
 
 \hline
G254.74\_EUC\_1 	 & 	60.7651 	 & 	$-$47.4641 	 & 	$1.81^{+0.01}_{-0.07}$ 	& 	3.1  & 	60.7815 	 & 	$-$47.4562 	 & 	$1.92^{+0.26}_{-0.26}$ 	 & 	5.3 & $1.71^{+0.07^*}_{-0.07}$ & 3.3 	\\ 
 
G254.74\_EUC\_2 	 & 	60.8219 	 & 	$-$47.4658 	 & 	$2.01^{+0.11}_{-0.09}$  & 	4.3  & 	60.7815 	 & 	$-$47.4562 	 & 	$1.92^{+0.26}_{-0.26}$ 	 & 	5.3 & $2.14^{+0.10^*}_{-0.10}$ & 2.1 	\\ 
& & & & & & & & & $1.79^{+0.08^*}_{-0.08}$ & 3.0 \\
 
 \hline
G257.45\_EUC\_1 	 & 	57.5446 	 & 	$-$48.7012 	 & 	$2.47^{+0.10}_{-0.05}$  & 	3.3 & $\cdots$ & $\cdots$ & $\cdots$ & $\cdots$ & $2.41^{+0.08}_{-0.08}$ & 4.3 \\ 
 
 \hline
G257.71\_EUC\_1 	 & 	59.6343 	 & 	$-$49.3654 	 & 	$1.39^{+0.08}_{-0.01}$ 	& 	3.4 & $\cdots$ & $\cdots$ & $\cdots$ & $\cdots$ & $1.23^{+0.08^*}_{-0.08}$ & 2.2 \\ 
 
G257.71\_EUC\_2 	 & 	59.7149 	 & 	$-$49.2533 	 & 	$1.79^{+0.06}_{-0.14}$ 	& 	4.0 & $\cdots$ & $\cdots$ & $\cdots$ & $\cdots$ & $1.85^{+0.06}_{-0.06}$ & 3.5\\ 
 
G257.71\_EUC\_3 	 & 	59.7054 	 & 	$-$49.3269 	 & 	$2.69^{+0.11}_{-0.08}$  & 	5.3 & $\cdots$ & $\cdots$ & $\cdots$ & $\cdots$ & $2.91^{+0.07^*}_{-0.07}$ & 3.5 \\ 
& & & & & & & & & $2.45^{+0.09^*}_{-0.09}$ & 4.4 \\
 
\hline
\end{tabular}
\tablefoot{
 Columns description: \Planck\ + \Euclid overdensity name, \texttt{DETECTIFz} RA and Dec in degrees, photometric redshift (based on \Euclid\ \texttt{NNPZ}) and signal-to-noise ratio of the overdensity detection, \texttt{DIANAS} RA and Dec in degrees, photometric redshift (based on \Euclid\ \texttt{PHZ}) and S/N of the closest detection to the corresponding \texttt{DETECTIFz} overdensity, \texttt{PPM} photometric redshift (based on \Euclid\ \texttt{NNPZ}) and S/N of the closest detection(s) to the corresponding \texttt{DETECTIFz} overdensity. Absence of data is denoted with ellipsis. An asterisk in the \texttt{PPM} photometric redshift columns denote cases for which \texttt{PPM} and \texttt{DETECTIFz} redshifts have more than 1$\sigma$ discrepancy. In three cases, there are two fairly equidistant \texttt{PPM} counterparts (G222.05\_EUC\_3, G254.74\_EUC\_2 and G257.71\_EUC\_3).}
\end{table*}

\begin{figure*}[htbp!]
    \includegraphics[width=\hsize]{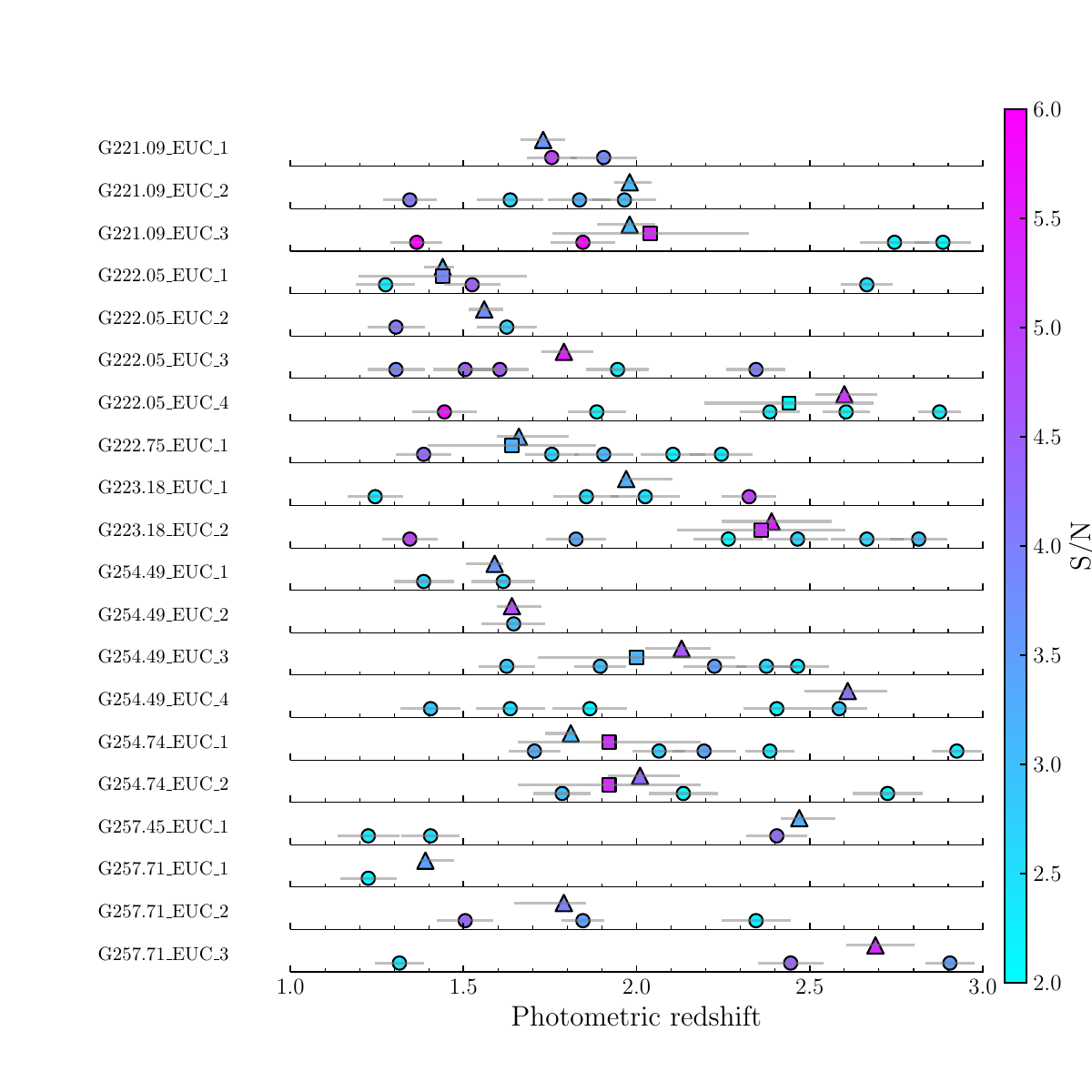}
    \caption{Triangles are photometric redshifts of the corresponding \texttt{DETECTIFz} detection, squares are photometric redshifts of overdensities detected by \texttt{DIANAS} in the angular radius of the \texttt{DETECTIFz} detection; circles are photometric redshifts of overdensities detected by \texttt{PPM} within the angular radius of the \texttt{DETECTIFz} detection. Colours represent S/N determined by each method.}
    \label{fig:ppm}
\end{figure*}

\section{\label{sect:merging}Merging of the two pairs of protocluster candidates?}

High SFRs of protocluster candidates could be the consequence of two structures merging. Indeed, we must consider the possibility that spatially close overdensities are in fact multiple dense regions of the same structure. This could be the case of the two pairs of \Euclid-\Planck\ protoclusters (G221.09\_EUC\_2; G221.09\_EUC\_3) and (G254.49\_EUC\_1; G254.49\_EUC\_2), since in both cases, photometric redshifts are compatible with the two subcomponents lying at the same distance on the line of sight. Their respective angular separations are \ang{;2.88;} and \ang{;4.12;}, which corresponds to physical distances of 1.5 and 2.2 Mpc. These distances are comparable to the virial radius of a galaxy cluster at $z=0$, so we might therefore wonder whether each of these pairs corresponds to one large protocluster \citep{muldrew2015}. 

We can try to quantify this point: let $M_{\rm pc}$ and $R_{\rm pc}$ be respectively the representative mass of a protocluster and the distance between two overdensities. The free-fall time $t_{\rm ff}$ for a protocluster at distance $R_{\rm pc}$ is then of order $(G\rho)^{-1/2} \simeq (GM_{\rm pc}/R_{\rm pc}^3)^{-1/2}$. We can consider that the two structures will have enough time to merge if this free fall time is lower than the Hubble time $t_{\rm H} = H(z)^{-1}$. This leads to a maximum physical separation between two structures that will eventually merge (Eq.~\ref{eq:ff_vs_hubble}):

\begin{equation}
    \label{eq:ff_vs_hubble}
    R_{\rm pc} < \left( \frac{GM_{\rm pc}}{H^2(z)} \right)^{1/3} \cdot
\end{equation}

Applying this criterion to (G221.09\_EUC\_2; G221.09\_EUC\_3) and (G254.49\_EUC\_1; G254.49\_EUC\_2) yields maximum proper separations of 0.9 and 1.3 Mpc respectively. Since these values have the same order of magnitude as the maximum separation, one should still consider that these two pairs might be two protoclusters. 

Under the hypothesis that (G221.09\_EUC\_2; G221.09\_EUC\_3) and (G254.49\_EUC\_1; G254.49\_EUC\_2) are single structures, their physical parameters $(M_*\:,\:{\rm SFR}\:,\:\Mh )$ would be as listed below (using the halo mass from \citetalias{shuntov2022}). These ‘merged' values would be more in line with the confirmed protoclusters from the literature, as seen in Figs.~\ref{fig:arthur_mh_vs_z} and \ref{fig:Mstar_vs_redshift_arthur}.
\vspace{-0.2cm}
\begin{eqnarray*}
    (1.8\expo{11} M_\odot\:,\: 520~ M_\odot \: {\rm yr}^{-1}\:,\:1.4\expo{13} M_\odot) &\textrm{ at }& z=1.98\\
    (3.4\expo{11} M_\odot\:,\:700 ~M_\odot \: {\rm yr}^{-1}\:,\:2.4\expo{13} M_\odot) &\textrm{ at }& z=1.6
\end{eqnarray*}

\end{appendix}

\label{LastPage}
\end{document}